\documentclass[11pt,letterpaper]{article}
\usepackage{sbm}
\usepackage{psmacros}
\usepackage{sos-macros}
\usepackage{tablefootnote}

\title{Optimal Robust Community Detection}
\author{
Samuel B. Hopkins\thanks{MIT EECS and CSAIL. Email: \texttt{samhop@mit.edu}. Supported by NSF award no. 2238080, MLA@CSAIL, FinTechAI@CSAIL, MIT Research Support Committee, MIT-Google Program for Computing Innovation.}
\and
Yiding Hua\thanks{ETH Zurich. Email: \texttt{yiding.hua@inf.ethz.ch}.}
\and
Sidhanth Mohanty\thanks{Northwestern University. Email: \texttt{sidhanth.mohanty@northwestern.edu}.}
\and
Amit Rajaraman\thanks{MIT EECS and CSAIL. Email: \texttt{amit\_r@mit.edu}.}
\and
David Steurer\thanks{ETH Zurich. Email: \texttt{dsteurer@inf.ethz.ch}.}
}
\date{\today}

\begin{document}
\hypersetup{pageanchor=false}
\maketitle

\begin{abstract}
We study community detection in the sparse two-community stochastic block model when an adversary may corrupt a small constant fraction of the edges of the observed graph.
For every choice of parameters above the Kesten--Stigum threshold and every $\eps > 0$, we give a nearly-linear-time algorithm which estimates the communities with accuracy within $\eps$ of the information-theoretically optimal accuracy, while tolerating a small constant fraction of adversarially corrupted edges.
Previously, efficient algorithms with optimal accuracy were known only in the absence of corruptions, while robust algorithms either achieved only weak recovery or approached optimal accuracy only in the limit of large signal-to-noise ratio.

Our algorithm is simple: approximate the second eigenvector of the nonbacktracking matrix, using a semidefinite program to cope with the corruptions, and use it to initialize belief propagation.
Along the way, we show that in the absence of corruptions, belief propagation initialized at this eigenvector achieves optimal accuracy, resolving an open problem of Zdeborov\'a in the two-community case.
The proof of this fact is short and independent of our robustness results---its key ingredient is a local characterization of the second eigenvector of the nonbacktracking matrix in terms of constant-depth nonbacktracking walks.
\end{abstract}

\thispagestyle{empty}
\setcounter{page}{0}
\newpage
\hypersetup{pageanchor=true}

\pagestyle{empty}

\setcounter{tocdepth}{2}
\tableofcontents

\thispagestyle{empty}

\newpage

\setcounter{page}{1}
\pagestyle{plain}

\section{Introduction}

The \emph{stochastic block model} is a long-studied random graph model for graphs with \emph{community structure}.
In the $n$-vertex, degree-$d$ two-community stochastic block model (SBM), $\text{SBM}(n,d,\lambda)$, we first assign to each vertex $u \in [n]$ a random community label $\bsigma_u^{\star} \sim \{\pm 1\}$, then add each possible edge $\{u,v\}$ independently with probability $\tfrac d n (1 + \lambda \bsigma_u^{\star} \bsigma_v^{\star})$ to form a graph $\bG$, where $\lambda \in (-1,1)$ captures the degree of (dis)affinity within communities.

The SBM arose independently in several research communities -- algorithms \cite{BCLS87, DF89, Bop87}, statistics and the social sciences \cite{HLL83}, statistical physics \cite{Sod02, DKMZ11prl, DKMZ11}, and mathematics \cite{BJR07}.
Across all of them, it is an important testbed for algorithmic and analytical techniques, since it provides a (relatively) simple model organism for a wide range of phenomena spanning clustering, spin glasses, social networks, and more.
In computer science, studying the SBM has led to new clustering algorithms \cite{DKMZ11, KMM+13, SKZ14, GV16, MS16, HS17, AS18}, new insights about average-case computational complexity \cite{DKMZ11, HS17, Hop18, BBK+21, KWB22}, novel understanding of the differences between random and semirandom input models in beyond-worst-case analysis \cite{MPW16}, and much more.

We focus on the sparse graph setting, with constant average degree $d$, where the most interesting algorithmic and probabilistic phenomena occur.
A long line of work has tackled the problem of estimating the community labels $\bsigma^*$ after observing the sparse graph $\bG$ via computationally efficient algorithms \cite{CO10, DKMZ11, KMM+13, Mas14, SKZ14, AS18, Abb18}.
A major theme has been the pursuit of algorithms which work for all $d,\lambda$ obeying $d \lambda^2 > 1$, the \emph{Kesten--Stigum threshold} -- this is the best that can be hoped for, since if $d \lambda^2 < 1$ estimating $\bsigma^{\star}$ with any constant advantage over oblivious guessing is information-theoretically impossible \cite{MNS15}.

Some highlights include \cite{Mas14, MNS18, BLM15}, which gave the first polynomial-time estimators to perform better than randomly guessing the labels (the \emph{weak recovery problem}) all the way down to the Kesten--Stigum threshold, \cite{MNS16, YP22} which give the first polynomial-time algorithm with information-theoretically optimal recovery rate (see also \cite{GP23} for more than two communities), and \cite{GV16, BMR21, DdNS21, DdHS23, DHLSS26, MRW24, LM22, MMV12, HWX16, CCdEIST23, BJKMMW24} which aim to design \emph{robust} algorithms to estimate $\bsigma^*$, that is, algorithms which can tolerate some adversarial corruption of the observed graph.

\parhead{Robust estimation of communities with optimal error rate}
We focus on the latter problem, estimating the community labels $\bsigma^*$ given an \emph{adversarial corruption} $G'$ of the SBM graph $\bG$.
We give the first polynomial-time algorithm which recovers the ground-truth labels with information-theoretically optimal accuracy for every $d$ and $\lambda$ while tolerating a constant fraction of adversarially corrupted edges.
More formally, for every $d,\lambda$, there is a number $\OPT_{d,\lambda} \in [0,1)$, defined as
\[
\OPT_{d,\lambda} = \lim_{n \rightarrow \infty} \sup_{\text{estimators $\wh{\bsigma}$}} \frac 1 n \E_{(\bsigma^{\star},\bG) \sim \SBM(n,d,\lambda)} | \iprod{\bsigma^{\star}, \wh{\bsigma}(\bG)} | \, .
\]
For any constant $d,\lambda$, $\OPT_{d,\lambda} < 1$ -- indeed, with high probability the graph $\bG$ will have isolated vertices, for which the graph structure gives no useful information about their label, though this is not the only reason for $\OPT < 1$; it is also strictly smaller than the size of the giant component.

Our main result is an algorithm which for every $d > 1$, and $\eps > 0$, and $\lambda \in [-1,1]$, obtains recovery accuracy $\OPT_{d,\lambda} - \eps$, even in the presence of a small constant fraction $\eta > 0$ of adversarial corruptions.

\begin{theorem}[Main theorem, follows from \Cref{thm:optimal-recovery,thm:robust-second-eigenvector}]
\label{thm:main}
For every $d > 1$, $\eps > 0$, and $\lambda \in [-1,1]$, there is $\eta > 0$ and a nearly-linear time algorithm which takes a graph $G$, returns a $\{\pm 1\}$-labeling $\wh{\bsigma}(G)$ of the vertices, and satisfies
\[
    \E_{(\bsigma^{\star},\bG) \sim \SBM(n,d,\lambda)} \min \Bigl\{  \frac 1 n \cdot | \iprod{\bsigma^{\star}, \wh{\bsigma}(G')} | \text{ s.t. } | E(\bG) \triangle E(G') | \leq \eta n \Bigr\} \geq \OPT_{d,\lambda} - \eps - o(1) \, .
\]
\end{theorem}

\begin{wrapfigure}{R}{0.37\textwidth}
\centering
\includegraphics[width=0.35\textwidth]{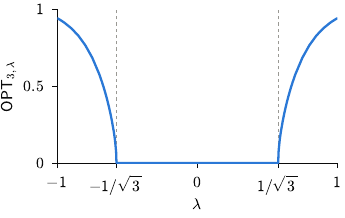}
\caption{$\OPT_{d,\lambda}$ plotted for $d=3$. The Kesten-Stigum threshold occurs at $1/\sqrt{3}$. Even at $\lambda = 1$ we have $\OPT_{d,1} < 1$ because of isolated vertices. The accuracy of our algorithm is described precisely by this curve.}
\label{fig:opt-d3}
\end{wrapfigure}
Here $E(\bG) \triangle E(G')$ denotes the symmetric difference between the edges of the graphs $\bG$ and $G'$, capturing that we allow an adversary to add or remove any $\eta n$ edges.

Prior efficient algorithms that obtained correlation $\OPT_{d,\lambda} - \eps$ but did not tolerate any adversarial corruption \cite{MNS16, YP22} were already a major achievement.
Prior efficient algorithms which do tolerate adversarial corruption either do not obtain information-theoretically optimal accuracy \cite{DdNS21, DdHS23, MRW24}, or obtain information-theoretically optimal accuracy only in the limit of large signal-to-noise ratio $d \lambda^2 \rightarrow \infty$ \cite{LM22, DHLSS26}, rather than for every $d,\lambda$ such that $d \lambda^2 > 1$---as we will see, the latter requires a substantially different algorithm.
Our algorithm runs in nearly-linear time.

\parhead{Why adversarial robustness?}
Algorithmic techniques designed for average-case models like the SBM can overfit to their models -- they may exploit features of the random input which have little to do with the structure they are meant to find, and fail when the input deviates even slightly from the model.
This concern is behind a long tradition of \emph{semirandom} models, in which the input is generated by a combination of random and adversarial choices \cite{BS95, FK01, Fei21}.
Aiming for robust algorithms often leads to fundamentally new algorithmic insights -- see, for instance, the revolution in high-dimensional algorithmic statistics sparked by the search for robust algorithms \cite{DK23}.
Our problem is especially ripe for this kind of development -- the only prior efficient algorithm which achieves accuracy $\OPT_{d,\lambda} - \eps$ for every $d,\lambda$  \cite{MNS16, YP22} is built on belief propagation, and prior to our work it was not known how to make belief-propagation-based algorithms robust to adversarial corruptions.

\parhead{Our algorithm: spectral initialization of BP}
Our algorithm is quite simple; we present an overview in \Cref{alg:non-robust}. 
The basic idea is to compute the second eigenvector of the \emph{nonbacktracking matrix} of the SBM graph $\bG$, or approximate this vector via a semidefinite program if given a corrupted graph $G'$, and then use this vector to initialize an algorithm called \emph{belief propagation} (BP).
BP aggregates the values that $y$ assigns to edges in a neighborhood of each vertex $u$ into a guess for the label of $u$.
Although our algorithm is simple, the interplay between the spectral initialization and BP is subtle, and our main technical contribution is an analysis which allows these two tools to be put together.

Belief propagation (BP) is a canonical algorithm for computing marginal probabilities in graphical models, based on a simple dynamic program.
Also known as the \emph{sum-product} algorithm, BP is fundamental enough to be commonly taught in undergraduate courses, and yet rigorous correctness analyses for BP on non-tree-like graphical models are very scarce. (BP is easily shown to compute marginal probabilities exactly in tree-structured graphical models.)
BP is the only tool we have which can lead to accuracy $\OPT_{d,\lambda}$ in the SBM, but because it is challenging to analyze, prior work \cite{MNS16} had to build a complicated initialization scheme to make the analysis tractable.
(We will discuss this further in \Cref{sec:techniques}.)

Zdeborov\'a \cite{ZdeborovaAIM17} therefore poses the open problem of whether the simple spectral initialization of BP we use here also leads to optimal-error community estimation.
We resolve this problem positively in the two-community case.
This gives us a nearly-linear-time algorithm and furnishes a rare example of a rigorous analysis of BP on a nontrivial family of non-tree graphs.
In the non-corrupted case, our analysis is not too complicated: if $\eta = 0$, the proof that \Cref{alg:non-robust} gives an optimal estimate of $\bsigma^{\star}$ will be done by the end of \cref{sec:optimal-recovery}.

\begin{figure}[t]
\refstepcounter{algocf}
\label{alg:non-robust}
\begin{tcolorbox}[title={Algorithm~\thealgocf: BP initialized at the nonbacktracking eigenvector}, before skip=0pt, after skip=0pt]
\textbf{Input:} a graph $G'$ on $[n]$ which differs from an SBM graph $\bG$ on at most $\eta n$ edges and parameter $t \geq 1$ depending on the target accuracy.
\begin{enumerate}
\item \emph{Compute the second eigenvector.}
Let $\vec E(\bG)$ be the set of directed edges of $\bG$, and let $B_{\bG} \in \{0,1\}^{\vec E(\bG) \times \vec E(\bG)}$ be the \emph{nonbacktracking matrix}, which has $(B_{\bG})_{u \to v, x \to w} = 1$ if $x = v$ and $w \neq u$, and $0$ otherwise.
If $\eta = 0$, compute a unit eigenvector $y$ of $B_{\bG}$ for eigenvalue of second-largest modulus.
If $\eta > 0$, use a semidefinite program to compute a vector $y$ with correlation $1-\eta'(\eta)$ to this unit eigenvector.
\item \emph{Round.}
Independently for each directed edge $e$, draw $\tau_e \in \{\pm 1\}$ with $\E \tau_e = \tanh \left(\beta \sqrt{|\vec E(G')|} \, y_e\right)$. ($\beta$ is a constant depending on $d$ and $\lambda$.)
\item \emph{Iterate BP update rule.}
Set $m^{(0)}_{u \to v} = \gamma \tau_{v \to u}$ for each directed edge $u \to v$ in $G'$, and for $s = 1,\ldots,t$ set $m^{(s)}_{u \to v} = \mathrm{BP}\bigl(\{ m^{(s-1)}_{w \to u} : w \in \partial u \setminus \{v\} \}\bigr)$, where $\partial u$ is the set of neighbors of $u$ in $G'$ and
\[
\mathrm{BP}(\{m_1,\ldots,m_k\}) = \frac{\prod_{i \leq k} (1 + \lambda m_i) - \prod_{i \leq k} (1 - \lambda m_i)}{\prod_{i \leq k} (1 + \lambda m_i) + \prod_{i \leq k} (1 - \lambda m_i)} \, .
\]
($\gamma$ is a constant depending on $d,\lambda$, and $t$.)
\item \emph{Round by signs.}
Output $\hat{\bsigma}_v = \operatorname{sign} \mathrm{BP}\bigl(\{ m^{(t)}_{w \to v} : w \in \partial v \}\bigr)$ for each vertex $v$.
\end{enumerate}
\end{tcolorbox}
\end{figure}

\subsection{Related work}
There is a vast literature on algorithms and information-theoretic limits for the SBM beyond the works we have already discussed -- see for instance \cite{CL15, CRV15, ABH16, MNS16b, BMNN16, MMV16, ZZ16, GMZZ17, DAM17, SM19, ABRS20, FC20, ACGP21,LMRRW24} and the survey \cite{Abb18}.
The Bethe Hessian of \cite{SKZ14} has also been analyzed rigorously: \cite{MRW24} construct its outlier eigenvectors above the Kesten--Stigum threshold, and Stephan and Zhu \cite{SZ25} show that its negative outliers detect the number of communities above the Kesten--Stigum threshold when $d \geq 2$.
Rigorous analyses of BP on graphs which are not trees are rare (though there are exceptions, e.g.\ \cite{COMV09}).
More broadly, algorithms for statistical estimation in the presence of adversarial contamination have flourished over the last decade; see \cite{DK23}.
We highlight a couple of works closely related to ours.

\parhead{Liu--Moitra robust community detection \cite{LM22}}
This work studies community detection with a stronger adversary than ours -- it can make an unbounded number of \emph{monotone} changes, adding edges inside communities and deleting edges connecting communities (when $\lambda > 0$).
The model is also nominally stronger in that $\eta n$ nodes can have their adjacency lists modified arbitrarily; however, \cite[Appendix~C]{DdHS23} show that this node-corruption model reduces to our edge-corruption model.

But, the \cite{LM22} approach requires $d \lambda^2$ to be at least a large constant, and achieves optimal accuracy only in the limit $d \lambda^2 \rightarrow \infty$.
More quantitatively, in our notation, the accuracy guarantee of \cite{LM22} even in the $\eta = 0$ case is $1 - (1-\OPT_{d,\lambda})^{1-o_{d\lambda^2 \rightarrow \infty}(1)}$.
By contrast, we obtain overlap $\OPT_{d,\lambda} - \eps$ for every $d,\lambda$ with $d \lambda^2 > 1$ and every $\eps > 0$.
This difference in guarantees corresponds to an important algorithmic difference.
The optimal accuracy as $d \lambda^2 \rightarrow \infty$ can be obtained by starting with a weak recovery algorithm and greedily flipping vertices whose labels disagree with a majority of their neighbors---this is sort of a ``poor man's belief propagation,'' whereas we need to use actual BP to achieve the $\OPT_{d,\lambda}$ accuracy guarantee.

\parhead{Robust approximate message passing \cite{IS24,IS25,IS26}}
A recent line of work by Ivkov and Schramm studies the robustness of approximate message passing (AMP), the dense-matrix analogue of BP, to adversarial corruption of a principal minor of the input matrix.
Especially relevant is \cite{IS26}, which studies $\mathbb{Z}_2$ synchronization, a dense analogue of the two-community block model.
Their algorithm has the same outline as ours: after spectral preprocessing, robustly estimate the top eigenvector to obtain a spectral initialization, then run the message-passing iteration, which turns out to be robust once its initialization is.
As is typical when generalizing from dense to sparse random matrix models, however, the analysis in our sparse setting is quite different.
Even in the non-corrupted setting, spectral initialization of the message-passing algorithm (in this case AMP) was well understood going back to \cite{MV21}.
In the sparse case the analogous result was not known until our work.

\parhead{Concurrent work}
In concurrent and independent work, Dong and McKenzie \cite{DM26} show that the outlier eigenvectors of the Bethe Hessian achieve weak recovery down to the Kesten--Stigum threshold, for every $d > 1$.
Our analysis of the corrected Bethe Hessian establishes a similar result.

\section{Technical overview}
\label{sec:techniques}
We introduce the key players in our algorithm -- the nonbacktracking matrix and belief propagation.
We then describe how our algorithm and analysis put these components together, and why it leads naturally to an adversarially robust algorithm.
The technical sections of our paper divide the implementation of this plan into two parts.
In \Cref{sec:optimal-recovery} we show that BP, initialized at the second eigenvector of the nonbacktracking matrix, succeeds at optimal community estimation.
Then in \Cref{sec:robust-estimation-of-second-eigenvector} we show how to estimate this vector in the presence of adversarial corruptions, with supporting lemmas proved in subsequent sections.

\parhead{Weak recovery and nonbacktracking walks}
\emph{Weak recovery} refers to finding $\wh{\bsigma}$ so that
\[
\frac 1 n \E_{(\bsigma^{\star},\bG) \sim \SBM(n,d,\lambda)} | \iprod{\wh{\bsigma},\bsigma^{\star}} | \geq \Omega(1) \, .
\]
This is information-theoretically possible so long as $d \lambda^2 > 1$, the \emph{Kesten--Stigum threshold} \cite{MNS15, Mas14, MNS18}.

The algorithm ``from the book'' for weak recovery in sparse SBM graphs assigns labels based on the second eigenvector of the \emph{nonbacktracking matrix}.
For a graph $G = (V,E)$, let $\vec{E}$ be the set of directed edges of $G$, so $|\vec{E}| = 2 |E|$.
The nonbacktracking matrix $B \in \R^{\vec{E} \times \vec{E}}$ has entries given by
\[
B_{(u,v),(w,z)} =
\begin{cases}
1 & \text{if } w = v \text{ and } z \neq u, \\
0 & \text{otherwise,}
\end{cases}
\]
The adjacency matrix of a sparse random graph has top eigenvectors dominated by unusually high degree vertices; they are not useful for finding global community structure in the graph.
The nonbacktracking matrix fixes this issue.
Concretely, \cite[Theorem~5]{BLM15} proves that with high probability, taking $\bxi$ to be an eigenvector associated to a second-largest-magnitude eigenvalue of $B$ (which w.h.p. has real entries), the algorithm which thresholds the value of $\sum_{u \sim v} \bxi_{(u,v)}$ at each vertex $v$ achieves weak recovery.
This eigenvector ``pokes out'' of the spectrum, creating a gap in the eigenvalue spectrum of $B$.
We will return to this second eigenvector $\bxi$ later, but for now we move on to the second key ingredient.

\parhead{Tree reconstruction and BP}
For our purposes, it is convenient to define BP through the lens of the \emph{broadcast tree reconstruction} problem.

\begin{definition}[Broadcasting on trees and tree reconstruction]
Let $T_{d,t}$ be a $d$-ary tree with height $t$.\footnote{Later we will instead take $T_{d,t}$ to be a Galton-Watson tree, that is, a tree where each vertex has a Poisson-distributed number of children.}
In the broadcast-on-trees process, a uniformly random label $\sigma_r \in \{ \pm 1\}$ is assigned to the root vertex $r$.
This label is then propagated down the tree by having each vertex independently choose to copy the label of its parent with probability $(1+\lambda)/2$ and otherwise flip the label, for a parameter $\lambda \in [-1,1]$.
At the end each vertex in the tree has a $\pm 1$ label.

The \emph{tree reconstruction problem} is to guess the value $\sigma_r$ given access to the signs $\sigma_t \in \{\pm 1\}^{d^t}$ of the leaf vertices.
In the \emph{noisy tree reconstruction} problem, the goal is to guess the value $\sigma_r$ given access to $\sigma_t'$, which is $\sigma_t$ with every bit flipped independently with some probability $\rho$.\footnote{Noisy tree reconstruction is often called \emph{robust} tree reconstruction, but in this section we reserve the word ``robust'' in this paper to refer to robustness against adversarial corruptions.}
\end{definition}

The Bayes-optimal algorithm for tree reconstruction is to output $\textrm{sign}( \E \sigma_r \, | \, \sigma_t)$, the sign of the expected value of the root conditioned on the leaves, or $\textrm{sign}( \E \sigma_r \, | \, \sigma_t')$ in the noisy case.
Let $\textrm{Tree-OPT}_{d,\lambda,\rho} \in [0,1]$ be the limit of the correlation of the root sign with this optimal guess as $t \rightarrow \infty$, which captures the optimal probability with which a tree reconstruction algorithm can correctly guess the sign of the root:
\[
\textrm{Tree-OPT}_{d,\lambda,\rho} = \lim_{t \rightarrow \infty} \E \Bigl[ \sigma_r \cdot \textrm{sign}\bigl( \E [\sigma_r \, | \, \sigma_t'] \bigr) \Bigr]
\]
where $\sigma_t' = \sigma_t$ when $\rho = 0$.

An important and surprising theorem we will need later is that the optimal probability of correctly guessing the root is actually independent of the noise rate $\rho$, as long as $\rho$ is not exactly $1/2$ (which would destroy all the signal about $\sigma_r$).

\begin{theorem}[\cite{MNS16, YP22}]
\label{thm:noisy-broadcast-intro}
  If $d > 1$, $\lambda \in (-1,1)$, and $\rho \neq 1/2$, then $\textrm{Tree-OPT}_{d,\lambda,\rho} = \textrm{Tree-OPT}_{d,\lambda,0}$.
\end{theorem}

BP is a dynamic program to compute the posterior expected value $\E[ \sigma_r \, | \, \sigma_t]$ or $\E[ \sigma_r \, | \, \sigma_t']$.
For concreteness we have described it explicitly in \Cref{alg:non-robust}, but actually all we will need to use about BP is that it computes this posterior expected value.

Algorithms for estimating $\bsigma^{\star}$ in the SBM with optimal error use BP as a subroutine, feeding it inputs which are not actually derived from a broadcast-on-tree process, but rather computed in some way from an SBM graph.
The key challenge is to analyze BP initialized with these ``off distribution'' inputs, by relating those inputs back to a tree process.

The basic relationship between the SBM and broadcasting on trees, pioneered in \cite{MNS15}, stems from taking a ``local view'' of the SBM.
For a vertex $u$, let $B_t(u)$ be a ball of radius $t$ in $\bG$ and look at the joint distribution $(\bsigma^{\star}_{B_t(u)}, B_t(u))$, where $\bsigma^{\star}_{B_t(u)}$ is the restriction of $\bsigma^{\star}$ to coordinates in $B_t(u)$.
It turns out that this distribution is very close to the broadcast-on-trees distribution.

\begin{lemma}[{Simplified version of \Cref{fact:local-law}, \cite[Proposition~36]{BLM15}}]
\label{lem:local-law-intro}
Let $f$ be a function mapping depth-$t$ trees whose vertices have $\pm 1$ labels to $[-1,1]$, let $d > 0$, and let $T_{d,t}$ be a depth-$t$ \emph{Galton-Watson tree} (where every vertex has $\Poi(d)$ children).
Then
\[
  \frac 1 n \sum_{u \in [n]} f\bigl(B_t(u), \bsigma^{\star}_{B_t(u)}\bigr) \longrightarrow \E f(T_{d,t}, \sigma)
\]
in probability as $n \to \infty$, where $\sigma$ is the broadcast labeling of $T_{d,t}$ with parameter $\lambda$, and $B_t(u)$ is rooted at $u$.
(With high probability $B_t(u)$ is a tree for all but $o(n)$ vertices $u$, so it does not matter how $f$ is defined on the others.)
\end{lemma}

A related fact, proved by \cite[Theorem~2.8]{MNS16}, is that $\textrm{Tree-OPT}_{d,\lambda} \geq \OPT_{d,\lambda}$, because even an estimator for $\bsigma^{\star}_u$ which is also given the true labels of all vertices at distance exactly $t$ from $u$ can do no better than tree reconstruction --- conditioned on those labels, the rest of the graph carries only vanishing information about $\bsigma^{\star}_u$, and by \Cref{lem:local-law-intro} the ball $B_t(u)$ together with its labels looks like a depth-$t$ broadcast tree.

\parhead{Idealized model for combining weak recovery and BP}
We are building towards an algorithm for estimating $\bsigma^{\star}$ which combines BP and nonbacktracking walks, and achieves correlation $\textrm{Tree-OPT}_{d,\lambda}$, the best possible in light of $\textrm{Tree-OPT}_{d,\lambda} \geq \OPT_{d,\lambda}$.
The basic idea is to initialize BP using the entries of the second eigenvector $\bxi$ of the nonbacktracking matrix.
At this point we have all the ingredients to describe a useful idealized model to explain why this kind of approach should work.

Imagine that in addition to $\bG$, our algorithm had access to an idealized weak recovery $\bsigma'$, which is $\bsigma^{\star}$ with every entry flipped independently with probability $\rho < 1/2$ (this is the block model with side information studied in \cite{MX16}).
The labeling $\bsigma'$ is itself a weak recovery of $\bsigma^{\star}$ because $\E |\iprod{\bsigma^{\star},\bsigma'}| \geq \Omega(n)$.

We could then ``boost'' the correlation of $\bsigma'$ with $\bsigma^{\star}$ by applying BP in a neighborhood of each vertex -- that is, for each vertex $u$, we could take $\wh{\bsigma}_u$ to be the sign of the output of BP initialized with $\bsigma'_{B_t(u)}$, the noisy signs of vertices in a depth-$t$ neighborhood (see also \cite{KMS16} on combining graph and node information).
\Cref{lem:local-law-intro} and \Cref{thm:noisy-broadcast-intro} together show that for this choice of $\wh{\bsigma}$, $\tfrac 1 n \E | \iprod{\wh{\bsigma},\bsigma^{\star}}| \rightarrow \OPT_{d,\lambda}$.

\parhead{The main challenge: the need for independence in the input to BP}
Departing from the idealized model, of course in reality we don't get access to $\bsigma'$ which is an independent noisy copy of the ground truth $\bsigma^{\star}$.
We would like to simulate such nice $\bsigma'$ to generate inputs to BP as before, but using a weak recovery algorithm we can actually run.

The major difficulty is that independence of the noise added to $\bsigma^{\star}$ to produce $\bsigma'$ in the idealized model is crucial to being able to apply \Cref{thm:noisy-broadcast-intro}.
Indeed, no analog of \Cref{thm:noisy-broadcast-intro} is possible if independent random flips of the leaf labels are replaced by adversarial flips -- the output of BP becomes useless in the presence of a vanishing fraction of adversarial bit-flips at the leaves (this is folklore; see, e.g., \cite{HL24}).
But the locations where the output $\bsigma_{\text{weak}}$ of a weak recovery algorithm differs from $\bsigma^{\star}$, which become the leaf-flip locations when we feed $\bsigma_{\text{weak}}$ to BP, will be correlated in an inscrutable way.

The prior work \cite{MNS16} tackles this issue by wrapping a weak recovery algorithm in a graph-splitting procedure which creates enough of the necessary independence to simulate $\bsigma'$ by ``repurposing'' some of the independent randomness in the edges of the SBM graph.
(This is sometimes called the ``cavity method.'')

Unlike \cite{MNS16}, we are looking for an algorithm that works in the presence of adversarial corruptions to the edge structure of the graph.
A first idea is to use the graph-splitting approach in conjunction with a robust weak recovery algorithm like \cite{DdNS21,MRW24} which still achieves weak recovery even with a small constant fraction of adversarial edge corruptions.
The problem is that the adversarial corruptions destroy the independence in the edges of the SBM graph which the graph-splitting approach relies on.
Thus, the edge-corrupting adversary might be able to arrange their corruptions so that the output $\bsigma_{\text{weak}}$ still has $|\iprod{\bsigma_{\text{weak}}, \bsigma^{\star}}| \geq \Omega(n)$, but for most vertices $u$, the locations among $B_t(u)$ where $\bsigma_{\text{weak}}$ and $\bsigma^{\star}$ disagree have a very non-random structure, which fools BP.
We need a different approach.

\parhead{Key idea: local characterization of the second eigenvector}
Let us return to the setting without adversarial corruptions to the graph, to design an algorithm for optimal-error community estimation without relying on graph splitting.
Our key insight is that a simple rounding scheme applied to the second eigenvector of the nonbacktracking matrix produces a sign vector which acts enough like the idealized $\bsigma'$ that we can use it directly to initialize BP.
In particular, we show that the locations of entries where $\bsigma^{\star}$ and the rounded vector differ are ``random enough'' -- this makes it unlike a generic weak recovery algorithm, where we wouldn't have control over the locations of these disagreements, only over their number.

The crucial technical ingredient is a new \emph{local} characterization of the second eigenvector of the nonbacktracking matrix.
Here ``local'' means that we show that the $(u,v)$ entry of the second eigenvector is essentially a function of a constant-radius local neighborhood of $(u,v)$ in $\bG$, together with the labels $\bsigma^\star$ in that local neighborhood.

\begin{lemma}[Local characterization of the second eigenvector, informal version of \Cref{lem:eigenvector-walk-score}]
\label{lem:local-char-intro}
Let $d \lambda^2 > 1$.
With high probability, the vector $W^{(\ell)} \in \R^{\vec{E}}$ given by aggregating $\bsigma^\star$ over depth-$\ell$ nonbacktracking walks
\[
W^{(\ell)}_{(u,v)} = \sum_{(u_0,\ldots,u_{\ell+1})} \bsigma^\star_{u_{\ell+1}} \, ,
\]
where the sum is over nonbacktracking walks with $u_0 = u$ and $u_1 = v$, has correlation $1 - o_{\ell \rightarrow \infty}(1)$ with the eigenvector of $B_{\bG}$ with eigenvalue of second-largest modulus.
\end{lemma}

\begin{figure}[ht]
\centering
\begin{tikzpicture}[
  vtx/.style={circle, draw, inner sep=1pt, minimum size=17pt},
  leaf/.style={circle, draw, fill=black!8, inner sep=1pt, minimum size=17pt},
  >={Stealth[length=2.2mm]}
]
\node[vtx] (u) at (0,0) {$u$};
\node[vtx] (v) at (2,0) {$v$};
\draw[->, thick] (u) -- (v);
\node[vtx] (a) at (4,1.3) {};
\node[vtx] (b) at (4,-1.3) {};
\draw (v) -- (a);
\draw (v) -- (b);
\node[leaf] (a1) at (6,2.0) {$+$};
\node[leaf] (a2) at (6,0.65) {$-$};
\node[leaf] (b1) at (6,-0.65) {$+$};
\node[leaf] (b2) at (6,-2.0) {$+$};
\draw (a) -- (a1);
\draw (a) -- (a2);
\draw (b) -- (b1);
\draw (b) -- (b2);
\node[align=left, anchor=west] at (6.6,0) {labels $\bsigma^\star$ at\\ depth $\ell$ from $v$};
\end{tikzpicture}
\caption{The walk score $W^{(\ell)}_{(u,v)}$, drawn for $\ell = 2$.
The nonbacktracking walks starting with the directed edge $(u,v)$ form a tree grown outwards from $v$, which does not include $u$, and $W^{(\ell)}_{(u,v)}$ sums the labels $\bsigma^\star$ at its leaves.}
\label{fig:walk-score}
\end{figure}
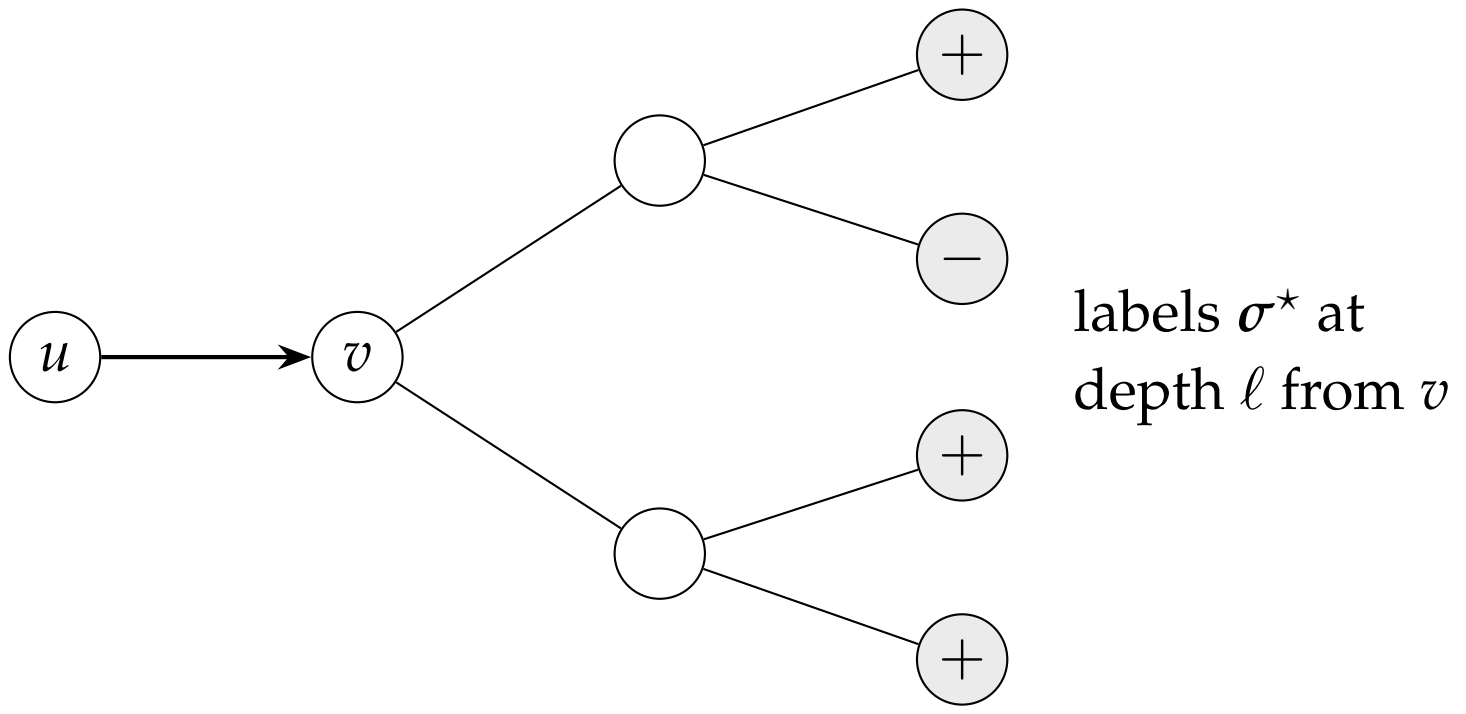

\subparhead{Intuition for \Cref{lem:local-char-intro}}
Because we are working with a sparse random graph and the non-Hermitian matrix $B$, proving \Cref{lem:local-char-intro} requires some care; for instance we cannot rely on an orthonormal basis of eigenvectors.
We are able to prove it by combining tools from \cite{BLM15}, especially  \Cref{lem:local-law-intro}.
But the intuition for why it ought to be true is simple.
Suppose we pretend that $B$ is a symmetric matrix with an orthonormal basis of eigenvectors such that:
\begin{itemize}
\item $B$ has constant size gaps in the spectrum: $|\lambda_1(B)| \geq |\lambda_2(B)| + \Omega(1) \cdot | \lambda_1(B)|$ and $|\lambda_2(B)| \geq |\lambda_3(B)| + \Omega(1) \cdot |\lambda_1(B)|$, where the eigenvalues are ordered by magnitude, and
\item the vector $\textrm{lift}(\bsigma^\star)\in \R^{\vec{E}}$ placing $\bsigma^\star_w$ on each directed edge $(x,w)$, has $\Omega(1)$ correlation with the second eigenvector, and is orthogonal to the top eigenvector.
\end{itemize}
This is roughly the picture established by \cite[Theorem~4]{BLM15}.
Then if we do $O(1)$ steps of power iteration initialized at $\textrm{lift}(\bsigma^\star)$, applying the spectral theorem we will arrive at a vector with correlation $0.99$ to the second eigenvector.
The walk score vector $W^{(\ell)}$ is exactly the result of $\ell$ steps of power iteration initialized at $\textrm{lift}(\bsigma^\star)$.

\subparhead{The heart of the argument -- consequences of \Cref{lem:local-char-intro} for spectral initialization of BP}
The local view of the second eigenvector means that we can pretend that our algorithm has direct access to the walk score vector $W^{(\ell)}$, for large constant $\ell$.
How can we use this vector to initialize BP in a way which simulates the idealized argument using $\bsigma'$?

Consider using BP to infer the label of vertex $u$ by initializing BP at each leaf $w$ of $B_t(u)$ with the sign of $W^{(\ell)}_{(\textrm{parent}(w),w)}$, where $\textrm{parent}(w)$ is the parent of $w$ in the tree $B_t(u)$.
The key idea is: when running BP on this depth-$t$ tree, the input $\{\textrm{sign}(W^{(\ell)}_{(\textrm{parent}(w),w)})\}_{w \in \textrm{leaves}(B_t(u))}$ essentially acts like $\bsigma'_{\textrm{leaves}(B_t(u))}$!

To see why, note that $W^{(\ell)}_{(\textrm{parent}(w),w)}$ only depends on the labels $\bsigma^\star$ at distance $\ell$ below $w$ -- that is, on the depth-$\ell$ subtree hanging off $w$, grown away from $u$ (see \Cref{fig:walk-score}).
So everything involved lives in $B_{t+\ell}(u)$, and by \Cref{lem:local-law-intro} we may pretend that this ball is a depth-$(t+\ell)$ broadcast tree.
In the broadcast process, once we condition on $\bsigma^\star_w$, the labels in the subtree below $w$ are independent of $\bsigma^\star_u$ and of the subtrees below all the other leaves, which do not overlap because of nonbacktracking.

All together, conditioned on the labels of the leaves of $B_t(u)$, the signs $\{ \textrm{sign}(W^{(\ell)}_{(\textrm{parent}(w),w)}) \}_{w \in \textrm{leaves}(B_t(u))}$ are independent, and by symmetry each one equals $\bsigma^\star_w$ with the same probability $1-\rho$.
This is exactly the noisy tree reconstruction problem with noise rate $\rho$.
And as long as $\rho < 1/2$, \Cref{thm:noisy-broadcast-intro} says that as $t \rightarrow \infty$, BP from these inputs is just as good as BP from the true labels of the leaves.
We just need to analyze the noise rate $\rho$, which is not so hard because the process which starts at a leaf $w$ and produces $W^{(\ell)}_{(\textrm{parent}(w),w)}$ is itself a broadcast process on a depth-$\ell$ tree.
We can show that $\rho$ stays bounded away from $1/2$ as $\ell \rightarrow \infty$ precisely when $d \lambda^2 > 1$ \cite{KS66b, EKPS00, MP03, MNS15}. 
This completes the non-robust version of our argument.
The formal version is captured by \Cref{thm:optimal-recovery}.

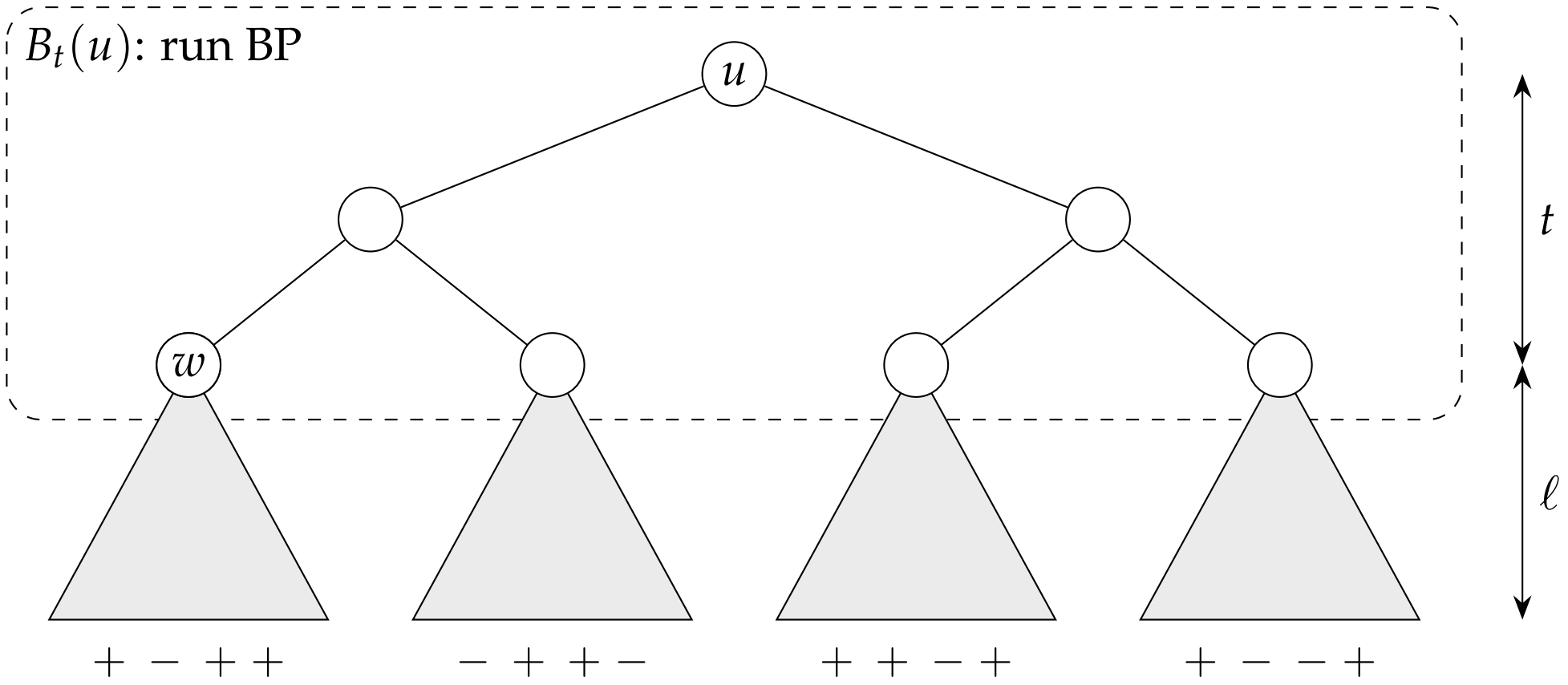
\begin{figure}[ht]
\centering
\begin{tikzpicture}[
  vtx/.style={circle, draw, fill=white, inner sep=1pt, minimum size=15pt},
  sub/.style={draw, fill=black!8},
  >={Stealth[length=2mm]}
]
\draw[dashed, rounded corners=8pt] (-6.0,0.55) rectangle (6.0,-2.85);
\node[anchor=north west] at (-6.0,0.55) {$B_t(u)$: run BP};
\node[vtx] (u) at (0,0) {$u$};
\node[vtx] (a) at (-3,-1.2) {};
\node[vtx] (b) at (3,-1.2) {};
\draw (u) -- (a);
\draw (u) -- (b);
\foreach \x/\p/\s in {-4.5/a/{+\;-\;+\;+}, -1.5/a/{-\;+\;+\;-}, 1.5/b/{+\;+\;-\;+}, 4.5/b/{+\;-\;-\;+}} {
  \draw[sub] (\x,-2.4) -- (\x-1.15,-4.5) -- (\x+1.15,-4.5) -- cycle;
  \node at (\x,-4.85) {\small $\s$};
  \draw (\p) -- (\x,-2.4);
  \node[vtx] at (\x,-2.4) {};
}
\node[vtx] at (-4.5,-2.4) {$w$};
\draw[<->] (6.5,0) -- node[right] {$t$} (6.5,-2.4);
\draw[<->] (6.5,-2.4) -- node[right] {$\ell$} (6.5,-4.5);
\end{tikzpicture}
\caption{The two-level ball $B_{t+\ell}(u)$, drawn for $t = 2$.
BP runs on the inner ball $B_t(u)$ (dashed).
Each leaf $w$ of $B_t(u)$ is initialized with $\textrm{sign}(W^{(\ell)}_{(\textrm{parent}(w),w)})$, which depends only on the labels $\bsigma^\star$ at the bottom of the depth-$\ell$ subtree hanging off $w$ (shaded).
These subtrees are disjoint, so conditioned on the labels of the leaves of $B_t(u)$, the inputs to BP are independent.}
\label{fig:two-level-ball}
\end{figure}

\parhead{Robustness of BP and the second-eigenvector initialization}
Now, what if instead of $\bG$, we get as input $G'$ for which $|E(\bG) \triangle E(G')| \leq \eta n$?
Our goal is to estimate $\bsigma^\star$ with accuracy $\OPT_{d,\lambda} - \eps$ while allowing $\eta(d,\lambda,\eps) > 0$.

A first observation is that for every $t$, only around $\eta d^t n$ vertices are distance $\leq t$ to an adversarially corrupted edge in $G'$.
So if we initialize depth-$t$ BP with the same vector in $\bG$ and $G'$, at all but $\eta d^t n$ vertices we will get the same output.
Furthermore, if the initializations were to differ at some $\delta n$ locations, it would introduce only an additional $\delta d^t n$ disagreements in the outputs.
We can tolerate this so long as $\delta, \eta \ll \eps d^{-t}$, which we can in turn ensure so long as $t$ is constant with respect to $n$, which it always is in the above arguments.

Thus, to robustify the whole algorithm, all we need to do is show that given $G'$ we can find a vector $y$ which has correlation $1- \delta^{O(1)}$ with the second eigenvector of $B_{\bG}$ (and hence also has correlation close to $1$ with the walk score vector $W^{(\ell)}$).
This suffices, because our procedure to round the second eigenvector to the signs we use to initialize BP produces sign vectors which differ on only $\delta n$ locations if given inputs which are $1-\delta^{O(1)}$ correlated.
(Ignoring dependence on $d$, which is constant anyway.)

As a first attempt, we could try directly using the second eigenvector of $B_{G'}$.
But it is well known that a small adversarial corruption of the graph can dramatically shift the eigenvectors of $B_{G'}$ (this argument is typically phrased for the adjacency matrix, but the phenomenon carries over to the nonbacktracking matrix).

\parhead{Robustly estimating the second eigenvector}
We design a new algorithm to obtain an estimate of the second eigenvector of $B_{\bG}$, captured by \Cref{thm:robust-second-eigenvector}.
For any $\delta > 0$ there is $\eta > 0$ such that our algorithm can find a vector whose correlation to the second eigenvector of $B_{\bG}$ is $1-\delta$, given a graph $G'$ with $\eta n$ edge corruptions.

Before describing our algorithm, we discuss the basic intuition for why enough information is still present after $\eta n$ edge corruptions to approximately recover the second eigenvector of $B_{\bG}$.
Let's briefly adopt the following simple toy model for how $B_{\bG}$ behaves.

\subparhead{Toy model and robust PCA}
Imagine that $M \in \R^{m \times m}$ is a \emph{symmetric} matrix with a spectral gap, having $\lambda_1(M) = 1$ and all other eigenvalues of $M$ at most $1/2$ in magnitude, and imagine further that the top eigenvector $v$ is delocalized; say $\|v\|_2 = 1$ and the maximum magnitude entry of $v$ satisfies $\|v\|_{\infty} \leq C/\sqrt{m}$.
$M$ is our stand-in for $B_{\bG}$ and $v$ is our stand-in for the second eigenvector of $B_{\bG}$.
When the adversary modifies $\eta n$ edges of the graph $\bG$, the matrix $B_{\bG}$ can change by at most $1$ in around $\eta n$ locations; in our toy model we can imagine we get access to a matrix $\widetilde{M}$ for which $M - \widetilde{M}$ is small in the entrywise $\ell_1$ norm $\|\cdot\|_{\textrm{ent},1}$.

It is well known that given access to such $\widetilde{M}$, it is possible to find a good approximation to $v$, and this can be done computationally efficiently \cite{GV16,MS16}; see also the robust PCA literature \cite{CLMW11,CSPW11}.
This can be done by solving a semidefinite programming (SDP) relaxation of the optimization problem
\[
\max \, \iprod{\widetilde{M}, xx^\top} \text{ s.t. } \|x\|_2 = 1 \, , \, \|x\|_{\infty} \leq C/\sqrt{m} \, .
\]
The SDP is closely related to the cut-norm/Grothendieck SDP.
The important feature of the analysis is that for any feasible solution, $\iprod{M - \widetilde{M}, xx^\top} \leq \|x\|_{\infty}^2 \|M - \widetilde{M}\|_{\textrm{ent},1}$ by H\"older's inequality, and this is preserved by the SDP relaxation.
The analysis relies on the spectral gap of $M$ and the delocalization of the top eigenvector.
Similar approaches were used in the context of the SBM in e.g. \cite{GV16}.

\subparhead{From the toy model back to the nonbacktracking matrix}
Our toy model fails to reflect the actual behavior of the nonbacktracking matrix $B_{\bG}$ in a few key ways, which our algorithm is designed to work around.
First, $B_{\bG}$ is not symmetric -- we could still try to use the above approach, but it would need a gap in the \emph{singular value} spectrum, which does not happen for $B_{\bG}$.
Second, we are interested in the second eigenvector of $B_{\bG}$, not the top one -- we will have to do something about the top eigenvalue so we can access the second one.
Third, the eigenvector we want to recover doesn't actually have $\ell_{\infty}$ norm $O(1/\sqrt{\dim(B_{\bG})})$.
We can see why by recalling the local characterization of the second eigenvector (\Cref{lem:local-char-intro}), a few unusually dense neighborhoods in the graph will contain a lot of nonbacktracking walks, driving up the magnitude of a few entries.
We address each of these issues in turn.

\subparhead{Passing to a symmetric matrix -- the corrected Bethe Hessian}
There is by now an established trick to obtain a symmetric matrix with a spectral gap whose eigenvectors can be transformed into good approximations to the eigenvectors of $B_{\bG}$.
This matrix is called the \emph{Bethe Hessian} \cite{SKZ14}.
It is actually a family of matrices, indexed by a real-valued parameter $t$
\[
H(t) = I - t A + t^2(D - I) \mcom
\]
where $A$ is the adjacency matrix and $D$ is the diagonal matrix of degrees.
This careful balancing of degree and adjacency matrices, for the right choices of $t$, regularizes the spectrum of $A$ to get rid of noisy localized eigenvectors, so that the community structure of $G$ can emerge.

More concretely, a technique pioneered by \cite{MRW24} shows that by choosing $t$ appropriately, when the underlying graph is $\bG \sim \SBM(n,d,\lambda)$, the matrix $H(t)$ has at most two negative eigenvalues, and the associated eigenvectors correspond roughly to the top and second eigenvectors of the nonbacktracking matrix.
Furthermore, $H(t)$ has spectral gaps and it cannot change much in entrywise $\ell_1$ norm when the adversary edits $\eta n$ edges, meaning we can hope to robustly recover those eigenvectors.
(Here ``correspond'' means that if we get access to the $H(t)$ eigenvectors we can in linear time construct good estimates of the corresponding eigenvectors of the nonbacktracking matrix, see \Cref{lem:robust-postprocessing}.)

Because we want to estimate the second eigenvector of $B_{\bG}$ to very high accuracy, we need the correspondence between $H(t)$'s eigenvectors and $B_{\bG}$'s to be quantitatively stronger than in prior work \cite{MRW24}.
To achieve this, we make two technical refinements on top of the ideas from \cite{MRW24}.
First, our algorithm computes a careful explicit rank-one correction to $H(t)$ to excise an uninformative eigenvector (the analog of the Perron-Frobenius eigenvector).
Second, to even better align the eigenvectors of $H(t)$ with those of $B_{\bG}$ we actually work with $A^{(\ell)} H(t) A^{(\ell)}$, where $A^{(\ell)}$ counts nonbacktracking walks of large constant length.
Importantly, these changes preserve the robustness to edge corruptions in the graph, if we first preprocess the graph by removing very high degree vertices, another standard trick \cite{FO05}.
The details are in \Cref{def:sandwiched-bethe-hessian,lem:bethe-hessian-structure,lem:robust-matrix-edit-stability}.

\subparhead{Tolerating approximate delocalization}
We turn to the last technical issue making our toy model differ from reality -- large entries of the second eigenvector.
(The solution would apply to $B_{\bG}$ or $H(t)$ or our modified version thereof so we don't distinguish carefully among them now.)
It turns out that the eigenvector we want to find still has a good enough delocalization property -- its $4$-norm is small.
Concretely, if we normalize so $\|v\|_2 = 1$, we are able to show that $\|v\|_4^4 \leq O(1/n)$, with high probability over $\bG$.
We adapt the standard robust PCA-style arguments to show that this $\ell_4$-style delocalization is still enough to approximately recover $v$ using an SDP relaxation of
\[
\max \iprod{ - A^{(\ell)} (H(t) - (\text{rank-one correction})) A^{(\ell)}, xx^\top} \text{ s.t. } \|x\|_2 = 1 \, , \, \|x\|_4^4 \leq O(1/n) \, ,
\]
even in the presence of adversarial edge corruptions.
The argument uses that $A^{(\ell)} (H(t) - \text{rank-one correction}) A^{(\ell)}$ is stable under edge corruptions in a slightly stronger norm than entrywise $\ell_1$.
Finally, we also show that we can solve this SDP in nearly-linear time via matrix multiplicative weights \cite{AK07}.

The details are in \Cref{lem:robust-matrix-edit-stability,lem:robust-alt-sdp}.

\section{Preliminaries}

\subsection{Stochastic block model and community detection}
\label{sec:prelim-model}

Fix $d>0$ and $\lambda\in(-1,1)$.
For all sufficiently large $n$, in the symmetric two-community stochastic block model $\SBM(n,d,\lambda)$, the hidden labels $\bsigma^\star\in\{\pm1\}^n$ are independent and uniform.
Conditional on the labels, each unordered pair of distinct vertices is joined independently with probability
\[
    \frac dn\bigl(1+\lambda\bsigma_u^\star\bsigma_v^\star\bigr).
\]
We work above the Kesten--Stigum threshold, namely $d\lambda^2>1$.
The problem of estimating $\bsigma^\star$ from the graph $G$ is called \emph{community detection}.
The performance of an estimator $\widehat{\bsigma}\in\{\pm1\}^n$ for community detection is measured by
\[
    \ov(\widehat{\bsigma},\bsigma^\star)
      =\frac1n\left|\left\langle
          \widehat{\bsigma},\bsigma^\star
        \right\rangle\right|.
\]
Let $\OPT_{d,\lambda}$ denote the information-theoretically optimal asymptotic overlap.
Section~\ref{sec:prelim-trees} recalls its equivalent characterization by reconstruction on a Poisson broadcast tree.

As a convention, random objects are written in boldface and their realizations are plaintext.

\subsection{Nonbacktracking matrix}
\label{sec:prelim-nonbacktracking}

For a graph $G$, let $\partial v$ denote the set of neighbors of a vertex $v$, let $\vec E(G)$ be its set of directed edges, containing both orientations $u\to v$ and $v\to u$ of every edge $\{u,v\}$.
The \emph{nonbacktracking matrix} \cite{Has89} $B_G\in\{0,1\}^{\vec E(G)\times\vec E(G)}$ has entries
\begin{equation*}
    (B_G)_{u\to v,\,x\to w}=\Ind[x=v]\,\Ind[w\ne u],
\end{equation*}
so a directed edge $u\to v$ may be followed by any edge leaving $v$ other than its reversal $v\to u$.
Equivalently, for $z\in\R^{\vec E(G)}$,
\begin{equation*}
    (B_Gz)_{u\to v}
      =\sum_{w\in\partial v\setminus\{u\}}z_{v\to w}.
\end{equation*}

\subsection{Broadcast trees, belief propagation, and optimal recovery}
\label{sec:prelim-trees}

The optimal overlap $\OPT_{d,\lambda}$ achieved by any community-detection algorithm is characterized by \emph{reconstruction on a Poisson broadcast tree} \cite[Theorem~2.8]{MNS16}, \cite[Theorem~38]{YP22}.

\begin{definition}[Broadcast process on Galton--Watson tree]
    \label{def:broadcast-tree}
    Fix $d>0$ and $\lambda\in(-1,1)$.
    Let $\bcalT_d$ be the rooted Galton--Watson tree \cite{AN72} in which every vertex has an independent $\Poi(d)$ number of children.
    Give its root $o$ a uniformly random spin $\bx_o\in\{\pm1\}$, and let each child independently copy its parent's spin with probability $(1+\lambda)/2$.
    Write $(\bcalT_{d,\lambda},\bx)$ for the resulting tree together with the vector $\bx$ of spins.

    Let $\bx_t$ consist of the spins at the depth-$t$ vertices.
    The depth-$t$ root posterior mean conditioned on the tree itself and the depth-$t$ spins is
    \[
        \bM_t=\E[\bx_o\mid (\bT,\bx_t)].
    \]
\end{definition}

The problem of computing $\bx_o$ given the tree $\bT$ and leaf spins $\bx_t$ is called \emph{broadcast tree reconstruction}.
The canonical algorithm for broadcast tree reconstruction is \emph{belief propagation}.

\begin{definition}[Belief propagation]
    \label{def:tree-bp}
    The \emph{belief propagation} (BP) algorithm on a depth-$t$ tree with $+1/-1$ labels $x_u$ on its leaves $u$ is defined as follows.
    Initialize every directed child-to-parent edge $u\to v$ at depth $t$ with a \emph{message} $m_{u\to v}=x_u$.
    Moving upward, after $u$ receives the messages from its children, it sends its parent $v$
    \begin{equation}
        m_{u\to v}
        =\mathrm{BP}\parens*{\{m_{w\to u}:w\in\operatorname{ch}(u)\}}
        :=\frac{
          \prod_{w\in\operatorname{ch}(u)}(1+\lambda m_{w\to u})
          -\prod_{w\in\operatorname{ch}(u)}(1-\lambda m_{w\to u})
        }{
          \prod_{w\in\operatorname{ch}(u)}(1+\lambda m_{w\to u})
          +\prod_{w\in\operatorname{ch}(u)}(1-\lambda m_{w\to u})
        },
        \label{eq:tree-bp-magnetization}
    \end{equation}
    where $\operatorname{ch}(u)$ is the set of children of $u$.
    Apply the same rule at the root to obtain $m_o$.
    The denominator is positive because $|\lambda|<1$ and every message lies in $[-1,1]$.
\end{definition}

The following fact is standard.

\begin{fact}[Exactness of tree BP]
    \label{fact:tree-bp-exactness}
    For every initialization $x_t$ on the vertices at depth $t$, $m_o$ is the exact posterior mean of the root.
    Equivalently, $\boldm_o=\bM_t$.
\end{fact}

\parhead{Robust tree reconstruction and Bayes-optimal algorithms.}
As pioneered by Mossel, Neeman, and Sly \cite{MNS16}, the algorithmic question of optimal recovery in block models is closely related to the problem of broadcast tree reconstruction after additional independent noise is applied to the leaves.
To be consistent with the literature, we call this the \emph{robust tree reconstruction} problem \cite{JM04}.
However, importantly, this problem does not involve \emph{adversarial} corruption of the observations in any way, making it quite distinct from the robust block model problem studied in this paper.
But, we will still make extensive use of this random-leaf-noise version of broadcast tree reconstruction.

\begin{definition}[Robust tree reconstruction]
    Run the broadcast process through generation $t$.
    Independently at every boundary vertex $v$, retain $\bx_v$ with probability $\gamma\in(0,1]$ and otherwise replace it by an independent uniform spin; denote the resulting noisy spin by $\boldsymbol\tau_v$, and let $\boldsymbol\tau_t$ consist of all these noisy spins at the depth-$t$ vertices.
    The root posterior mean conditioned on the noisy spins is
    \[
        \bM_t^{(\gamma)}
          =\E[\bx_o\mid (\bT,\boldsymbol\tau_t)].
    \]
    Robust reconstruction asks how well the root can be recovered as $t\to\infty$.
\end{definition}

\begin{remark}[BP for robust reconstruction]
    Belief propagation also computes $\bM_t^{(\gamma)}$ exactly; only the initialization changes.
    In \Cref{def:tree-bp}, the message $m_{u\to v}$ that a vertex $u$ sends its parent $v$ is the posterior mean of $x_u$ given the observations in the subtree below $u$, under a uniform prior on $x_u$.
    For a depth-$t$ vertex $u$, the only observation is $\tau_u$, which equals $x_u$ with probability $(1+\gamma)/2$; hence $\E[x_u\mid\tau_u]=\gamma\tau_u$.
    So we initialize $m_{u\to v}=\gamma\tau_u$ at every depth-$t$ vertex $u$, in place of $m_{u\to v}=x_u$, and run the same update \eqref{eq:tree-bp-magnetization}.
    The argument of \Cref{fact:tree-bp-exactness} then gives $m_o=\bM_t^{(\gamma)}$.
\end{remark}

Given the tree, $\bx_{t+1}$ is obtained from $\bx_t$ by one more step of broadcasting, so $\E|\bM_{t+1}|\le\E|\bM_t|$.
Define
\[
    \OPT_{d,\lambda}=\lim_{t\to\infty}\E|\bM_t|.
\]
We justify momentarily the collision of notation between the tree reconstruction problem and the previously-defined optimal overlap for the stochastic block model.

Following Mossel, Neeman, and Sly \cite{MNS16}, who proved this when $d\lambda^2$ is sufficiently large, Yu and Polyanskiy showed that the noise in robust reconstruction does not change this limit.\footnote{Yu and Polyanskiy work with log-likelihood ratios.
In their notation, the tree has offspring law $\Poi(d)$ and edge channels $\mathrm{BSC}_{(1-\lambda)/2}$, and the noisy leaves are observed through $\mathrm{BSC}_{(1-\gamma)/2}$, which is nontrivial since $\gamma>0$.
Their Theorem~34 says that the root log-likelihood ratio $\bR_t$ has the same limiting law as $t\to\infty$ with noisy leaves as with exact leaves.
Since $|\bM_t|=\tanh(|\bR_t|/2)$ is a bounded continuous function of $\bR_t$, \eqref{eq:noisy-boundary-optimality} follows \cite[Remark~8]{YP22}.}

\begin{fact}[Robust reconstruction {\cite[Theorem~34 and Proposition 48(7)]{YP22}}]
    \label{fact:yp-robust-reconstruction}
    For every $d>0$, $\lambda\in(-1,1)$, and $\gamma\in(0,1]$,
    \begin{equation}
        \lim_{t\to\infty}\E|\bM_t^{(\gamma)}|=\OPT_{d,\lambda}.
        \label{eq:noisy-boundary-optimality}
    \end{equation}
    Further, for any $t \ge 0$ and $0\le\gamma' \le \gamma\le 1$, we have $\E\abs*{ \bM_t^{(\gamma)} } \ge \E\abs*{ \bM_t^{(\gamma')} }$.
\end{fact}

The following theorem relates the (robust) broadcast tree reconstruction problem to the problem of recovering the community labels in the stochastic block model.

\begin{theorem}[Tree characterization of optimal recovery {\cite{MNS16,YP22}}]
    \label{thm:optimal-recovery-tree-characterization}
    For $(\bG,\bsigma^\star)$ distributed according to the stochastic block model with parameters $d,\lambda$, and every estimator $\widehat{\bsigma}(\bG)$, and every $\varepsilon>0$,
    \[
       \Pr\bigl(\ov(\widehat{\bsigma},\bsigma^\star)
                    >\OPT_{d,\lambda}+\varepsilon\bigr)\longrightarrow0.
    \]
    Conversely, for every $\varepsilon>0$ there is an estimator, computable in time $n^{1+o(1)}$, whose overlap is at least $\OPT_{d,\lambda}-\varepsilon$ with high probability.
\end{theorem}

The fact that no estimator in the block model setting performs better than tree reconstruction is Mossel, Neeman, and Sly \cite[Theorem~2.8]{MNS16}.
Yu and Polyanskiy showed there is an estimator for the communities in the block model which obtains overlap $\OPT_{d,\lambda}$, for every choice of $d,\lambda$, by improving the analysis of the algorithm of Mossel, Neeman, and Sly \cite[Theorem~38]{YP22}.

\section{Initialization of belief propagation at the second eigenvector of the nonbacktracking matrix}
\label{sec:optimal-recovery}

In this section we show that belief propagation, initialized from the second eigenvector of the nonbacktracking matrix, gives an optimal recovery algorithm in the stochastic block model.
We also show that this is robust in two senses:
\begin{enumerate}
  \item It suffices to initialize BP at any vector close to this eigenvector, and
  \item Running BP on a corrupted graph (with a small constant fraction of adversarial edge corruptions) also produces a near-optimal solution, with a loss that depends on the fraction of corrupted edges.
\end{enumerate}

We will need to compare vectors indexed by the directed edges of different graphs on $[n]$.
To make this meaningful, we view $\R^{\vec E(G)}$, for every graph $G$ on $[n]$, as the subspace of $\R^{n^2-n}$ consisting of the vectors indexed by ordered pairs of distinct vertices that vanish off $\vec E(G)$.
Inner products and distances between vectors on the directed edges of different graphs are taken in $\R^{n^2-n}$, extending vectors in $\R^{\vec E(G)}$ by padding with zeros as necessary.

\begin{definition}[BP from an initialization]
    \label{def:bp-from-score}
    Given a graph $G$, an initialization $Y\in\R^{\vec E(G)}$, an integer $t\ge1$, and $\gamma\in(0,1)$, the procedure $\mathsf{BP}_{t,\gamma}(G,Y)$ does the following.
    \begin{enumerate}
        \item Independently for each directed edge $e$, draw $\boldsymbol\tau_e\in\{\pm1\}$ with $\E\boldsymbol\tau_e=\tanh Y_e$.
        \item Set $m^{(0)}_{u\to v}=\gamma\boldsymbol\tau_{v\to u}$ and, for $s=1,\dots,t$, $m^{(s)}_{u\to v}=\mathrm{BP}\bigl(\{m^{(s-1)}_{w\to u}: w\in\partial u\setminus\{v\}\}\bigr)$.
        \item Output $\widehat{\bsigma}_v=\operatorname{sign}\bigl(\mathrm{BP} (\{m^{(t)}_{w\to v}:w\in\partial v\})\bigr)$.
    \end{enumerate}
\end{definition}

The first, randomized step is a technical convenience that makes it easier to apply the results of \cite{YP22}; we expect that with a more delicate analysis a deterministic initialization procedure mapping $Y_e$ to initial messages would also work.

Our main theorem for this section is:

\begin{theorem}[BP from the second eigenvector is optimal, robustly]
    \label{thm:optimal-recovery}
    Let $d\lambda^2>1$ and $(\bG,\bsigma^\star)\sim\SBM(n,d,\lambda)$, and let $\bxi$ be a unit eigenvector for the eigenvalue of $B_{\bG}$ of second-largest modulus.
    For every $\varepsilon>0$ there are $t\ge1$, $\gamma\in(0,1)$, $\beta>0$, and $\eta>0$ such that the following holds with high probability.
    For every graph $G'$ on $[n]$ with $|E(\bG)\triangle E(G')|\le\eta n$ and every unit $y\in\R^{\vec E(G')}$ such that $\langle y,\bxi\rangle^2\ge1-\eta$, the output $\widehat{\bsigma}$ of $\mathsf{BP}_{t,\gamma}\bigl(G',\beta\sqrt{|\vec E(G')|}\, y\bigr)$ satisfies
    \[
        \E\,\ov(\widehat{\bsigma},\bsigma^\star)\ge\OPT_{d,\lambda}-\varepsilon,
    \]
    where the expectation is over the random rounding in $\mathsf{BP}_{t,\gamma}$, and the procedure uses $O_{d,\lambda,\varepsilon}(n)$ arithmetic operations.
\end{theorem}

We state the main lemmas from which \Cref{thm:optimal-recovery} follows and then put them together to prove \Cref{thm:optimal-recovery}.

The most important lemma establishes a \emph{local} characterization of the second eigenvector of the nonbacktracking matrix -- it shows that the $(u,v)$ entry of the second eigenvector is essentially a function of the ground-truth labels $\bsigma^\star$ in a constant-radius neighborhood of the edge $(u,v)$.

Setting up some notation, for a blockmodel graph $\bG$ with ground-truth labeling $\bsigma^\star$ and for $\ell\ge0$, the \emph{walk score} $W^{(\ell)}\in\R^{\vec E(\bG)}$ is
\[
    W^{(\ell)}_{u\to v}=(d\lambda)^{-\ell}\sum_{(u_0,\dots,u_{\ell+1})}\bsigma^\star_{u_{\ell+1}},
\]
where the sum is over nonbacktracking walks in $\bG$ with $u_0=u$, $u_1=v$, and $u_{i+1}\ne u_{i-1}$ for $1\le i\le\ell$.
The normalization by $(d \lambda)^{-\ell}$ is chosen so that when translated to the broadcast tree, $W^{(\ell)}$ is an unbiased estimator of $\bsigma^\star_v$.

\begin{restatable}[The eigenvector is close to a walk score]{lemma}{lemeigenvectorwalkscore}
    \label{lem:eigenvector-walk-score}
    Let $d\lambda^2>1$, let $(\bG,\bsigma^\star)\sim\SBM(n,d,\lambda)$, and let $\bxi$ be as in \Cref{thm:optimal-recovery}.
    For every fixed $\ell\ge0$, $\langle\bxi,W^{(\ell)}\rangle^2/\|W^{(\ell)}\|^2\to1-O\parens*{(d\lambda^2)^{-\ell}}$ in probability.
\end{restatable}

We prove the above in \Cref{sec:eigenvector-walk-score}.


The next lemma shows that these walk scores provide a good initialization for BP on the SBM graph.
Some notation is needed:
\begin{itemize}
\item For a graph $G$, a vector $y\in\R^{\vec E(G)}$, $t\ge1$, and $\gamma\in(0,1)$, and a vector $\bsigma^* \in \{ \pm 1 \}^n$ which will be clear from context, let $\mathrm{Ov}_{t,\gamma}(G,y)$ be the expected overlap $\E\,\ov(\widehat{\bsigma},\bsigma^\star)$ of the output $\widehat{\bsigma}$ of $\mathsf{BP}_{t,\gamma}\bigl(G,\beta\sqrt{|\vec E(G)|}\,y\bigr)$, where the expectation is over the randomized rounding.
\item let $\beta=\frac12\sqrt{1-1/(d\lambda^2)}$, and
\item returning to the broadcast tree \Cref{def:broadcast-tree}, let $\mathbf W_\ell=(d\lambda)^{-\ell}\sum_u\bx_u$ be the sum over the spins of the depth-$\ell$ vertices of the broadcast tree, and let $\gamma_\ell=\E\bigl[\bx_o\tanh\bigl(\beta\mathbf W_\ell/(\E\mathbf W_\ell^2)^{1/2}\bigr)\bigr]$. Equivalently, if $\boldsymbol\tau\in\{\pm1\}$ is drawn with $\E[\boldsymbol\tau\mid\mathbf W_\ell]=\tanh\bigl(\beta\mathbf W_\ell/(\E\mathbf W_\ell^2)^{1/2}\bigr)$, which is how $\mathsf{BP}_{t,\gamma}$ rounds a score, then $\gamma_\ell=\E[\bx_o\boldsymbol\tau]$ is the correlation of the root spin with the rounded, normalized depth-$\ell$ walk score.
\end{itemize}

\begin{restatable}[BP initialized from constant-depth walk score is near-optimal]{lemma}{lembpwalkscore}
    \label{lem:bp-walk-score}
    Let $d\lambda^2>1$ and $(\bG,\bsigma^\star)\sim\SBM(n,d,\lambda)$.
    For every $\varepsilon>0$ there is $t=t(d,\lambda,\varepsilon)\ge1$ such that for every $\ell\ge0$, with high probability, $\mathrm{Ov}_{t,\gamma_\ell}\bigl(\bG,W^{(\ell)}/\|W^{(\ell)}\|\bigr) \ge\OPT_{d,\lambda}-\varepsilon$ and the same holds for $-W^{(\ell)} / \|W^{(\ell)}\|$.
\end{restatable}

We prove the above in \Cref{sec:bp-walk-score}.
Because $\ell$ is constant, we can prove \Cref{lem:bp-walk-score} by comparing each tree-like neighborhood of $\bG$ to a broadcast tree.

The next two lemmas, proved in \Cref{sec:bp-stability}, establish the robustness aspects of \Cref{thm:optimal-recovery}.
Both can be established by simple direct analysis of the BP operation.

\begin{restatable}[BP is stable under changes to the initialization]{lemma}{lembpscorestability}
    \label{lem:bp-score-stability}
    Let $d\lambda^2>1$ and $(\bG,\bsigma^\star)\sim\SBM(n,d,\lambda)$.
    For every $t\ge1$ there is $C=C(d,\lambda,t)>0$ such that with high probability, for all $\gamma\in(0,1)$ and all $y,y'\in\R^{\vec E(\bG)}$,
    \[
        \bigl|\mathrm{Ov}_{t,\gamma}(\bG,y)-\mathrm{Ov}_{t,\gamma}(\bG,y')\bigr|
        \le C\|y-y'\|.
    \]
\end{restatable}

\begin{restatable}[BP is stable under edge edits]{lemma}{lembpeditstability}
    \label{lem:bp-edit-stability}
    Let $d\lambda^2>1$ and $(\bG,\bsigma^\star)\sim\SBM(n,d,\lambda)$.
    For every $t\ge1$ there is $C=C(d,\lambda,t)>0$ such that with high probability the following holds for all $\eta,\gamma\in(0,1)$.
    Let $G'$ be a graph on $[n]$ with $|E(\bG)\triangle E(G')|\le\eta n$, let $y\in\R^{\vec E(G')}$ have $\|y\|\le1$, and let $y|_{\vec E(\bG)}$ be the restriction of $y$ to $\vec E(\bG)$, that is, the vector in $\R^{\vec E(\bG)}$ that agrees with $y$ on the directed edges of both graphs and vanishes elsewhere.
    Then $\bigl|\mathrm{Ov}_{t,\gamma}(G',y)-\mathrm{Ov}_{t,\gamma}(\bG,y|_{\vec E(\bG)})\bigr| \le C\sqrt\eta$.
\end{restatable}

The proofs of \Cref{lem:bp-score-stability,lem:bp-edit-stability} use only that with high probability $\bG$ has $O_d(n)$ edges and its radius-$(t+1)$ neighborhoods are not too large on average.

We are ready to prove \Cref{thm:optimal-recovery}.

\begin{proof}[Proof of \Cref{thm:optimal-recovery}]
We need to show how to choose $t,\gamma$ and $\eta > 0$ such that there is a high-probability event under $(\bG,\bsigma^\star) \sim \SBM(n,d,\lambda)$ for which $\mathrm{Ov}_{t,\gamma}(G',y) \geq \OPT_{d,\lambda} - \varepsilon$, for every $G',y$ allowed by the theorem statement.

First, for any $t,\gamma$ we eventually choose, by \Cref{lem:bp-edit-stability} there is a high probability event such that for every $G',y$ allowed by the theorem, we have
\[
  \mathrm{Ov}_{t,\gamma}(G',y) \geq \mathrm{Ov}_{t,\gamma}(\bG,y|_{\vec E(\bG)}) - O_{d,\lambda,t}(\sqrt\eta) \, ,
\]
where $y|_{\vec E(\bG)}$ is the restriction of $y$ to $\vec E(\bG)$, as in \Cref{lem:bp-edit-stability}.

Next, for any $\ell$ we eventually choose, write $\widehat W=W^{(\ell)}/\|W^{(\ell)}\|$, where $W^{(\ell)}$ is the walk score vector of the SBM graph $\bG$.
By \Cref{lem:bp-score-stability}, applied to the pairs of vectors $y|_{\vec E(\bG)}$ and $\pm\widehat W$, there is a high-probability event on which, for every $G',y$ allowed by the theorem,
\[
  \mathrm{Ov}_{t,\gamma}(\bG,y|_{\vec E(\bG)}) \geq \min_{s=\pm1}\mathrm{Ov}_{t,\gamma}(\bG,s\widehat W) - O_{d,\lambda,t}\Bigl(\min_{s=\pm1}\|y|_{\vec E(\bG)}-s\widehat W\|\Bigr)\,.
\]

Next, by \Cref{lem:eigenvector-walk-score}, with high probability $\min_{s=\pm1}\|\bxi-s\widehat W\|^2=2-2|\langle\bxi,\widehat W\rangle|\le2(d\lambda^2)^{-(\ell+1)}+o(1)$, where $o(1)$ tends to $0$ as $n\to\infty$ with all other parameters fixed.
Since $\|y\|\le1$ and $\langle y,\bxi\rangle^2\ge1-\eta$, also $\min_{s=\pm1}\|y-s\bxi\|^2\le2-2\sqrt{1-\eta}\le2\eta$.
As vectors in $\R^{n^2-n}$, $y|_{\vec E(\bG)}$ is obtained from $y$ by setting its entries off $\vec E(\bG)$ to zero, and $\bxi$ and $\widehat W$ vanish off $\vec E(\bG)$, so $\|y|_{\vec E(\bG)}-s\widehat W\|\le\|y-s\widehat W\|$ for both signs $s$.
So by the triangle inequality
\[
  \min_{s=\pm1}\|y|_{\vec E(\bG)}-s\widehat W\| \leq \min_{s=\pm1}\|y-s\widehat W\| \leq \sqrt{2\eta}+\sqrt{2(d\lambda^2)^{-(\ell+1)}}+o(1)\,.
\]

Finally, take $t=t(d,\lambda,\varepsilon/3)$ from \Cref{lem:bp-walk-score} and $\gamma=\gamma_\ell$ (defined just before \Cref{lem:bp-walk-score}).
Then for every $\ell$, with high probability,
\[
  \min_{s=\pm1}\mathrm{Ov}_{t,\gamma}(\bG,s\widehat W) \geq \OPT_{d,\lambda}-\varepsilon/3\,.
\]

Putting these together, with high probability, for every $G',y$ allowed by the theorem,
\[
  \mathrm{Ov}_{t,\gamma}(G',y) \geq \OPT_{d,\lambda}-\varepsilon/3-O_{d,\lambda,t}\bigl(\sqrt\eta+(d\lambda^2)^{-(\ell+1)/2}\bigr)-o(1)\,.
\]
Since $t$ depends only on $d,\lambda,\varepsilon$, we can choose $\ell$ large and then $\eta$ small, also depending only on $d,\lambda,\varepsilon$, so that the $O(\cdot)$ term is at most $\varepsilon/3$; for large $n$ the $o(1)$ term is also at most $\varepsilon/3$.

    Each round of $\mathsf{BP}_{t,\gamma}$ takes $O(|\vec E(G')|)=O(n)$ operations: each vertex $u$ computes $\prod_{v\in\partial u}(1\pm\lambda m_{v\to u})$ once and divides out one factor per outgoing message.
    This establishes the running time and completes the proof.
\end{proof}

In the sections that follow, we will use the following statement proved in \cite{BLM15}, which shows that the local neighborhoods of a graph drawn from the stochastic block model resemble broadcast processes on Galton--Watson trees.
\begin{fact}[Local law {\cite[Proposition~36]{BLM15}}]
    \label{fact:local-law}
    Let $q\le L+1$.
    Let $f$ and $F$ be functions of a rooted graph whose vertices carry $\pm1$ labels, depending only on the radius-$q$ neighborhood of the root, with $|f|\le F$ and $F$ nondecreasing under the addition of edges.
    Write $f(\bG,v)$ for $f$ evaluated on $\bG$ rooted at $v$ with the labels $\bsigma^\star$, and $f(\bcalT_{d,\lambda},o)$ for $f$ evaluated on the broadcast tree with its spins $\bx$.
    Condition on labels $\bsigma^\star$ for which both communities have size $n/2\pm n^{3/4}$.
    Then
    \begin{align*}
        &\E\Bigl|\frac1n\sum_{v\in[n]}f(\bG,v)-\E f(\bcalT_{d,\lambda},o)\Bigr|\\
        &\qquad\le C\,d^{q/2}n^{-1/8}\log n\,\max\Bigl\{\bigl(\E\max_{v}F(\bG,v)^4\bigr)^{1/4},\,\bigl(\E F(\bcalT_{d,\lambda},o)^2\bigr)^{1/2}\Bigr\},
    \end{align*}
    where $C=C(d,\lambda)$.
\end{fact}

\subsection{The eigenvector is close to a constant-depth walk score}
\label{sec:eigenvector-walk-score}

In this section, we establish \Cref{lem:eigenvector-walk-score}, restated for convenience.

\lemeigenvectorwalkscore*

Throughout, we fix a sufficiently small constant $\kappa > 0$ possibly depending on $d$ and $\lambda$, and $L\coloneqq\kappa\log_d n$.
We first show that $\bxi$ is close to the walk score $W^{(L)}$ of logarithmic depth, and then compare $W^{(L)}$ with $W^{(\ell)}$.

We will need the following two ingredients.
The first shows that the second eigenvector of the nonbacktracking matrix is roughly equal to the logarithmic-depth walk scores, and can be found in \cite[Eq. (5.1) and Section 5.2, Page 226]{SM22}.
\begin{lemma}[The logarithmic-depth walk score is the eigenvector]
    \label{lem:log-depth-walk-score}
    Let $d\lambda^2>1$, let $(\bG,\bsigma^\star)\sim\SBM(n,d,\lambda)$, and let $\bxi$ be as in \Cref{thm:optimal-recovery}.
    Then $\min_{s=\pm1}\bigl\|\bxi-sW^{(L)}/\|W^{(L)}\|\bigr\|\to0$ in probability.
\end{lemma}

The second is a bound on the difference between consecutive walk scores from \cite[Eq. (8.11)]{SM22}.
\begin{lemma}   \label{lem:bound-consecutive-scores}
    There exists a constant $C > 0$ depending on $d$ and $\lambda$ such that with high probability over $\bG$, we have for every $\ell \le L$:
    \[
        \norm*{ W^{(\ell+1)} - W^{(\ell)} } \le \sqrt{n} \cdot \parens*{ C(\lambda^2 d)^{-\ell/2} + \frac{C^t \log^7 n}{\sqrt{n}}} \mper
    \]
\end{lemma}


\begin{proof}[Proof of \Cref{lem:eigenvector-walk-score}]
    Without loss of generality, assume $s = +1$ in \Cref{lem:log-depth-walk-score}.
    Then we have:
    \begin{align*}
        \norm*{ \bxi - \frac{W^{(\ell)}}{\norm*{W^{(\ell)}}} } &\le \norm*{ \bxi - \frac{W^{(L)}}{\norm*{W^{(L)}}} } + \norm*{ \frac{W^{(L)}}{\norm*{W^{(L)}}} - \frac{W^{(\ell)}}{\norm*{W^{(\ell)}}} } \\
        &\le o(1) + 2\frac{ \norm*{ W^{(L)} - W^{(\ell)} } }{ \norm*{ W^{(L)} } }
    \end{align*}
    where we bounded the first term with \Cref{lem:log-depth-walk-score}.
    The numerator of the second term can be bounded by $(C (d \lambda^2)^{-\ell/2} + o(1))\cdot \sqrt{n}$ by triangle inequality and \Cref{lem:bound-consecutive-scores}.
    The denominator is lower bounded by $\Omega(\sqrt{n})$ by \cite[Proposition 10, Eq. (8.9)]{SM22}, and therefore the bound above translates to $O\parens*{(d\lambda^2)^{-\ell/2}} + o(1)$.
    Rearranging this bound yields the desired statement.
\end{proof}

\subsection{BP from the walk score is near-optimal}
\label{sec:bp-walk-score}

In this section, we establish \Cref{lem:bp-walk-score}.

\lembpwalkscore*
\begin{proof}
    We will prove the desired statement by comparing the following pair of experiments we perform after conditioning on $(\bG, \bsigma^{\star})$.
    We will choose $t$ sufficiently large depending on $\eps$.
    
    \parhead{Experiment 1.}
    Pick a uniformly random vertex $\bo\sim V(\bG)$.
    Let $\bT_{\bo}$ be the radius-$t+\ell+1$ ball around $\bo$, and let $\bx$ be a labeling of the vertices of $\bT$ according to its true labeling $\bsigma^{\star}$.
    Denote by $\calD_1$ the law of $(\bT,\bx)$ arising from this experiment.
    
    \parhead{Experiment 2.}
    Draw a tree $\bT$ rooted at a vertex $o$ along with a labeling $\bx$ of the vertices from the depth $t+\ell+1$ broadcast process tree with parameters $d$ and $\lambda$ as in \Cref{def:broadcast-tree}.
    Denote by $\calD_2$ the law of $(\bT,\bx)$ arising from this experiment.
    
    Let $(\bT,\bx)$ be a pair arising from Experiment 1 or 2 where $\bT$ is rooted at a vertex $o$.
    Now, consider the following process.
    \begin{itemize}
        \item For each vertex $v$ at depth-$t$, and each child $w$ of $v$, define $\bW^{(\ell,\bT)}$ as the vector
        \[
            \bW^{(\ell,\bT)}_{v\to w} = (d\lambda)^{-\ell} \sum_{\substack{a\text{ descendant of }w \\ \text{at depth $t+\ell+1$}}} \bx_a\mper
        \]
        \item For each such $w\to v$, let $\by_{w\to v}$ be drawn from $\{\pm\gamma_{\ell}\}$ such that $\E\by_{w\to v} = \gamma_{\ell} \tanh \left( \frac{\beta \bW^{(\ell,\bT)}_{v\to w}}{ \sqrt{ \E \bW_{\ell}^2 } } \right)$.
        \item Let $\wh{\bx}_o$ be the outcome of running $t$ rounds of BP with initial messages given by $\by$.
    \end{itemize}
    Observe that $\Ov_{t,\gamma_{\ell}}\left(\bG,\bW^{(\ell)}/\|\bW^{(\ell)}\|\right) = \E_{(\bT,\bx)\sim\calD_1} \bx_{\bo} \cdot \wh{\bx}_{\bo}$.
    
    A consequence of \Cref{fact:local-law} is the fact that $d_{\mathrm{TV}}(\calD_1,\calD_2) = o(1)$ with high probability over $\bG$.
    
    Therefore, we have:
    \[
        \E_{(\bT,\bx)\sim\calD_1} \bx_{\bo} \cdot \wh{\bx}_{\bo} \ge \E_{(\bT,\bx)\sim\calD_2} \bx_{\bo} \cdot \wh{\bx}_{\bo} - o(1) \ge \OPT_{d,\lambda} - o(1)\mcom
    \]
    where the first inequality used closeness in total variation distance.
    The second inequality follows from \Cref{fact:yp-robust-reconstruction}; we elaborate on details below.
    Under Experiment 2, conditional on the labels at depth-$t$, the labels on the subtrees rooted at each of the depth-$t$ vertices are independent.
    Thus, $\frac{\bW^{(\ell,\bT)}}{ \sqrt{ \E \bW_{\ell}^2 } }$ restricted to the depth-$t$ leaves can be construed as passing each label $\bsigma_v$ through an independent noisy channel.
    It can also be verified that there is an absolute constant $c > 0$ such that $\gamma_{\ell} > c$, and therefore the hypothesis of \Cref{fact:yp-robust-reconstruction} applies.
    In particular, we may choose $t$ independent of $\ell$.
\end{proof}

\subsection{Stability of BP}
\label{sec:bp-stability}

In this subsection we prove \Cref{lem:bp-score-stability,lem:bp-edit-stability}, starting with \Cref{lem:bp-score-stability}.
Throughout, for a graph $G$, a vertex $v$ and $s\ge0$, we write $|B_G(v,s)|$ for the number of vertices of $G$ within distance $s$ of $v$.

\subsubsection{Proof of \Cref{lem:bp-score-stability}}

\lembpscorestability*

The idea behind \Cref{lem:bp-score-stability} is that after $t$ rounds of BP, the output at a vertex $v$ depends only on the initialization $Y$ at directed edges near $v$.
Furthermore, if we round two initializations $Y$ and $Y'$ using the same randomness, the two outputs can differ at $v$ only if the two roundings differ at some directed edge $e$ near $v$, which happens with probability at most $|Y_e-Y'_e|/2$.
Summing over $v$, the expected number of vertices at which the outputs differ is at most $\|Y-Y'\|$ times a measure of the size of a typical neighborhood.
We isolate this deterministic part of the argument in the following lemma.

\begin{lemma}[BP outputs from nearby scores can be coupled]
    \label{lem:bp-coupling}
    Let $G$ be a graph on $[n]$, let $t\ge1$ and $\gamma\in(0,1)$, and let $Y,Y'\in\R^{\vec E(G)}$.
    Then the outputs $\widehat{\bsigma}$ and $\widehat{\bsigma}'$ of $\mathsf{BP}_{t,\gamma}(G,Y)$ and $\mathsf{BP}_{t,\gamma}(G,Y')$ can be coupled so that
    \begin{equation}
        \E\,\bigl|\{v:\widehat{\bsigma}_v\ne\widehat{\bsigma}'_v\}\bigr|
        \le\frac12\Bigl(\sum_{v\in[n]}|B_G(v,t+1)|^3\Bigr)^{1/2}\|Y-Y'\|.
        \label{eq:bp-coupling}
    \end{equation}
\end{lemma}

The second ingredient is that neighborhoods in the block model are small on average.
This is an immediate corollary of \Cref{fact:local-law}, and that the moments of $|B_{\bT_{d,\lambda}}(o,s)|$ and $|B_{\bG}(v,s)|$ are finite.
\begin{lemma}[Neighborhoods are small on average]
    \label{lem:neighborhood-cubes}
    Let $d\lambda^2>1$ and $(\bG,\bsigma^\star)\sim\SBM(n,d,\lambda)$.
    For every $s\ge1$ there is $C=C(d,\lambda,s)$ such that, with high probability,
    \[
        \sum_{v\in[n]}|B_{\bG}(v,s)|^3\le Cn.
    \]
\end{lemma}

We now put these together, and then prove \Cref{lem:bp-coupling}.

\begin{proof}[Proof of \Cref{lem:bp-score-stability}]
    Let $C_0=C(d,\lambda,t+1)$ be the constant of \Cref{lem:neighborhood-cubes}, and condition on the high-probability event that $\sum_v|B_{\bG}(v,t+1)|^3\le C_0n$.
    On this event also $|\vec E(\bG)|\le C_0n$.

    Fix $\gamma\in(0,1)$ and $y,y'\in\R^{\vec E(\bG)}$.
    Let $\widehat{\bsigma}$ and $\widehat{\bsigma}'$ be the outputs of $\mathsf{BP}_{t,\gamma}$ initialized at $Y=\beta\sqrt{|\vec E(\bG)|}\,y$ and $Y'=\beta\sqrt{|\vec E(\bG)|}\,y'$, respectively, coupled as in \Cref{lem:bp-coupling}.
    Changing the output at one vertex changes the overlap by at most $2/n$.
    Indeed, since $\ov(\widehat{\bsigma},\bsigma^\star)=\frac1n|\langle\widehat{\bsigma},\bsigma^\star\rangle|$ and $\bigl||x|-|x'|\bigr|\le|x-x'|$ for real numbers $x,x'$,
    \begin{align*}
        \bigl|\ov(\widehat{\bsigma},\bsigma^\star)-\ov(\widehat{\bsigma}',\bsigma^\star)\bigr|
        &=\frac1n\Bigl|\bigl|\langle\widehat{\bsigma},\bsigma^\star\rangle\bigr|-\bigl|\langle\widehat{\bsigma}',\bsigma^\star\rangle\bigr|\Bigr|\\
        &\le\frac1n\bigl|\langle\widehat{\bsigma}-\widehat{\bsigma}',\bsigma^\star\rangle\bigr|
        \le\frac2n\,\bigl|\{v:\widehat{\bsigma}_v\ne\widehat{\bsigma}'_v\}\bigr|.
    \end{align*}
    The quantities $\mathrm{Ov}_{t,\gamma}(\bG,y)$ and $\mathrm{Ov}_{t,\gamma}(\bG,y')$ are the expectations of the two overlaps, whatever the coupling.
    So taking expectations and applying \eqref{eq:bp-coupling} with $\|Y-Y'\|=\beta\sqrt{|\vec E(\bG)|}\,\|y-y'\|$,
    \begin{align*}
        \bigl|\mathrm{Ov}_{t,\gamma}(\bG,y)-\mathrm{Ov}_{t,\gamma}(\bG,y')\bigr|
        &\le\frac2n\,\E\,\bigl|\{v:\widehat{\bsigma}_v\ne\widehat{\bsigma}'_v\}\bigr|\\
        &\le\frac1n\Bigl(\sum_{v}|B_{\bG}(v,t+1)|^3\Bigr)^{1/2}\beta\sqrt{|\vec E(\bG)|}\,\|y-y'\|\\
        &\le\frac1n\sqrt{C_0n}\cdot\beta\sqrt{C_0n}\,\|y-y'\|
        =\beta C_0\,\|y-y'\|.
    \end{align*}
    This proves \Cref{lem:bp-score-stability} with $C=\beta C_0$, which depends only on $d$, $\lambda$ and $t$.
\end{proof}

\begin{proof}[Proof of \Cref{lem:bp-coupling}]
    We first couple the random rounding in \Cref{def:bp-from-score}.
    For each directed edge $e$, let $\mathbf U_e$ be uniform on $[-1,1]$, independently over $e$, and set
    \[
        \boldsymbol\tau_e=
        \begin{cases}
            1&\text{if }\mathbf U_e\le\tanh Y_e,\\
            -1&\text{otherwise},
        \end{cases}
    \]
    and define $\boldsymbol\tau'_e$ in the same way from $Y'_e$, with the same $\mathbf U_e$.
    Then $\E\boldsymbol\tau_e=\tanh Y_e$ and $\E\boldsymbol\tau'_e=\tanh Y'_e$, as required, and $\boldsymbol\tau_e\ne\boldsymbol\tau'_e$ only if $\mathbf U_e$ lies between $\tanh Y_e$ and $\tanh Y'_e$, an interval of length $|\tanh Y_e-\tanh Y'_e|$ on which $\mathbf U_e$ has density $1/2$.
    Since $\tanh$ is $1$-Lipschitz,
    \begin{equation}
        \Pr(\boldsymbol\tau_e\ne\boldsymbol\tau'_e)=\frac12\,|\tanh Y_e-\tanh Y'_e|\le\frac12\,|Y_e-Y'_e|.
        \label{eq:rounding-coupling}
    \end{equation}
    Let $\widehat{\bsigma}$ and $\widehat{\bsigma}'$ be the outputs of the remaining steps of $\mathsf{BP}_{t,\gamma}$ run on $\boldsymbol\tau$ and on $\boldsymbol\tau'$.

    Since $t$ rounds of BP only propagate information a distance $t$, the output $\widehat{\bsigma}_v$ depends only on $\boldsymbol\tau_e$ for directed edges $e$ whose head is within distance $t+1$ of $v$, and likewise for $\widehat{\bsigma}'_v$ and $\boldsymbol\tau'$.

    So $\widehat{\bsigma}_v\ne\widehat{\bsigma}'_v$ only if $\boldsymbol\tau_e\ne\boldsymbol\tau'_e$ for some directed edge $e$ whose head is within distance $t+1$ of $v$.
    The head $v$ of a directed edge $u\to v$ is within distance $t+1$ of exactly $|B_G(v,t+1)|$ vertices.
    So a union bound over these edges, followed by \eqref{eq:rounding-coupling}, gives
    \[
        \E\,\bigl|\{v:\widehat{\bsigma}_v\ne\widehat{\bsigma}'_v\}\bigr|
        \le\sum_{u\to v}|B_G(v,t+1)|\,\Pr(\boldsymbol\tau_{u\to v}\ne\boldsymbol\tau'_{u\to v})
        \le\frac12\sum_{u\to v}|B_G(v,t+1)|\,|Y_{u\to v}-Y'_{u\to v}|.
    \]
    By the Cauchy--Schwarz inequality, and since each vertex $v$ is the head of $\deg{v}$ directed edges,
    \begin{align*}
        \sum_{u\to v}|B_G(v,t+1)|\,|Y_{u\to v}-Y'_{u\to v}|
        &\le\Bigl(\sum_{u\to v}|B_G(v,t+1)|^2\Bigr)^{1/2}\|Y-Y'\|\\
        &=\Bigl(\sum_{v\in[n]}\deg{v}\,|B_G(v,t+1)|^2\Bigr)^{1/2}\|Y-Y'\|.
    \end{align*}
    Finally $\deg{v}\le|B_G(v,t+1)|$, which gives \eqref{eq:bp-coupling}.
\end{proof}

\subsubsection{Proof of \Cref{lem:bp-edit-stability}}

\lembpeditstability*

The idea behind the above is again locality: if a vertex $v$ is far from every corrupted edge, then $t$ rounds of BP compute the same output at $v$ on both graphs if initialized at the same vector.
By \Cref{lem:neighborhood-cubes}, most vertices have no corrupted edges nearby.
Apart from this, BP on the corrupted graph and non-corrupted graph differ only in the normalization $\beta\sqrt{|\vec E(G)|}$ of the initialization, which changes by a factor $1+O(\eta)$ and is handled by \Cref{lem:bp-score-stability}.
We isolate the deterministic part of the argument in the following lemma.

\begin{lemma}[BP only sees nearby edits]
    \label{lem:bp-edits-local}
    Let $G$ and $G'$ be graphs on $[n]$, let $t\ge1$ and $\gamma\in(0,1)$, and let $A$ be the set of endpoints of the edges in $E(G)\triangle E(G')$.
    Let $y\in\R^{n^2-n}$.
    Then the outputs $\widehat{\bsigma}$ of $\mathsf{BP}_{t,\gamma}(G,y|_{\vec E(G)})$ and $\widehat{\bsigma}'$ of $\mathsf{BP}_{t,\gamma}(G',y|_{\vec E(G')})$ can be coupled so that with probability $1$ over the randomized rounding of the initialization,
    \[
        \bigl|\{v:\widehat{\bsigma}_v\ne\widehat{\bsigma}'_v\}\bigr|\le\sum_{a\in A}|B_G(a,t)|.
    \]
\end{lemma}

We now prove \Cref{lem:bp-edit-stability}, and then \Cref{lem:bp-edits-local}.

\begin{proof}[Proof of \Cref{lem:bp-edit-stability}]
    Let $C_0$ be the constant of \Cref{lem:neighborhood-cubes} with $s=t+1$, and let $C_1=C(d,\lambda,t)$ be the constant of \Cref{lem:bp-score-stability}.
    We work on the intersection of the high-probability events of these two lemmas and the event $|\vec E(\bG)|\ge dn/2$, which has high probability because $|\vec E(\bG)|/n\to d$ in probability.
    Fix $\eta,\gamma\in(0,1)$, and $G'$ and $y$ as in the statement, and let
    \[
        c=\Bigl(\frac{|\vec E(G')|}{|\vec E(\bG)|}\Bigr)^{1/2}.
    \]
    We compare $\mathrm{Ov}_{t,\gamma}(G',y)$ and $\mathrm{Ov}_{t,\gamma}(\bG, y |_{\vec E(\bG)})$ with $\mathrm{Ov}_{t,\gamma}(\bG,c\,y|_{\vec E(\bG)})$.

    \emph{Changing the graph.}
    We apply \Cref{lem:bp-edits-local} with $\beta\sqrt{|\vec E(G')|}\,y$ in place of $y$.
    Its restriction to $\vec E(\bG)$ is $\beta\sqrt{|\vec E(G')|}\,y|_{\vec E(\bG)}=\beta\sqrt{|\vec E(\bG)|}\,c\,y|_{\vec E(\bG)}$.
    So the expected overlaps of the two outputs in \Cref{lem:bp-edits-local} are $\mathrm{Ov}_{t,\gamma}(G',y)$ and $\mathrm{Ov}_{t,\gamma}(\bG,c\,y|_{\vec E(\bG)})$.
    As in the proof of \Cref{lem:bp-score-stability}, changing the output at one vertex changes the overlap by at most $2/n$, so \Cref{lem:bp-edits-local} gives
    \[
        \bigl|\mathrm{Ov}_{t,\gamma}(G',y)-\mathrm{Ov}_{t,\gamma}(\bG,c\,y|_{\vec E(\bG)})\bigr|\le\frac2n\sum_{a\in A}|B_{\bG}(a,t)|,
    \]
    where $A$ is the set of endpoints of the edges in $E(\bG)\triangle E(G')$, so that $|A|\le2\eta n$.
    Then we have
    \[
        \sum_{a\in A}|B_{\bG}(a,t)|
        \le|A|^{1/2}\Bigl(\sum_{v\in[n]}|B_{\bG}(v,t)|^2\Bigr)^{1/2}
        \le|A|^{1/2}\Bigl(\sum_{v\in[n]}|B_{\bG}(v,t+1)|^3\Bigr)^{1/2}
        \le(2\eta n)^{1/2}(C_0n)^{1/2}
        =\sqrt{2C_0\eta}\;n \, ,
    \]
    where we used
     Cauchy--Schwarz, then $1\le|B_{\bG}(v,t)|\le|B_{\bG}(v,t+1)|$, and finally \Cref{lem:neighborhood-cubes}.

    \emph{Changing the normalization.}
    By \Cref{lem:bp-score-stability}, and since $\|y|_{\vec E(\bG)}\|\le\|y\|\le1$,
    \[
        \bigl|\mathrm{Ov}_{t,\gamma}(\bG,c\,y|_{\vec E(\bG)})-\mathrm{Ov}_{t,\gamma}(\bG,y|_{\vec E(\bG)})\bigr|\le C_1\,|c-1|\,\|y|_{\vec E(\bG)}\|\le C_1\,|c-1|.
    \]
    Each edge in $E(\bG)\triangle E(G')$ accounts for two directed edges, so $\bigl||\vec E(G')|-|\vec E(\bG)|\bigr|\le2\eta n$.
    Since $|\sqrt x-1|\le|x-1|$ for $x\ge0$ and $|\vec E(\bG)|\ge dn/2$,
    \[
        |c-1|\le\frac{\bigl||\vec E(G')|-|\vec E(\bG)|\bigr|}{|\vec E(\bG)|}\le\frac{2\eta n}{dn/2}=\frac{4\eta}d.
    \]

    Putting the two steps together, and using $\eta\le\sqrt\eta$,
    \[
        \bigl|\mathrm{Ov}_{t,\gamma}(G',y)-\mathrm{Ov}_{t,\gamma}(\bG,y|_{\vec E(\bG)})\bigr|
        \le2\sqrt{2C_0\eta}+\frac{4C_1\eta}d
        \le\Bigl(2\sqrt{2C_0}+\frac{4C_1}d\Bigr)\sqrt\eta.
    \]
    This proves \Cref{lem:bp-edit-stability} with $C=2\sqrt{2C_0}+4C_1/d$, which depends only on $d$, $\lambda$ and $t$.
\end{proof}

\begin{proof}[Proof of \Cref{lem:bp-edits-local}]
    We couple the random rounding of the initialization in the two runs of BP as in the proof of \Cref{lem:bp-coupling}.
    That is, we use the same uniform variable $\mathbf U_e$ for each directed edge $e$ that belongs to both graphs, and independent uniform variables for all other directed edges.
    Since the score of such an edge $e$ is $y_e$ in both runs, the rounded spins agree on every directed edge of both graphs.

    Now, if the distance-$t$ neighborhood of a vertex $v$ in $G$ is disjoint from $A$, then every vertex within distance $t$ of $v$ has the same neighbors in $G$ and in $G'$.
    So the vertices within distance $t+1$ of $v$, and the edges incident to vertices within distance $t$ of $v$, are the same in the two graphs.
    The output at $v$ after $t$ rounds of BP is computed from these edges and the rounded spins on them, so $\widehat{\bsigma}_v=\widehat{\bsigma}'_v$.
    Every other vertex is within distance $t$ of some $a\in A$ in $G$, and there are at most $\sum_{a\in A}|B_G(a,t)|$ such vertices.
\end{proof}


\begingroup
\renewcommand*{\theHtheorem}{bethe-hessian.\arabic{theorem}}
\renewcommand*{\theHlemma}{bethe-hessian.\arabic{theorem}}
\renewcommand*{\theHfact}{bethe-hessian.\arabic{theorem}}

\section{Robustly estimating the second eigenvector of the nonbacktracking matrix}
\label{sec:robust-estimation-of-second-eigenvector}

We prove the following main theorem for this section.

\begin{theorem}[Robust estimation of the informative eigenvector]
    \label{thm:robust-second-eigenvector}
    For every $d \geq 1$, $\lambda \in (-1,1)$ with $d \lambda^2 > 1$, and $\eps > 0$ there is $\eta > 0$ and a nearly-linear time algorithm with the following guarantee.
    Let $(\bG,\bsigma^\star)\sim\SBM(n,d,\lambda)$
    and let $\bxi$ be a unit eigenvector for the eigenvalue of $B_{\bG}$ of second-largest modulus.
    With high probability over $(\bG,\bsigma^\star)$,
    for every graph $F$ satisfying $|E(F)\triangle E(\bG)|\le\eta n$,
    with high probability over its internal randomness the algorithm outputs $\widehat\xi(F)$ with
    \[
        \min_{a\in\{\pm1\}}
            \|\widehat\xi(F)-a\bxi\|\le\varepsilon.
    \]
\end{theorem}

We will gather the lemmas we need, capturing several algorithmic subroutines, and then prove \Cref{thm:robust-second-eigenvector}.

A key challenge in designing and analyzing an algorithm for \Cref{thm:robust-second-eigenvector} is the asymmetry of the nonbacktracking matrix---in particular, $\bxi \in \R^{\vec E(\bG)}$ does not live in an orthonormal basis of eigenvectors.
To design an algorithm which proves \Cref{thm:robust-second-eigenvector}, we therefore take a somewhat indirect approach, inspired by \cite{MRW24}.
We reduce to the problem of recovering, in the presence of corruptions, a certain $n$-dimensional vector which counts signed nonbacktracking walks, defined below.
We will be able to postprocess this vector in a simple (linear time) way to recover $\bxi$.

\begin{definition}
   For a graph $G$ on $[n]$ and $\ell \in \N$, let $A^{(\ell)} \in \R^{n \times n}$ count length-$\ell$ nonbacktracking walks in $G$.
   We write $A_G^{(\ell)}$ if $G$ is not clear from context, and by convention let $A^{(0)} = I$.
   Moreover, for $(\bG,\bsigma^*) \sim \SBM$, let $\bsigma^{(s)} = (d \lambda)^{-s} A^{(s)} \bsigma^*$.
\end{definition}

The choice of $(d\lambda)^{-s}$ as a normalization factor will ensure that $\bsigma^{(s)}$ converges to a nonzero vector as $s \rightarrow \infty$.

For $(\bG,\bsigma^*) \sim \SBM$, the following lemma, proved in \Cref{subsec:alg-postprocessing}, lets us reduce to the problem of recovering $\bsigma^{(s)}$ given an adversarial corruption of $\bG$.

\begin{restatable}[Robust postprocessing]{lemma}{lemrobustpreprocessing}
    \label{lem:robust-postprocessing}
    Let $(\bG,\bsigma^\star)\sim\SBM(n,d,\lambda)$
    and $\bxi$ be a unit eigenvector for the eigenvalue of $B_{\bG}$ of second-largest modulus.
    For every $d \geq 1$, $\lambda \in (-1,1)$ with $d \lambda^2 > 1$, and $\varepsilon > 0$,
    there are $\eta > 0$, $\delta > 0$, $s_0\in\N$, and a linear-time algorithm with the following guarantee for every $s\ge s_0$:
    with high probability over $(\bG,\bsigma^\star)$, given input $y \in \R^n$ satisfying
    \[
        \min_{a\in\{\pm1\}}
        \left\|y-a
            \frac{\bsigma^{(s)}}
                 {\|\bsigma^{(s)}\|}\right\|\le\delta,
    \]
    and an arbitrary graph $F$ satisfying $|E(F)\triangle E(\bG)|\le\eta n$, the algorithm outputs a vector $\widehat\xi(F)$ such that
    \[
        \min_{a\in\{\pm1\}}
            \|\widehat\xi(F)-a\bxi\|\le\varepsilon.
    \]
\end{restatable}

Our strategy to robustly recover $\bsigma^{(s)}$ has two parts.
First, we show that if we had access to the non-corrupted graph $\bG$, a spectral algorithm based on the \emph{Bethe Hessian} matrix (defined below) would recover $\bsigma^{(s)}$ with approximation error going to zero as $s \rightarrow \infty$.
Then we show that even given access to an adversarially corrupted graph, there is a simple preprocessing step together with a semidefinite program which ``robustifies'' this spectral algorithm and hence still recovers $\bsigma^{(s)}$.
The ability to robustify this spectral algorithm in a straightforward way is our reason for working with the $n \times n$ symmetric matrices rather than directly with the nonbacktracking matrix.

To state the main lemmas capturing these algorithms we need to define the key matrices involved, which we denote $M_s = M_s(G)$ where $G$ is an underlying graph and $s \in \N$.

\begin{definition}[Bethe Hessian]
Our starting point is the \emph{Bethe Hessian}~\cite{SKZ14}, a one-parameter family of $n \times n$ matrices associated to a graph $G$.
\[
    H(t)=I-tA+t^2(D-I).
\]
\end{definition}
The \emph{Ihara--Bass formula}~\cite{Iha66,Bass92} gives a correspondence between $t$ which make $H(t)$ singular and eigenvalues of the nonbacktracking matrix.
We will use something more general, that at some carefully chosen values of $t$ where $H(t)$ is nonsingular, $H(t)$ has a small number of negative eigenvectors and the vector we want, $\bsigma^{(s)}$, lies close to their span, after applying a simple linear transformation depending on $\bG$.
This strategy was also followed by \cite{MRW24}.
We refine it here by (approximately) isolating $\bsigma_{\bG}^{(s)}$ in the span of these negative eigenvectors by applying a delicate rank-one correction to $H(t)$.
This leads to more accurate recovery of $\bsigma_{\bG}^{(s)}$ via the Bethe Hessian than was previously achieved.
We package this into an analysis of a family of matrices we call ``corrected Bethe Hessian matrices.''

\begin{definition}[Corrected Bethe-Hessian Matrices]
\label{def:sandwiched-bethe-hessian}
For each $s \in \N$ and $n$-vertex graph $G$, we define a matrix $M_s(G) \in \R^{n \times n}$.
When the block-model parameters $d,\lambda$ are clear from context, let $t = t(d,\lambda) = \frac{1}{d\lambda}\Bigl(1+\frac{\sqrt d-1}{d}\,(d\lambda^2-1)\Bigr)$.
\footnote{Although, a range of choices of $t$ in $\left[ \frac{1}{\lambda d}, \frac{1}{\sqrt{d}} \right)$ would work, and specific details of this choice are not so important.}
Since $t$ is always now fixed to the quantity above, we just write $H=H(t)$ for the Bethe Hessian going forward.

Our choice of $t$ results in $2$ eigenvalues of $H$ below zero (when $\lambda > 0$).
Roughly, these correspond to the top two eigenvalues (by modulus) of the nonbacktracking matrix.
We can remove one of those eigenvalues as follows.
Let
\[
    b=\mathbf1+\sum_{j=1}^L t^j(A^{(j)} -dA^{(j-1)}) \mathbf 1,
\]
where $L=L(d,\lambda)\ge2$ will be a sufficiently large constant (see \Cref{lem:bh-tree}(i)).

We will eventually see that, when $\lambda>0$, the chosen $b$ satisfies $b^\top H b<0$ with high probability.
To align an eigenvector of $H$ with $\bsigma^{(s)}$ we also need to multiply by $A_{\bG}^{(s)}$.
All together, we get the corrected Bethe Hessian:
\[
M_s(G)=(d\lambda)^{-2s}\cdot
\begin{cases}
    A^{(s)}\Bigl(H+\dfrac{2(H b)(H b)^\top}{-b^\top H b}\Bigr)A^{(s)}\,,&\lambda>0\,,\\[2ex]
    A^{(s)}HA^{(s)}\,,&\lambda<0\,.
\end{cases}
\]

\end{definition}

Now we can state the key properties of the $M_s$ matrices when $\bG \sim \SBM$.
First is that $\bsigma^{(s)}$ can be recovered from its (unique) negative eigenvector; for making this recovery algorithm robust it is also important that this vector is delocalized in the sense of having small $\ell_4$ norm.

\begin{restatable}[Informative negative eigenvector]{lemma}{lembethehessianstructure}
    \label{lem:bethe-hessian-structure}
    Let $d \lambda^2 > 1$.
    There is a constant $g(d,\lambda)>0$ such that for every $\eps > 0$ there is $s_0(\eps)$ such that for every $s > s_0$, with high probability $M_s$ has exactly one negative eigenvalue, and $\bsigma^{(s)}$ is approximately an associated eigenvector, in that
    \[
    \bigl\|(M_s+gI)\bsigma^{(s)}\bigr\|\le\eps\,\|\bsigma^{(s)}\| \, .
    \]
    Moreover, there is $C=C(d,\lambda)$, not depending on $s$, such that with high probability, $\bsigma^{(s)}$ is delocalized, in that
    \[
        \frac{\|\bsigma^{(s)}\|_4^4}{\|\bsigma^{(s)}\|^4}\le\frac Cn\,.
    \]
\end{restatable}

We prove the above in \Cref{sec:bethe-hessian-negative-spectrum}. In light of \Cref{lem:bethe-hessian-structure}, our goal will be to recover the negative eigenvector of the corrected Bethe Hessian $M_s(\bG)$, given access only to an adversarially corrupted graph $F$.
A first ingredient is to define a norm in which the adversarial corruptions cannot change $M_s$ too much (allowing for some light preprocessing).

\begin{definition}[Row-sum norms]
    \label{def:row-sum-norm}
    For $E\in\R^{n\times n}$,
    \[
    \|E\|_{1,2}=\left(\sum_{i\in[n]}\left(\sum_{j\in[n]}|E_{ij}|\right)^2\right)^{1/2} \quad \text{ and } \quad \|E\|_{1,\infty}= \max_{i \in [n]} \sum_{j \in [n]} |E_{ij}| \, .
    \]
\end{definition}

A simple preprocessing routine -- removing high degree vertices -- can use the adversarially-corrupted graph $F$ and arrive at a matrix $\widetilde{M}$ which is close to $M_s(\bG)$.

\begin{restatable}[Preprocessing for robustness]{lemma}{lemrobustmatrixeditstability}
    \label{lem:robust-matrix-edit-stability}
    Let $(\bG,\bsigma^\star)\sim\SBM(n,d,\lambda)$ and $d\lambda^2>1$.
    For every $s \in \N$ and $\eps > 0$, there is a linear-time algorithm, \(C\ge 1\) and $\eta > 0$ which with high probability over $\bG$, given as input any $n$-vertex graph $F$ such that     $|E(F)\triangle E(\bG)|\le \eta n$, produces a symmetric matrix $\widetilde{M} \in \R^{n \times n}$ such that
    \[
        \|\widetilde{M}-M_{s}(\bG)\|_{1,2}
            \le \eps \sqrt n,
        \qquad
        \|\widetilde{M}\|_{1,\infty}\le C \, .
    \]
    Moreover, matrix-vector products by $\widetilde{M}$ can be computed in linear time.
\end{restatable}

We prove the above in \Cref{subsec:alg-preprocessing}. Finally, we arrive at the key algorithm: a convex program which approximately recovers a delocalized negative eigenvector from a matrix, 
tolerating perturbations in $\| \cdot \|_{1,2}$ norm.
Its analysis is presented in \Cref{sec:sdp-recovery-implementation}.

\begin{restatable}[Robust recovery of a delocalized negative eigenvector]{lemma}{lemrobustaltsdp}
    \label{lem:robust-alt-sdp}
    There is a randomized algorithm with the following properties.
    Let \(\varepsilon,g >0\) and $M,\widetilde M\in\R^{n\times n}$ be symmetric, with $M$ having exactly one negative eigenvalue $-g$.
    Suppose there exists a unit vector $w\in\R^n$ satisfying
    \[
        \|w\|_4^4\le\frac1{\varepsilon n}
        \quad\text{and}\quad
        w^\top Mw\le-(1-\varepsilon)g\,.
    \]
    Suppose also that
    \(
        \|\widetilde M-M\|_{1,2}\le\varepsilon g\sqrt n.
    \)
    Then, given matrix-vector access to $\widetilde M$, an upper bound \(C\ge \|\widetilde M\|_{1,\infty} +1\), and \(\varepsilon\) (but not \(g\)) as inputs, the algorithm returns a unit vector $\hat {\mathbf w}$ such that with probability at least $1-n^{-10}$ over the internal randomness of the algorithm,
    \[
        \langle\hat {\mathbf  w}, w\rangle^2\ge1-O(\sqrt\varepsilon).
    \]
    On the same event, the algorithm performs at most \(
        \left(1+\frac{C\log n}{\varepsilon g}\right)^{O(1)}
    \) matrix-vector products with \(\widetilde M\) and at most \(n\) times as many additional operations.
\end{restatable}

The algorithm applies matrix entropy mirror descent (that is, the matrix multiplicative weights update) to approximately minimize the following convex function \(F\) over the set of density matrices \(\{X\in\R^{n\times n}:X\succeq0,\ \Tr X=1\}\) 
\[
 F(X)=\langle\widetilde M,X\rangle
       +C\cdot\operatorname{dist}_1(\diag X,\mathcal D_{1/\varepsilon}),
\]
where $\mathcal D_{1/\varepsilon}=\{q\in\R^n:\|q\|_2^2\le1/(\varepsilon n)\}$.
\medskip

We are ready to prove \Cref{thm:robust-second-eigenvector}.

\begin{proof}[Proof of \Cref{thm:robust-second-eigenvector}]
The algorithm runs the preprocessing algorithm of \Cref{lem:robust-matrix-edit-stability} on $F$, gives the resulting matrix $\widetilde M$ to the algorithm of \Cref{lem:robust-alt-sdp}, and runs the postprocessing algorithm of \Cref{lem:robust-postprocessing} on the resulting unit vector $\hat w$ and on $F$.

We choose the parameters in the following order.
Let $\eta_1$, $\delta$, and $s_0$ be given by \Cref{lem:robust-postprocessing} with accuracy $\eps$.
Let $g$ and $C$ be the constants of \Cref{lem:bethe-hessian-structure}, and let $\eps'>0$ be small enough that $\eps'\le1/C$ and the $O(\sqrt{\eps'})$ in \Cref{lem:robust-alt-sdp} is at most $\delta^2/2$.
Fix $s\ge s_0$ large enough to apply \Cref{lem:bethe-hessian-structure} with accuracy $\eps'g/4$.
Let $\eta_2$ and $C'$ be given by \Cref{lem:robust-matrix-edit-stability} for this $s$ and accuracy $\eps'g/2$, and let $\eta=\min\{\eta_1,\eta_2\}$.
The algorithm of \Cref{lem:robust-alt-sdp} is run with accuracy $\eps'$ and upper bound $C'+1$.

With high probability over $(\bG,\bsigma^\star)$, the conclusions of \Cref{lem:robust-postprocessing,lem:bethe-hessian-structure,lem:robust-matrix-edit-stability} all hold.
Fix such $(\bG,\bsigma^\star)$ and a graph $F$ with $|E(F)\triangle E(\bG)|\le\eta n$.
Let $M=M_s(\bG)$, let $-g'$ be its unique negative eigenvalue, and let $w=\bsigma^{(s)}/\|\bsigma^{(s)}\|$, so that $\|(M+gI)w\|\le\eps'g/4$.
Expanding $w$ in an eigenbasis of $M$ shows that $M$ has an eigenvalue within $\eps'g/4$ of $-g$.
That eigenvalue is negative, so it is $-g'$, and $|g'-g|\le\eps'g/4$; in particular $g'\ge g/2$.
So the hypotheses of \Cref{lem:robust-alt-sdp} hold for $M$, $\widetilde M$, $w$, and $\eps'$:
\begin{align*}
    \|w\|_4^4&\le\frac Cn\le\frac1{\eps'n}\,,\\
    w^\top Mw&=w^\top(M+gI)w-g\le-g'+\frac{\eps'g}2\le-(1-\eps')g'\,,\\
    \|\widetilde M-M\|_{1,2}&\le\frac{\eps'g}2\sqrt n\le\eps'g'\sqrt n\,,
\end{align*}
and $\|\widetilde M\|_{1,\infty}\le C'$.
Hence with probability at least $1-n^{-10}$ over its internal randomness, the algorithm returns $\hat w$ with $\iprod{\hat w,w}^2\ge1-\delta^2/2$, and then
\[
    \min_{a\in\{\pm1\}}\|\hat w-aw\|^2=2-2|\iprod{\hat w,w}|\le2\bigl(1-\iprod{\hat w,w}^2\bigr)\le\delta^2\,.
\]
So \Cref{lem:robust-postprocessing} gives $\min_{a\in\{\pm1\}}\|\widehat\xi(F)-a\bxi\|\le\eps$.

For the running time, preprocessing and postprocessing take linear time.
Since $C'$, $\eps'$, and $g'\ge g/2$ are constants, \Cref{lem:robust-alt-sdp} uses $(\log n)^{O(1)}$ matrix-vector products with $\widetilde M$, each taking linear time by \Cref{lem:robust-matrix-edit-stability}, and $n(\log n)^{O(1)}$ further operations.
\end{proof}

\section{Negative eigenvalues of the Bethe Hessian}
\label{sec:bethe-hessian-negative-spectrum}

In this section we prove \Cref{lem:bethe-hessian-structure}.

\lembethehessianstructure*

Throughout this section, $d\lambda^2>1$, and $t$, $H=H(t)$, $b$, and $M_s$ are as in \Cref{def:sandwiched-bethe-hessian}.
We have yet to choose the constant $L(d,\lambda)$ in the definition of $b$ but we will do so in this section.

We split the proof into two lemmas.

\begin{lemma}[Negative eigenvalue bound for the Bethe Hessian]
    \label{lem:robust-bh-inertia}
    With high probability,
    \[
        n_-(H)\le
        \begin{cases}2,&\lambda>0,\\1,&\lambda<0.\end{cases}
    \]
    where $n_-(H)$ counts the number of strictly negative eigenvalues of $H$.
\end{lemma}

\Cref{lem:robust-bh-inertia} is a quick consequence of \cite{BLM15} and \cite{MRW24} so we discharge its proof now.

\begin{proof}[Proof of \Cref{lem:robust-bh-inertia}]
Recall that $B=B_{\bG}$ is the nonbacktracking matrix.
By \cite[Lemma 5.4]{MRW24}, $n_-(H)$ is at most the number of real eigenvalues $\theta$ of $B$ with $t\theta>1$, counted with multiplicity.
By \cite[Theorem 4]{BLM15}, with high probability all eigenvalues of $B$ other than $d+o(1)$ and $d\lambda+o(1)$ have modulus at most $\sqrt d+o(1)$.
Since $|t|\sqrt d<1$, none of these other eigenvalues satisfies $t\theta>1$, so $n_-(H)\le2$.
When $\lambda<0$, also $t<0$, so $t\theta>1$ forces $\theta<0$, which rules out the eigenvalue near $d$; hence in this case $n_-(H)\le1$.
\end{proof}

The second main lemma shows that several scalar inequalities related to $H$, $M_s$, and $\bsigma^{(s)}$ hold with high probability in the SBM.
We will prove it in subsequent subsections.
The notation is as in \Cref{def:sandwiched-bethe-hessian,lem:bethe-hessian-structure}.

\begin{lemma}[Estimates for the corrected Bethe Hessian, see \Cref{subsec:bh-reduction}]
    \label{lem:bh-estimates}
    \leavevmode
    \begin{enumerate}[(a)]
        \item If $\lambda>0$, then $\frac1n b^\top Hb$ converges in probability to a negative constant.
        \item There is a constant $g=g(d,\lambda)>0$ such that for every $\eps>0$ there is $s_0$ such that for every $s\ge s_0$, with high probability,
        \[
            \bigl\|(M_s+gI)\bsigma^{(s)}\bigr\|\le\eps\,\|\bsigma^{(s)}\|\,.
        \]
        \item There is $C=C(d,\lambda)$ such that for every $s$, with high probability 
        \[
            \|\bsigma^{(s)}\|_4^4\le\frac Cn\,\|\bsigma^{(s)}\|^4\,.
        \]
    \end{enumerate}
\end{lemma}

Now we can prove \Cref{lem:bethe-hessian-structure}.

\begin{proof}[Proof of \Cref{lem:bethe-hessian-structure}]
The bound on $\left\| \parens*{ M_s + g I }\bsigma^{(s)} \right\|$ and the delocalization bounds in \Cref{lem:bethe-hessian-structure} are \Cref{lem:bh-estimates}(b) and (c) respectively.
It remains to count negative eigenvalues of $M_s$.
Let $g$ be the constant in \Cref{lem:bh-estimates}(b).
Fix $\eps>0$, without loss of generality we can assume $\eps < g/2$. Let $s_0$ be given by \Cref{lem:bh-estimates}(b) applied with $\min\{\eps,g/2\}$ in place of $\eps$.
Fix $s>s_0$.
Everything below holds on the intersection of the high-probability events in \Cref{lem:robust-bh-inertia,lem:bh-estimates}.

First we show that $M_s$ has at most one negative eigenvalue.
If $\lambda < 0$ then $M_s = (d \lambda)^{-2s} A^{(s)} H A^{(s)}$.
Multiplication on right and left by $A^{(s)}$ cannot increase the number of negative eigenvalues, and by \Cref{lem:robust-bh-inertia} $H$ has at most one negative eigenvalue.

If $\lambda > 0$, then $H$ could have $2$ negative eigenvalues.
It will be enough to show that the rank-one corrected matrix
\[
H' = H +\frac{2(Hb)(Hb)^\top}{-b^\top Hb}
\]
has at most one negative eigenvalue.
This uses only simple linear algebra to show that such a rank-one correction of a matrix with $2$ negative eigenvalues leads to one with a single negative eigenvalue.
We construct an $(n-1)$-dimensional subspace on which $H'$ is PSD.
Let $U$ be the span of the eigenvectors of $H$ with nonnegative eigenvalues, so that $\dim U\ge n-2$ by \Cref{lem:robust-bh-inertia}.
By \Cref{lem:bh-estimates}(a), $-b^\top Hb>0$, so $b\notin U$ and $U \oplus \operatorname{span}\{b\}$ has dimension at least $n-1$.
For $y=u+\alpha b$ with $u\in U$, expanding and completing the square gives
\[
    y^\top H'y=u^\top Hu-b^\top Hb\,\Bigl(\alpha+\frac{b^\top Hu}{b^\top Hb}\Bigr)^2-\frac{(b^\top Hu)^2}{b^\top Hb}\ge0\,.
\]

It remains now to show that $M_s$ also has at least one negative eigenvalue.
Let $u=\bsigma^{(s)}/\|\bsigma^{(s)}\|$.
By \Cref{lem:bh-estimates}(b), $\|(M_s+g I)u\|\le g/2$, so
\[
    u^\top M_su=u^\top(M_s+gI)u-g\le\|(M_s+gI)u\|-g<0\,. \qedhere
\]
\end{proof}

\subsection{Reduction to the broadcast tree}
\label{subsec:bh-reduction}

In this section we prove \Cref{lem:bh-estimates} by assembling a number of facts about the broadcasting on trees process.
We start with some definitions.

\begin{definition}[Local statistics]
An \emph{$R$-local statistic} is a real-valued function $f$ of a rooted graph whose vertices carry $\pm1$ labels such that, for some $R$, $f$ depends only on the radius-$R$ neighborhood of the root, whose magnitude is always at most $(1+N)^R$, where $N$ is the number of vertices of that neighborhood.
For a labeled graph which will be clear from context and a vertex $v$, we write $f(v)$ for $f$ evaluated on the graph rooted at $v$.
\end{definition}

Local statistics form a vector space -- it's easy to check that if $f,h$ are both $R$-local then $c \cdot f$ is $(R + \log |c|)$-local and $f+h$ is $(R+1)$-local.
(This fact depends on using $(1+N)^R$ in the definition rather than the more-natural-looking $N^R$.)
Furthermore, the broadcast process induces an inner product on local statistics.
Let $\iprod{f,h} = \E_{\bT, x}[ f(o) h(o)]$ where $(\bT,x)$ is a labeled broadcast tree with parameters $d,\lambda$ whose root is $o$, and let $\|f\|_{L^p} = (\E |f(o)|^p)^{1/p}$.
We write $\|f\|$ for $\|f\|_{L^2}$.

Mildly abusing notation, for each of the 
matrices involved in \Cref{def:sandwiched-bethe-hessian} we introduce a linear operator on local statistics with the same name.
In particular,
\[
    (A^{(j)}f)(v)=\sum_{(v_0,\dots,v_j)}f(v_j)\,,
    \qquad
    (Df)(v)=\deg{v} \cdot f(v)
\]
where the sum is over nonbacktracking walks with $v_0=v$, and the notation $f(v_j)$ means that $f$ is evaluated on the same underlying labeled graph which contains $v$, but moving the root to vertex $v_j$.
$H$ is defined analogously.
Locality is well behaved under these linear transformations: $A^{(j)}$ maps $R$-local statistics to $(R+j)$-local statistics, and $D$ maps $R$-local statistics to $(R+1)$-local statistics.

In particular, continuing to abuse notation by conflating vectors in $\R^n$ and local statistics, $\bsigma^{(s)}$ is the local statistic $(d \lambda)^{-s} A^{(s)} x$, and $b$ and $Hb$ are local statistics.
Products of local statistics are again local statistics.

We will repeatedly use the following immediate consequence of \Cref{fact:local-law}, which justifies our abuse of notation -- it says that as far as local statistics are concerned, the inner product spaces induced by the SBM graph and by the broadcast tree act the same.

\begin{lemma}[Transfer to the broadcast tree, corollary of \Cref{fact:local-law}]
    \label{lem:bh-transfer}
    For every $R \in \N$ and every pair of $R$-local statistics $f,h$,
    if $(\bG,\bsigma^*) \sim \SBM(n,d,\lambda)$,
    \[
        \frac1n\sum_{v\in[n]}f(v)h(v)\xrightarrow{\Pr}\iprod{f,h}_{\bcalT}\,,
    \]
    where in the sum $v$ ranges over vertices of $\bG$.
\end{lemma}

The following lemma collects the estimates on the broadcast tree that we need; we prove it in the next subsection.
Importantly, each of the quantities in the lemma concerns a local statistic.

\begin{lemma}[Tree estimates, proved in \Cref{subsec:bh-tree-proofs}]
    \label{lem:bh-tree}
    \leavevmode
    \begin{enumerate}[(i)]
        \item If $\lambda>0$ and $L$ is large enough, then $\iprod{b,Hb}_{\bcalT}<0$.
        \item If $\lambda>0$, then $\iprod{Hb,A^{(s)}\bsigma^{(s)}}_{\bcalT}=0$ for every $s$.
        \item For every $s$, $\|\bsigma^{(s)}\|_{L^2}\ge1$, and $\sup_s\|\bsigma^{(s)}\|_{L^4}<\infty$.
        \item There is a constant $g=g(d,\lambda)>0$ such that, as $s\to\infty$,
        \[
            \bigl\|(d\lambda)^{-2s}A^{(s)}HA^{(s)}\bsigma^{(s)}+g\,\bsigma^{(s)}\bigr\|_{L^2}\to0\,.
        \]
    \end{enumerate}
\end{lemma}

\begin{proof}[Proof of \Cref{lem:bh-estimates}]
Part (a) is immediate from \Cref{lem:bh-transfer} and \Cref{lem:bh-tree}(i).
Part (c) is immediate from \Cref{lem:bh-transfer} and \Cref{lem:bh-tree}(iii).

We turn to part (b).
Most of the work is to show that the rank-one correction term has negligible interaction with $\bsigma^{(s)}$.
For notational convenience, let
\[
    r_s=(d\lambda)^{-2s}A^{(s)}HA^{(s)}\bsigma^{(s)}+g\,\bsigma^{(s)}
\]
be the local statistic in \Cref{lem:bh-tree}(iv).
(As usual we sometimes abuse notation and also write $r_s$ for the $n$-dimensional vector associated to an SBM graph.)

Fix $\eps>0$.
By \Cref{lem:bh-tree}(iv), there is $s_0$ such that $\|r_s\|_{L^2}\le\eps/4$ for every $s\ge s_0$; fix such an $s$.
By \Cref{lem:bh-transfer}, $\frac1n\|r_s\|^2$ converges in probability to a limit at most $\eps^2/16$, so with high probability $\|r_s\|\le\frac\eps3\sqrt n\le\frac\eps2\|\bsigma^{(s)}\|$, using $\|\bsigma^{(s)}\|^2\ge n/2$.
If $\lambda<0$, then $M_s=(d\lambda)^{-2s}A^{(s)}HA^{(s)}$, so $(M_s+gI)\bsigma^{(s)}=r_s$, so we are done with the $\lambda < 0$ case.
If $\lambda>0$, then
\[
    (M_s+gI)\bsigma^{(s)}=r_s+(d\lambda)^{-2s}A^{(s)}\,\frac{2(Hb)(Hb)^\top}{-b^\top Hb}\,A^{(s)}\bsigma^{(s)}\,.
\]
The last term is $A^{(s)}Hb$ times the scalar $2(Hb)^\top A^{(s)}\bsigma^{(s)}/((d\lambda)^{2s}(-b^\top Hb))$, so divided by $\sqrt n$ it has norm
\[
    \frac{2\,\|A^{(s)}Hb\|}{(d\lambda)^{2s}\sqrt n}\cdot\frac{\bigl|\frac1n(Hb)^\top A^{(s)}\bsigma^{(s)}\bigr|}{-\frac1nb^\top Hb}\,.
\]
By \Cref{lem:bh-transfer}, the first factor converges to $2(d\lambda)^{-2s}\|A^{(s)}Hb\|_{L^2}<\infty$, the numerator converges to $\iprod{Hb,A^{(s)}\bsigma^{(s)}}_{\bcalT}=0$ by \Cref{lem:bh-tree}(ii), and the denominator converges to $-\iprod{b,Hb}_{\bcalT}>0$ by \Cref{lem:bh-tree}(i).
So with high probability the last term has norm at most $\frac\eps3\sqrt n\le\frac\eps2\|\bsigma^{(s)}\|$, and the triangle inequality gives (b).
\end{proof}

\subsection{Proof of the tree estimates}
\label{subsec:bh-tree-proofs}

We now prove \Cref{lem:bh-tree}.
We prove them roughly in order of difficulty, (ii), (iii), (i), and finally (iv), building up the necessary infrastructure about broadcast trees along the way.

\begin{definition}[Odd and even local statistics]
Call a local statistic \emph{odd} if reversing all labels negates it, and \emph{even} if reversing all labels leaves it unchanged.
For example, the root label $x$ is odd, $\mathbf1$ is even, and $A^{(j)}$, $D$, and $H$ preserve parity, since they do not look at labels.
\end{definition}

Since the law of the Galton-Watson tree is invariant under flipping the signs of the labels, even and odd local statistics are mutually orthogonal.
Now we can discharge the proof of (ii).

\begin{proof}[Proof of \Cref{lem:bh-tree}(ii)]
The statistic $Hb$ is even and $A^{(s)}\bsigma^{(s)}$ is odd.
\end{proof}

Now we assemble a few more tools needed for (iii).
For $m\ge0$, let $\mathcal F_m$ be the $\sigma$-algebra generated by the broadcast tree and its labels up to depth $m$, and let $Z_m$ be the number of vertices at depth $m$, so that $\E Z_m=d^m$.

\begin{lemma}[Convergence of $\bsigma^{(s)}$]
    \label{lem:bh-shells}
    As $r\to\infty$, $\sup_{s\ge r}\|\bsigma^{(s)}-\bsigma^{(r)}\|_{L^4}\to0$.
\end{lemma}

\begin{proof}
Writing $\bsigma^{(s)}-\bsigma^{(r)}=\sum_{j=r+1}^s(\bsigma^{(j)}-\bsigma^{(j-1)})$ as a telescoping sum, it suffices to show that $\|\bsigma^{(j)}-\bsigma^{(j-1)}\|_{L^4}$ is summable in $j$.
Given $\mathcal F_{j-1}$, each vertex $u$ at depth $j-1$ has independent $\Poi(d(1+\lambda)/2)$ and $\Poi(d(1-\lambda)/2)$ numbers of children with signs $\bx_u$ and $-\bx_u$, and expanding $\bsigma^{(j)}$ and $\bsigma^{(j-1)}$ each evaluated at the root we get
\[
    (\bsigma^{(j)}-\bsigma^{(j-1)})(o)= (d \lambda)^{-j} \cdot \sum_{u \text{ at depth $j-1$} } \Bigl(\sum_{c\text{ child of }u}\bx_c-d\lambda\bx_u\Bigr)\,.
\]
Given $\mathcal F_{j-1}$, the terms in the sum $\sum_{u \text{ at depth $j-1$}} (\cdots)$ are independent and mean zero, and the terms in the sum all have the same fourth moment, which depends only on $d,\lambda$ and is independent of $j$.
There are $Z_{j-1}$ terms in the sum.
So
\[
\E [ (\bsigma^{(j)} - \bsigma^{(j-1)})(o)^4  \, | \, \mathcal F_{j-1} ] \leq (d \lambda)^{-4j} \cdot O_{d,\lambda}( Z_{j-1}^2 ) \, .
\]
Since $\E Z_m^2 \leq O_{d} (d^{2m})$, taking expectations,
\[
    \|\bsigma^{(j)}-\bsigma^{(j-1)}\|_{L^4}^4\le(d\lambda)^{-4j}\cdot O_{d,\lambda}(d^{2j})=O_{d,\lambda}\bigl((d\lambda^2)^{-2j}\bigr)\,,
\]
so $\|\bsigma^{(j)}-\bsigma^{(j-1)}\|_{L^4}=O_{d,\lambda}((d\lambda^2)^{-j/2})$, which is summable in $j$ since $d\lambda^2>1$.
\end{proof}

Now we can prove \Cref{lem:bh-tree}(iii).

\begin{proof}[Proof of \Cref{lem:bh-tree}(iii)]
A quick calculation shows $\iprod{x, \bsigma^{(s)}} = 1$, and since $\|x\| = 1$, by Cauchy-Schwarz we must have $\|\bsigma^{(s)}\| \geq 1$.
The fourth moment bound is immediate from \Cref{lem:bh-shells}.
\end{proof}

Next we prove \Cref{lem:bh-tree}(i), using a similar telescoping-sum strategy.

The only computation needed is the following truncation of the generating function of nonbacktracking walks.
It is an identity of matrices on any graph, and of operators on local statistics.

\begin{fact}[Nonbacktracking generating function]
    \label{fact:bh-nb-generating}
    For every $m\ge1$,
    \[
        H\sum_{j=0}^mt^jA^{(j)}=(1-t^2)I-t^{m+1}A^{(m+1)}+t^{m+2}(D-I)A^{(m)}\,.
    \]
\end{fact}

\begin{proof}
Expand $H=I-tA+t^2(D-I)$ and apply $AA^{(j)}=A^{(j+1)}+(D-I)A^{(j-1)}$ for $j\ge2$ and $AA^{(1)}=A^{(2)}+(D-I)+I$.
All terms other than the displayed ones cancel in pairs.
\end{proof}

\begin{proof}[Proof of \Cref{lem:bh-tree}(i)]
By definition, $b=F_L-dtF_{L-1}$, where $F_m=\sum_{j=0}^mt^jA^{(j)}\mathbf1$.
Writing $\Delta_j=A^{(j)}\mathbf1-dA^{(j-1)}\mathbf1$, \Cref{fact:bh-nb-generating}, applied with $m=L$ and $m=L-1\ge1$, gives
\begin{align*}
    Hb&=HF_L-dt\,HF_{L-1}\\
    &=(1-t^2)\mathbf1-t^{L+1}A^{(L+1)}\mathbf1+t^{L+2}(D-I)A^{(L)}\mathbf1\\
    &\qquad-dt\Bigl((1-t^2)\mathbf1-t^{L}A^{(L)}\mathbf1+t^{L+1}(D-I)A^{(L-1)}\mathbf1\Bigr)\\
    &=(1-t^2)(1-dt)\mathbf1-t^{L+1}\Delta_{L+1}+t^{L+2}(D-I)\Delta_L\,.
\end{align*}
Using the same telescoping expansion as in the proof of (iii) applied to  $\Delta_j$ shows $\|\Delta_j\| \leq O(d^{j/2})$ and $\| (D-I) \Delta_j\| \leq O(d^{j/2})$.
Hence we have
\[
  Hb = (1-t^2)(1-dt) \mathbf 1 - O_d( (dt^2)^{L/2}) \, .
\]
Furthermore, $b=\mathbf1+\sum_{j=1}^Lt^j\Delta_j$ has $\iprod{b,\mathbf1}=1$ and $\|b\|^2=\sum_{j=0}^L(dt^2)^j=O(1)$ where we used $dt^2 < 1$.
Hence, $\iprod{b,Hb} = (1 - t^2)(1 - dt) + O((dt^2)^{L/2})$, which is negative for $L$ large enough, since $dt^2<1<dt$.
\end{proof}

We turn to the proof of (iv), which is a bit more involved.
Write $C^{(j)}=(d\lambda)^{-j}A^{(j)}$, so that $\bsigma^{(j)}=C^{(j)}x$.
The key ingredient for (iv) is the following lemma.

\begin{restatable}[$\bsigma^{(s)}$ is an approximate eigenvector of $C^{(s)}$]{lemma}{lemsigmasapproxeigenvecofCs}
    \label{lem:bh-approx-eigen}
    Let $c=d\lambda^2/(d\lambda^2-1)$.
    As $s\to\infty$, for each $s'\in\{s-1,s,s+1\}$,
    \begin{enumerate}[(a)]
        \item $\|C^{(s)}\bsigma^{(s')}-c\,\bsigma^{(s)}\|\to0$,
        \item $\|C^{(s)}(D-I)\bsigma^{(s')}-(dc+1)\,\bsigma^{(s)}\|\to0$, and
        \item $\|C^{(s)}H(C^{(s)}\bsigma^{(s)}-c\,\bsigma^{(s)})\|\to0$.
    \end{enumerate}
\end{restatable}

The basic idea of this lemma is simple: once you are summing up the signs at depth $s$, there isn't much additional information in signs at depth $2s$.
The main difficulty in proving \Cref{lem:bh-approx-eigen} is that as $s \rightarrow \infty$ both $C^{(s)}$ and $\bsigma^{(s)}$ ``see'' more of the tree at the same time; if instead we only needed to analyze $C^{(s)} \bsigma^{(a)}$ for a fixed $a$ as $s \rightarrow \infty$ it would be much easier.
We prove \Cref{lem:bh-approx-eigen} in \Cref{app:bh-approx-eigen}, and turn to the proof of \Cref{lem:bh-tree}(iv).

\begin{proof}[Proof of \Cref{lem:bh-tree}(iv)]
By \Cref{lem:bh-approx-eigen}(c),
\[
    \bigl\|C^{(s)}HC^{(s)}\bsigma^{(s)}-c\,C^{(s)}H\bsigma^{(s)}\bigr\|\to0\,,
\]
as $s \rightarrow \infty$, so it suffices to show that $C^{(s)}H\bsigma^{(s)}$ is close to a negative multiple of $\bsigma^{(s)}$.
For $s\ge2$, the recurrence $AA^{(s)}=A^{(s+1)}+(D-I)A^{(s-1)}$ gives $A\bsigma^{(s)}=d\lambda\,\bsigma^{(s+1)}+(d\lambda)^{-1}(D-I)\bsigma^{(s-1)}$, so
\[
    C^{(s)}H\bsigma^{(s)}=C^{(s)}\bsigma^{(s)}-d\lambda t\,C^{(s)}\bsigma^{(s+1)}-\frac{t}{d\lambda}\,C^{(s)}(D-I)\bsigma^{(s-1)}+t^2\,C^{(s)}(D-I)\bsigma^{(s)}\,.
\]
By \Cref{lem:bh-approx-eigen}(a), the first two terms on the right are close to $c\,\bsigma^{(s)}$ and $-d\lambda t\,c\,\bsigma^{(s)}$, and by \Cref{lem:bh-approx-eigen}(b), the last two are close to $-\frac{t}{d\lambda}(dc+1)\,\bsigma^{(s)}$ and $t^2(dc+1)\,\bsigma^{(s)}$.
Adding these, $\|C^{(s)}H\bsigma^{(s)}-q\,\bsigma^{(s)}\|\to0$, where
\[
    q=(1-d\lambda t)\,c+\Bigl(t^2-\frac{t}{d\lambda}\Bigr)(dc+1)=(1-d\lambda t)\Bigl(c-\frac{t(dc+1)}{d\lambda}\Bigr)\,.
\]
Hence $\|C^{(s)}HC^{(s)}\bsigma^{(s)}+g\,\bsigma^{(s)}\|\to0$ with $g=-c\,q$, and $C^{(s)}HC^{(s)}=(d\lambda)^{-2s}A^{(s)}HA^{(s)}$.
Finally, the choice of $t$ in \Cref{def:sandwiched-bethe-hessian} ensures that $d\lambda t>1$ and $t(dc+1)/(d\lambda)<c$, so $q<0$ and $g>0$.
\end{proof}

\endgroup




\section{Analysis of degree-cutoff preprocessing}
\label{subsec:alg-preprocessing}

\paragraph{Algorithm (Degree-Cutoff Preprocessing).}
Choose a sufficiently large fixed integer $k$.
Delete every edge incident to a vertex whose degree in \(F\) exceeds \(k\), retaining all vertices.
Let the degree truncated graph be $F_k$ and return $\widetilde M=M_s(F_k)$.

Since all the statistics of $M_s(\cdot)$ can be computed by fixed-radius neighborhoods, the \emph{Degree-Cutoff Preprocessing} algorithm can be implemented in linear time.
Now, we show that the output $\widetilde M$ of \emph{Degree-Cutoff Preprocessing} satisfies \Cref{lem:robust-matrix-edit-stability}.

\lemrobustmatrixeditstability*

\paragraph{Notation.} In this section, we use subscripts to denote the graph on which the matrix is computed, e.g., $A_{\bG}^{(s)}$ and $A_{F_k}^{(s)}$ denote the length-$s$ nonbacktracking adjacency matrix of $\bG$ and $F_k$ respectively.
For notational simplicity, we define, for any graph $G$,
\[
    C_G^{(s)} \coloneq (d \lambda)^{-s} A_G^{(s)},\qquad
    a_G  \coloneq -b_G^\top H_Gb_G,\qquad
    h_G  \coloneq C_G^{(s)}H_Gb_G.
\]

\begin{proof}[Proof of \Cref{lem:robust-matrix-edit-stability}]
Let $\bG_k$ be the degree-truncated graph of $\bG$.
By the triangle inequality,
\begin{equation}
\begin{aligned}
    \|\widetilde M-M_s(\bG)\|_{1,2}
       &\le\|M_s(F_k)-M_s(\bG_k)\|_{1,2}\\
       &\quad+\|M_s(\bG_k)-M_s(\bG)\|_{1,2}.
\end{aligned}
    \label{eq:algo-12norm-preprocessing-triangle}
\end{equation}
The first term is the edge corruption error and the second is the
cutoff error. We bound them separately.

\paragraph{Edge corruption error.}
Let $S$ be the endpoints of edited edges.
Notice that $\eta n$ edge edits can change the degrees of at most $2 \eta n$ vertices.
Only these vertices can change cutoff status between $F_k$ and $\bG_k$
and each contributes at most $k$ edge differences. Hence,
\[
    |E(F_k)\triangle E(\bG_k)|\le(2k+1) \eta n.
\]
Since both $F_k$ and $\bG_k$ have maximum degree $k$, each vertex starts at most $k^s$ nonbacktracking walks of length $s$. Moreover, each row of the Bethe Hessian has at most $k$ off-diagonal entries of magnitude $|t|$ and diagonal entries of magnitude at most $1+t^2(k+1)$.
Therefore,
\[
    \|C_{F_k}^{(s)}\|_{1,\infty},\,
    \|C_{\bG_k}^{(s)}\|_{1,\infty}
    \le |d\lambda|^{-s}k^s,
    \qquad
    \|H_{F_k}\|_{1,\infty},\,
    \|H_{\bG_k}\|_{1,\infty}
    \le 1+|t|k+t^2(k+1).
\]
It follows that
\[
    \|C_{F_k}^{(s)}H_{F_k}C_{F_k}^{(s)}\|_{1,\infty}
    \leq \|C_{F_k}^{(s)}\|_{1,\infty}^2 \|H_{F_k}\|_{1,\infty}
    =O(1).
\]
Observe that the $(u,v)$ entry of the matrix $C_{G}^{(s)}H_{G}C_{G}^{(s)}$
is a normalized weighted sum over pairs of length-$s$ nonbacktracking walks in $G$,
one starting at $u$ and one at $v$, whose endpoints coincide or are adjacent.
Therefore, only rows within distance $2s+1$ of $S$ can differ between
$C_{F_k}^{(s)}H_{F_k}C_{F_k}^{(s)}$ and $C_{\bG_k}^{(s)}H_{\bG_k}C_{\bG_k}^{(s)}$.
Hence, the row-sum bounds give
\[
    \|C_{F_k}^{(s)}H_{F_k}C_{F_k}^{(s)}
       -C_{\bG_k}^{(s)}H_{\bG_k}C_{\bG_k}^{(s)}\|_{1,2}
       =O(\sqrt{\eta n}).
\]
This concludes the bound for $\lambda<0$.
For $\lambda>0$, we need to bound the error in the rank-one corrections.
Recall that, by the definition of $b_{\bG_k}$, it is an $L$-neighbourhood local statistic.
Since the degrees are bounded by $k$ and $L, s$ are fixed constants,
the entries of $h_{F_k}$ and $h_{\bG_k}$ are bounded.
The differences in $h$ between the two graphs are supported on distance $L+s+1$ neighbourhood
of vertices in $S$, and, hence, the support of $h_{F_k}-h_{\bG_k}$ is of size at most $O((2k)^{L+s+1} \eta n)$.
Thus,
\[
    \|h_{F_k}\|_{\infty},\|h_{\bG_k}\|_{\infty}=O(1),\qquad
    \|h_{F_k}-h_{\bG_k}\|_1=O(\eta n),\qquad
    \|h_{F_k}-h_{\bG_k}\|_2=O(\sqrt{\eta n}).
\]
Hence,
\begin{align*}
    \|h_{F_k}h_{F_k}^{\top}-h_{\bG_k}h_{\bG_k}^{\top}\|_{1,2}
        & \leq\|(h_{F_k}-h_{\bG_k})h_{F_k}^{\top}\|_{1,2}
            +\|h_{\bG_k}(h_{F_k}-h_{\bG_k})^{\top}\|_{1,2}\\
        & \leq\|h_{F_k}-h_{\bG_k}\|_2\|h_{F_k}\|_1 +\|h_{\bG_k}\|_2\|h_{F_k}-h_{\bG_k}\|_1 \\
        & = O\!\left((\eta + \sqrt{\eta}) n^{3/2} \right)
        = O\!\left(\sqrt{\eta} n^{3/2} \right).
\end{align*}
Following similar local analysis, we can show that
\[
    |a_{F_k}-a_{\bG_k}|=O(\eta n).
\]
By \Cref{lem:bh-estimates}, we have $a_{\bG} = \Omega(n)$ with high probability.
Taking a large enough $k$ and using similar local analysis on $a_{\bG_k}$ yields the lower bound $a_{\bG_k} = \Omega(n)$.
Therefore, we can get $a_{F_k},a_{\bG_k} = \Omega(n)$ with high probability.
Now, we can bound the rank-one correction differences as
\begin{align*}
    \left\|\frac{2h_{F_k}h_{F_k}^{\top}}{a_{F_k}}-\frac{2h_{\bG_k}h_{\bG_k}^{\top}}{a_{\bG_k}}\right\|_{1,2}
    & = \left\|\frac{2h_{F_k}h_{F_k}^{\top}}{a_{F_k}} - \frac{2h_{\bG_k}h_{\bG_k}^{\top}}{a_{F_k}}
    +\frac{2h_{\bG_k}h_{\bG_k}^{\top}}{a_{F_k}} -\frac{2h_{\bG_k}h_{\bG_k}^{\top}}{a_{\bG_k}}\right\|_{1,2} \\
    & \le \frac{2}{a_{F_k}} \left\|h_{F_k}h_{F_k}^{\top}-h_{\bG_k}h_{\bG_k}^{\top}\right\|_{1,2}
             +\frac{2|a_{F_k}-a_{\bG_k}|}{a_{F_k}a_{\bG_k}}\|h_{\bG_k}\|_1 \|h_{\bG_k}\|_2 \\
    &=O\!\left((\eta + \sqrt{\eta}) \sqrt{n}\right)
    = O\!\left(\sqrt{\eta n}\right).
\end{align*}
Thus, by picking $\eta$ small enough, we can get
\begin{align}
     \|\widetilde M-M_s(\bG_k)\|_{1,2}
    &\le
       \|C_{F_k}^{(s)}H_{F_k}C_{F_k}^{(s)}
         -C_{\bG_k}^{(s)}H_{\bG_k}C_{\bG_k}^{(s)}
       \|_{1,2} \notag \\
    &\quad+
       \left\|
       \frac{2h_{F_k}h_{F_k}^{\top}}{a_{F_k}}
       -\frac{2h_{\bG_k}h_{\bG_k}^{\top}}{a_{\bG_k}}
       \right\|_{1,2} \notag \\
    &=O\!(\sqrt{\eta n}) \leq \frac{\varepsilon}{2} \sqrt{n},
\label{eq:degree-12norm-cutoff-L1-1}
\end{align}
and,
\begin{equation}
    \|\widetilde M\|_{1,\infty}
       \le\|C_{F_k}^{(s)}H_{F_k}C_{F_k}^{(s)}\|_{1,\infty}
          +\frac{2 \|h_{F_k}\|_{\infty} \|h_{F_k}\|_{1}}{a_{F_k}}
        = O(1).
    \label{eq:degree-12norm-cutoff-operator-norm}
\end{equation}

\paragraph{Cutoff error.}
Let $r_k(i)$ denote the absolute row sum for row $i$ of $C_{\bG_k}^{(s)}H_{\bG_k}C_{\bG_k}^{(s)}-C_{\bG}^{(s)}H_{\bG}C_{\bG}^{(s)}$.
Define $r_k(o)$ analogously for the Poisson tree.
Since the row sums $r_k(i)$ are local statistics of radius $2s+2$,
the walk-counting bound on the Poisson tree gives
\[
    r_k(o)^2
       \le C^2(1+|B_{2s+2}(o)|)^{4s+2}
       \Ind\!\left[
           \exists v\in B_{2s+1}(o):\operatorname{deg}(v)>k
       \right],
\]
where $B_r(o)$ is the radius $r$ neighbourhood of $o$ and $C$ is a constant independent of $k$.
The indicator tends to zero almost surely as $k\to\infty$.
Since fixed-radius Poisson-tree neighborhoods have moments of every order,
by dominated convergence, we have $\E[r_k(o)^2]\to0$.
Hence, by the local law \Cref{fact:local-law}, it follows that
\[
    \frac1n
    \left\|
        C_{\bG_k}^{(s)}H_{\bG_k}C_{\bG_k}^{(s)}
          -C_{\bG}^{(s)}H_{\bG}C_{\bG}^{(s)}
    \right\|_{1,2}^2
       =\frac1n\sum_i r_k(i)^2
       \xrightarrow{\Pr}\E[r_k(o)^2].
\]
Consequently, for every $\delta>0$ and every sufficiently large
fixed $k$, with high probability,
\[
    \|C_{\bG_k}^{(s)}H_{\bG_k}C_{\bG_k}^{(s)}
       -C_{\bG}^{(s)}H_{\bG}C_{\bG}^{(s)}\|_{1, 2}
       \le\delta \sqrt{n}.
\] 
This concludes the cutoff bound for $\lambda<0$.
For $\lambda>0$, we can apply the same argument to bound the squared coordinate differences of $h_{\bG_k}$ and $h_{\bG}$ and $|a_{\bG_k}-a_{\bG}|$.
Therefore, using the same analysis as in \emph{edge corruption error}, it follows that, for every $\delta>0$ and every sufficiently large fixed $k$, with high probability,
\[
    \left\|\frac{2h_{\bG}h_{\bG}^\top}{a_{\bG}}
          -\frac{2h_{\bG_k}h_{\bG_k}^\top}{a_{\bG_k}}\right\|_{1, 2}
        \le \delta \sqrt{n}.
\]
Combining both parts and choosing $\delta$ sufficiently small yields
\begin{equation}
    \|M_s(\bG_k)-M_s(\bG)\|_{1, 2}
        \le \frac{\varepsilon}{2} \sqrt{n}.
    \label{eq:degree-12norm-cutoff-L1-2}
\end{equation}

\paragraph{Conclusion}
Plugging \eqref{eq:degree-12norm-cutoff-L1-1} and \eqref{eq:degree-12norm-cutoff-L1-2} into \eqref{eq:algo-12norm-preprocessing-triangle},
it follows that, with high probability,
\[
    \|\widetilde M-M_s(\bG)\|_{1, 2}
        \le \varepsilon \sqrt{n}.
\]
Combining this with the bound for $\|\widetilde M\|_{1, \infty}$ in \eqref{eq:degree-12norm-cutoff-operator-norm} finishes the proof.
\end{proof}

\section{SDP recovery and implementation}
\label{sec:sdp-recovery-implementation}

Let $n$ be an integer with $n\ge2$.
For $K\ge1$, define
\[
 \mathcal S\defeq\{X\in\mathbb R^{n\times n}:X\succeq0,\ \operatorname{Tr}X=1\},
 \qquad
 \mathcal D_K\defeq\{q\in\mathbb R^n:\|q\|_2^2\le K/n\}.
\]
For matrices $E$, write $\|E\|_{1,p}\defeq(\sum_{i}(\sum_j|E_{ij}|)^{p})^{1/p}$.
For matrices $B,C\in\mathbb R^{n\times n}$, write $\langle B,C\rangle=\operatorname{Tr}(B^{T}C)$.
For $p\in\mathbb R^n$ and $D\subseteq\mathbb R^n$, write $\operatorname{dist}_1(p,D)=\inf_{q\in D}\|p-q\|_1$.

\medskip

The following theorem shows that almost optimal solutions of a particular convex optimization problem over density matrices \(\mathcal S\) are close to the projector to our target vector's span.
Its proof in \Cref{subsec:sdp-seven-recovery-analysis} combines ideas from earlier recovery algorithms for stochastic block models \cite{GV16} and robust principal component analysis \cite{CLMW11}.

\begin{theorem}[Matrix recovery]
  \label{thm:sdp-seven-recovery}
Let $K\ge1$, $g>0$, and $\epsilon\ge0$.
Suppose $M,E\in\mathbb R^{n\times n}$ are symmetric and $v,w\in\mathbb R^n$ are unit vectors satisfying
\[
 M\succeq-gvv^T,\qquad n\|w\|_4^4\le K,\qquad w^TMw\le-(1-\epsilon)\cdot g.
\]
Set $A=M+E$ and let $P\ge\|A\|_{1,\infty}$.
Define $F(X)=\langle A,X\rangle+P\cdot \operatorname{dist}_1(\operatorname{diag}X,\mathcal D_K)$.
Then
\[
   \Abs{\tfrac 1 g\min_{\mathcal S}F + 1} \le \epsilon + \sqrt{K/n}\|E\|_{1,2}/g,
\]
and every $X\in\mathcal S$ with $F(X)\le\min_{\mathcal S}F+\epsilon g$ satisfies
\[
  \|X-ww^T\|_F^2 \le 20\left( \epsilon+\sqrt{K/n}\|E\|_{1,2}/g \right).
\]
\end{theorem}

The next theorem shows that the convex program from the previous theorem statement admits nearly-linear time algorithms assuming that the input matrix supports nearly-linear time matrix-vector operations.
Its proof in \Cref{subsec:sdp-seven-implementation} adapts ideas from \cite{AK07}.

\begin{theorem}[Fast optimization]\label{thm:sdp-seven-factorized-optimization}
  Given $K\ge1$, $0<\delta\le P$, a symmetric matrix $A\in\mathbb R^{n\times n}$ through a matrix-vector multiplication oracle, and a bound $P\ge\|A\|_{1,\infty}+1$, a randomized algorithm computes $\mathbf Z\in\mathbb R^{n\times r}$ such that
\[
 \mathbf Z\mathbf Z^T\in\mathcal S,\qquad
 \mathbb E F(\mathbf Z\mathbf Z^T)\le\min_{\mathcal S}F+\delta,\qquad
 r\le O(1+P^4\log n/\delta^4),
\]
where $F(X)=\langle A,X\rangle+P\cdot\operatorname{dist}_1(\operatorname{diag}X,\mathcal D_K).$
The algorithm uses $O((1+P/\delta)^5(\log n)^2)$ matrix-vector multiplications with $A$ and at most $n$ times as many additional operations.
\end{theorem}

Together the above theorems imply the main algorithmic lemma from \Cref{sec:robust-estimation-of-second-eigenvector}.

\lemrobustaltsdp*

\begin{proof}
Assume $n\ge2$ and $\varepsilon$ is sufficiently small; otherwise
the conclusion is trivial. Set $A=\widetilde M$, $P=C$, and
$K=1/\varepsilon$.
Since \( \sqrt{K/n}\|A-M\|_{1,2}\le\sqrt\varepsilon\,g\),
\Cref{thm:sdp-seven-recovery} gives \( F^{*}\defeq \min_{\mathcal S}F\in[-2g,-g/2]\), so \(g/2\le\abs{F^*}\le2g\).

At stage $j\in \set{0,1,\ldots }$, run the fast solver from \Cref{thm:sdp-seven-factorized-optimization} with $\delta_j=C2^{-j}$.
Retain the best output \(X_{j}\) among
$\lceil\log_2(1/\rho_j)\rceil$ independent runs, where
$\rho_j=n^{-10}/(10j+10)^{2}$.
Markov's inequality gives, simultaneously at all stages except
with probability $\sum_{j} \rho_{j} < n^{-10}/4$,
\[
 0\le F(X_j)-F^{*}\le2\delta_j.
\]
Stop when $4\delta_j\le\varepsilon(-F(X_j))$.\footnote{
  In order to check this condition, we need to evaluate \(F(X_{j})\).
  The linear part of \(F\) corresponds to a small number of matrix-vector products with \(\widetilde M\).
  Moreover, we can compute the \(\ell_{1}\) distance of a vector to a centered Euclidean ball in time \(O(n\log n)\).
}
This condition holds once $\delta_j\le\varepsilon\abs{F^*}/8$, and at stopping
\[
 F(X_j)- F^{*}\le2\delta_j\le\varepsilon\abs{F^*}/2\le\varepsilon g.
\]
Hence the recovery theorem gives
$\|X_j-ww^T\|_F^2=O(\sqrt\varepsilon)$.

Write $X_j=\lambda uu^T+R$, where $u$ is its leading unit
eigenvector. Then $\lambda\ge1/2$,
$\|R\|_F^2=O(\sqrt\varepsilon)$, and
$1-\langle u,w\rangle^2=O(\sqrt\varepsilon)$.
For a uniform random sign vector $\xi$,
\[
 \Pr\bigl[|\langle u,\xi\rangle|\ge1/\sqrt2\bigr]\ge1/12,
 \qquad
 \mathbb E\|R\xi\|^2=\|R\|_F^2.
\]
Thus, with constant probability, normalizing $X_j\xi$ gives
squared correlation $1-O(\sqrt\varepsilon)$ with $u$.
Repeat $O(\log n)$ times with independent randomness and retain the
normalized nonzero vector with largest Rayleigh quotient for \(X_{j}\).
In this way, the selected vector has squared correlation $1-O(\sqrt\varepsilon)$ with $u$,
except with probability $n^{-10}/2$, and therefore also  squared correlation $1-O(\sqrt\varepsilon)$ with $w$ as claimed.

The search uses $O(1+\log(C/(\varepsilon g)))$ stages,
all with $\delta_j=\Omega(\varepsilon g)$.
The fast-solver invocations, objective evaluations, and the matrix-vector products $X_j\xi=Z_j(Z_j^T\xi)$ therefore are within the desired
bounds on the number of matrix-vector products with \(\widetilde M\) and the number of additional operations.
The total failure probability is at most $n^{-10}$.
\end{proof}

\subsection{Recovery analysis}\label{subsec:sdp-seven-recovery-analysis}

To prove \Cref{thm:sdp-seven-recovery}, we show that an $\epsilon g$-approximate minimizer $X\in\mathcal S$ of the penalized objective $F$ is close to $ww^T$ in Frobenius norm. Here $w$ is the unit witness vector from the theorem, and $E$ is the perturbation in $A=M+E$. Recall that $M\succeq-gvv^T$, $w^TMw\le-g+\epsilon g$, and $F(X)=\langle A,X\rangle+P\operatorname{dist}_1(\operatorname{diag}X,\mathcal D_K)$ with $A=M+E$ and $P\ge\|A\|_{1,\infty}$.

\begin{proof}[Proof of \Cref{thm:sdp-seven-recovery}]
Put $r=\sqrt{K/n}$ and $b=r\|E\|_{1,2}$.
For every $U\succeq0$, symmetry and
$|U_{ij}|\le(U_{ii}+U_{jj})/2$ give
\[
 |\langle E,U\rangle|
 \le\sum_i U_{ii}\sum_j|E_{ij}|
 \le\|\operatorname{diag}U\|_2\|E\|_{1,2}.
\]

Let $X\in\mathcal S$ and write $d=\operatorname{diag}X$.
If $\|d\|_2\le r$, set $q=d$.
Otherwise choose $t>0$ such that
$q_i=\min(d_i,t)$ satisfies $\|q\|_2=r$.
Then
\[
 h\defeq\sum_i(d_i-q_i)
 =\operatorname{dist}_1(d,\mathcal D_K).
\]
Indeed, in the latter case $y=q/t$ satisfies
$\|y\|_\infty\le1$, $\langle y,d-q\rangle=h$, and
$\langle y,z\rangle\le\langle y,q\rangle$ for every
$z\in\mathcal D_K$. Hence $\|d-z\|_1\ge h$.

Let $D_{ii}=\sqrt{q_i/d_i}$ when $d_i>0$, and $D_{ii}=1$
otherwise, and put $Y=DXD$.
Then $Y\succeq0$, $\operatorname{diag}Y=q\in \mathcal D_{K}$, and
$\operatorname{Tr}Y\le1$. Moreover,
\[
 |X_{ij}-Y_{ij}|
 \le\sqrt{d_id_j}-\sqrt{q_iq_j}
 \le\frac{d_i-q_i+d_j-q_j}{2}.
\]
The last inequality follows from the AM-GM inequality when both coordinates
are clipped; when only $d_i$ is clipped, use
$\sqrt{d_j}\le\sqrt t$ and
$2\sqrt t(\sqrt{d_i}-\sqrt t)\le d_i-t$.
By symmetry of $A$ and the entrywise bound,
\[
 |\langle A,X-Y\rangle|
 \le\sum_i(d_i-q_i)\sum_j|A_{ij}|
 \le Ph.
\]
Thus the penalty pays for clipping:
\[
 F(X)=\langle A,X\rangle+Ph\ge\langle A,Y\rangle.
\]
Since $Y\succeq0$, $M\succeq-gvv^T$, and
$\|\operatorname{diag}Y\|_2\le r$,
\[
 \langle A,Y\rangle
 \ge-gv^TYv-|\langle E,Y\rangle|
 \ge-gv^TYv-b
 \ge-g-b.
\]
where the last inequality uses $v^TYv\le\operatorname{Tr}Y\le1$.
On the other hand, $\operatorname{diag}(ww^T)\in\mathcal D_K$,
so
\[
 \min_{\mathcal S}F\le F(ww^T)\le-(1-\epsilon)g+b.
\]
Together these prove the first assertion.

Now suppose $F(X)\le\min_{\mathcal S}F+\epsilon g$.
The preceding bounds imply
\[
 -gv^TYv-b\le F(X)\le-g+2\epsilon g+b,
 \qquad
 1-v^TYv\le2\epsilon+2b/g.
\]
It remains to transfer alignment from $Y=DXD$ to $X$.
Put $C=2D-I$, so $\|C\|_{\mathrm{op}}\le1$.
Since $0\preceq X\preceq I$,
\[
 v^TYv
 =\tfrac14\|X^{1/2}v+X^{1/2}Cv\|_2^2
 \le\tfrac12 v^TXv+\tfrac12 v^TCXCv
 \le\tfrac12 v^TXv+\tfrac12.
\]
Therefore, using $\operatorname{Tr}(X^2)\le1$,
\[
 \|X-vv^T\|_F^2
 =\operatorname{Tr}(X^2)+1-2v^TXv
 \le4(1-v^TYv)
 \le8\epsilon+8b/g.
\]
The witness assumption also gives
$(v^Tw)^2\ge1-\epsilon$, hence
$\|vv^T-ww^T\|_F^2\le2\epsilon$.
The squared triangle inequality now yields
\[
 \|X-ww^T\|_F^2
 \le2\|X-vv^T\|_F^2+2\|vv^T-ww^T\|_F^2
 \le20\epsilon+16b/g.
 \qedhere
\]

\end{proof}

\subsection{Fast implementation}\label{subsec:sdp-seven-implementation}

For a convex function $F$ on all real symmetric matrices and a real symmetric matrix $X$, let $\partial F(X)$ denote the set of symmetric matrices $H$ satisfying
$F(U)\ge F(X)+\langle H,U-X\rangle$ for every real symmetric matrix $U$.

The following theorem is the starting point for our fast implementation.
At iteration $t$, we compute $\widetilde X_t\in\mathcal S$ and $H_t\in\partial F(\widetilde X_t)$.
We will work with a low-rank representation of \(\widetilde X_{t}\) in order to be able to efficiently compute \(H_{t}\).
The theorem shows that if $\widetilde X_t$ approximates in direction $H_t$ a particular matrix $X_t$, defined as a function of \(H_{1}+ \cdots + H_{t-1}\), then the average of the computed iterates approximately minimizes $F$ over~$\mathcal S$.
Its proof in \Cref{app:sdp-fast-deferred} is adapted from \cite{AHK12}.

\begin{restatable}[Nearby-point entropy descent]{theorem}{thmsdpsevenentropy}
    \label{thm:sdp-seven-entropy}
    Let $F$ be convex on all real symmetric matrices and attain its minimum over $\mathcal S$.
    Let $\eta>0$, let $T\ge1$ be an integer, and let $H_1,\ldots,H_T\in\mathbb R^{n\times n}$ be symmetric matrices. For $1\le t\le T$, define
    \[
     S_t=-\eta\sum_{s<t}H_s,\qquad
     X_t=\frac{e^{S_t}}{\operatorname{Tr}e^{S_t}}.
    \]
    Suppose $\widetilde X_1,\ldots,\widetilde X_T\in\mathcal S$ satisfy $H_t\in\partial F(\widetilde X_t)$ for $1\le t\le T$.
    Then $\bar X=T^{-1}\sum_{t=1}^T\widetilde X_t$ satisfies
    \[
     F(\bar X)-\min_{\mathcal S}F\le\frac{\log n}{\eta T}+2\eta\max_{1\le t\le T}\|H_t\|_{\mathrm{op}}^2
     +\frac1T\sum_{t=1}^T\langle H_t,\widetilde X_t-X_t\rangle.
    \]
\end{restatable}

Recall that $F(X)=\langle A,X\rangle+P\operatorname{dist}_1(\operatorname{diag}X,\mathcal D_K)$. The linear part of $F$ contributes the constant subgradient $A$. For the $\ell_1$-distance term to $\mathcal D_K$, the following fact bounds the entries of subgradients of the distance function.

\begin{fact}[Distance to a convex set]\label{fact:sdp-seven-distance}
Let $C\subseteq\mathbb R^n$ be nonempty and convex. The function
\[
 f_C(x)=\inf_{q\in C}\|x-q\|_1\qquad(x\in\mathbb R^n)
\]
is finite, convex~\cite[Example~3.16, p.~88]{BV04}, and $1$-Lipschitz in $\ell_1$. At every $x\in\mathbb R^n$, its subdifferential satisfies
$\varnothing\ne\partial f_C(x)\subseteq[-1,1]^n$.
\end{fact}

The above fact directly gives us the form of the subgradients of our objective function $F$, recorded in the following lemma.

\begin{lemma}[Subgradients of the penalized objective]\label{lem:sdp-seven-penalty}
Let $K\ge1$, $P\ge0$, and $A\in\mathbb R^{n\times n}$ be symmetric. Let $F(X)=\langle A,X\rangle+P\operatorname{dist}_1(\operatorname{diag}X,\mathcal D_K)$ on all real symmetric matrices. Every subgradient $H$ of $F$ at a point of $\mathcal S$ has the form
\[
 H=A+\operatorname{Diag}(z),\qquad z\in\mathbb R^n,\quad\|z\|_\infty\le P.
\]
In particular,
\[
 \|H\|_{\mathrm{op}}\le\|A\|_{\mathrm{op}}+P.
\]
\end{lemma}

The following lemma lets us replace the matrix exponential in \Cref{thm:sdp-seven-entropy} by a low-degree matrix polynomial.
(For completeness, we include its standard proof in \Cref{app:sdp-fast-deferred}.)
The degree of this polynomial depends linearly on the spectral norm of the exponent \(S_{t}\).
Since the number of iterations of the algorithm is logarithmic in \(n\), we have a bound on this spectral norm that is logarithmic in \(n\).
Hence matrix-vector multiplications for \(Q\) have the same cost as for \(S_{t}\) up to this logarithmic factor.
However, explicitly computing \(Q\) may take quadratic time even for sparse matrices \(S_{t}\).

\begin{restatable}[Approximate matrix exponential]{lemma}{lemsdpsevenexponentialaction}
    \label{lem:sdp-seven-exponential-action}
    Let $S\in\mathbb R^{n\times n}$ be symmetric with $\|S\|_{\mathrm{op}}\le L$, where $L\ge0$, and let $0<\alpha<1$. Set $m=\left\lceil8(L+\log(3/\alpha)+1)\right\rceil$ and
    \[
     Q=\sum_{j=0}^{m-1}\frac{(S/2)^j}{j!}.
    \]
    Then
    \[
     (1-\alpha)e^S\preceq Q^2\preceq(1+\alpha)e^S.
    \]
    If one product with $S$ costs $T_S$ operations, then, for any $u\in\mathbb R^n$, computing $Qu$ costs $O((L+\log(3/\alpha)+1)(T_S+n))$ operations.
\end{restatable}

As a final ingredient, we sketch PSD matrices \(X\), given access to a factorization $X=VV^{T}$, by a low-rank matrix $ZZ^T$ while controlling quadratic forms in expectation in each fixed direction.

\begin{lemma}[Random-sign quadratic-form estimate]\label{lem:sdp-seven-diagonal}
Let $V\in\mathbb R^{n\times n}$ and let $k\ge1$ be an integer.
Draw $G\in\{-1,1\}^{n\times k}$ with independent uniform entries and set
$Z=k^{-1/2}VG$. For every fixed $u\in\mathbb R^n$,
\[
 \mathbb E\left|\left\langle u,(ZZ^T-VV^T)u\right\rangle\right|
 \le\sqrt{2/k}\left\langle u,VV^Tu\right\rangle.
\]
\end{lemma}
\begin{proof}
We bound the variance of the scalar error. Put $a=V^Tu$ and let $g_j$ be the columns of $G$, so
\[
 u^T(ZZ^T-VV^T)u=k^{-1}\sum_{j=1}^k((a^Tg_j)^2-\|a\|_2^2).
\]

Independence of the signs gives $\mathbb E(a^Tg_j)^2=\|a\|_2^2$ and
\[
 \operatorname{Var}((a^Tg_j)^2)
 =4\sum_{i<\ell}a_i^2a_\ell^2\le2\|a\|_2^4.
\]
Independence of the columns and Cauchy--Schwarz therefore give $\mathbb E|u^T(ZZ^T-VV^T)u|\le\sqrt{2/k}\|a\|_2^2$, the claimed bound since $\|a\|_2^2=u^TVV^Tu$.
\end{proof}

The following lemma combines the preceding two lemmas to efficiently construct a low-rank approximation $\widetilde X$ of $X$, with error controlled in the subgradient directions relevant to minimizing $F$.
We defer its proof to \Cref{app:sdp-fast-deferred}.

\begin{restatable}[Exponential matrix density sketch]{lemma}{lemsdpsevenexponentialdensitysketch}
    \label{lem:sdp-seven-exponential-density-sketch}
    There is a randomized algorithm that, given a symmetric matrix
    $S\in\mathbb R^{n\times n}$ through a matrix--vector product oracle,
    a bound $L\ge\|S\|_{\mathrm{op}}$, and $0<\alpha<1$, returns
    $\mathbf Z\in\mathbb R^{n\times k}$ with $\|\mathbf Z\|_F=1$ and
    $k\le O(\alpha^{-2})$, such that
    $\widetilde X=\mathbf Z\mathbf Z^T\in\mathcal S$ satisfies, for every
    symmetric $A\in\mathbb R^{n\times n}$ and every $P\ge0$,
    \[
     \mathbb E\sup_{\|d\|_\infty\le P}
     \left|\left\langle A+\operatorname{Diag}(d),\widetilde X-X\right\rangle\right|
     \le\alpha(\|A\|_{\mathrm{op}}+P),
    \]
    where $X=e^S/\operatorname{Tr}e^S$.
    The algorithm uses $O(\alpha^{-2}(1+L+\log(1/\alpha)))$
    matrix-vector products with $S$ and $n$ times as many additional operations.
\end{restatable}

In order to efficiently compute subgradients for \(F\) at the points \(\widetilde X_{t}\), we also need the following lemma.

\begin{restatable}[Ball distance oracle]{lemma}{lemsdpsevenoracle}
    \label{lem:sdp-seven-oracle}
    There is a deterministic algorithm that, given $K\ge1$ and $p\in\mathbb R^n$, computes $y\in\partial f(p)$, where $f(x)=\operatorname{dist}_1(x,\mathcal D_K)$ for $x\in\mathbb R^n$.
    The algorithm uses $O(n\log n)$ arithmetic operations.
\end{restatable}

\begin{proof}[Proof of \Cref{thm:sdp-seven-factorized-optimization}]
Set
\[
 \eta=\frac{\delta}{32P^2},\qquad
 T=\left\lceil\frac{128P^2\log n}{\delta^2}\right\rceil,\qquad
 \alpha=\frac{\delta}{16P}.
\]

Initialize $S_1=0$. At iteration $t$, apply
\Cref{lem:sdp-seven-exponential-density-sketch} with fresh
independent randomness and the bound $L_t=2\eta(t-1)P$ to obtain
$\widetilde X_t=\mathbf Z_t\mathbf Z_t^T$.
Compute $\operatorname{diag}\widetilde X_t$ from the squared
row norms of $\mathbf Z_t$.
Writing $f(p)=\operatorname{dist}_1(p,\mathcal D_K)$,
\Cref{lem:sdp-seven-oracle} computes
$y_t\in\partial f(\operatorname{diag}\widetilde X_t)$
in $O(n\log n)$ operations. Thus
\[
 H_t=A+P\operatorname{Diag}(y_t)
 \in\partial F(\widetilde X_t).
\]
By \Cref{lem:sdp-seven-penalty},
$\|H_t\|_{\mathrm{op}}\le\|A\|_{\mathrm{op}}+P\le2P$.
Set $S_{t+1}=S_t-\eta H_t$.
In this way, the iterations maintain the invariant $\|S_t\|_{\mathrm{op}}\le L_t$.

Conditioned on the preceding iterations, $S_t$ is fixed.
Although $y_t$ depends on the current sketch, $\|y_t\|_\infty\le1$,
so the supremum over diagonal perturbations in
\Cref{lem:sdp-seven-exponential-density-sketch} gives
\[
 \mathbb E\langle H_t,\widetilde X_t-X_t\rangle
 \le\alpha(\|A\|_{\mathrm{op}}+P)\le2\alpha P,
 \qquad
 X_t=\frac{e^{S_t}}{\operatorname{Tr}e^{S_t}}.
\]
Consequently, \Cref{thm:sdp-seven-entropy} yields
\[
 \mathbb E F\left(\frac1T\sum_{t=1}^T\widetilde X_t\right)
 -\min_{\mathcal S}F
 \le\frac{\log n}{\eta T}+8\eta P^2+2\alpha P
 \le\delta.
\]
Return
$\mathbf Z=T^{-1/2}[\mathbf Z_1\ \cdots\ \mathbf Z_T]$.
Its Gram matrix is the average above and belongs to $\mathcal S$.

It remains to bound the cost. Maintain
$b_t=\sum_{s<t}y_s$ incrementally, so
\[
 S_tx=-\eta\bigl((t-1)Ax+P\operatorname{Diag}(b_t)x\bigr).
\]
Each product with $S_t$ therefore uses one product with $A$
and $O(n)$ additional operations.
Since $0<\delta\le P$,
\[
 T=O((P/\delta)^2\log n),\qquad
 1+L_t+\log(1/\alpha)=O((P/\delta)\log n).
\]
The sketch lemma thus gives a total of
$O((P/\delta)^5(\log n)^2)$ products with $A$ and
\[
 r=O(T\alpha^{-2})=O(P^4\log n/\delta^4)
\]
output columns.
Computing the diagonals, forming the output factor,
maintaining $b_t$, and the $T$ ball-oracle calls all fit
$n$ times this product bound.
\end{proof}


\section{Analysis of postprocessing by lifting}
\label{subsec:alg-postprocessing}

In this section we prove \Cref{lem:robust-postprocessing}.

\lemrobustpreprocessing*

We first describe the algorithm.
For a graph $Q$ on $[n]$ and $x\in\R^n$, let $R_Qx\in\R^{\vec E(Q)}$ lift to a vector on directed edges:
\[
    (R_Qx)_{u\to v}=x_v-\frac{x_u}{d\lambda}\,.
\]
For $J\ge1$ let $Q_J$ be the graph obtained from $Q$ by deleting every edge with an endpoint of degree more than $J$ in $Q$.
As in \Cref{sec:optimal-recovery}, vectors on the directed edges of different graphs are compared as vectors indexed by all ordered pairs, in $\R^{n^2-n}$.
Given $y$ and $F$, the postprocessing algorithm $\mathcal A_{\text{post}}$ outputs $\widehat\xi(F)=R_{F_J}y/\|R_{F_J}y\|$, for a constant $J=J(d,\lambda,\varepsilon)$ chosen below; it runs in time $O(n+|E(F)|)$.

We put together a few lemmas to prove \Cref{lem:robust-postprocessing}.
First, recall the ``walk score vector'' from \Cref{sec:optimal-recovery},
\[
    W^{(\ell)}_{u\to v}=(d\lambda)^{-\ell}\sum_{\substack{\text{nonbacktracking }(u_0,\dots,u_{\ell+1})\\u_0=u,\ u_1=v}}\bsigma^\star_{u_{\ell+1}}\,,
\]
which is close to the second eigenvector $\bxi$ of the nonbacktracking matrix for SBM graphs.

The first lemma shows that the lifting operation $R_{\bG}$ lifts the vector $A^{(s)} \bsigma^*$ to a vector close to $W^{(s)}$ (up to rescaling).
We prove it at the end of this section.

\begin{lemma}[Lifting the vertex walk score]
    \label{lem:post-lift-identity}
    Let $(\bG,\bsigma^\star)\sim\SBM(n,d,\lambda)$ with $d\lambda^2>1$, and let $W^{(\ell)}$ be the walk scores of \Cref{sec:optimal-recovery}.
    For every $\varepsilon>0$ there is $s_0=s_0(d,\lambda,\varepsilon)$ such that for every fixed $s\ge s_0$, with high probability,
    \[
        \Bigl\|\frac{(d\lambda)^2}{(d\lambda)^2-1}\,\cdot \, R_{\bG}\frac{A_{\bG}^{(s)}\bsigma^\star}{(d\lambda)^s}-W^{(s)}\Bigr\|
        \le\varepsilon\sqrt n\,.
    \]
\end{lemma}

The second lemma captures robustness of the lifting operation -- it's not too harmful that the algorithm uses the corrupted graph $F$ to do the lifting. 
We prove it at the end of this section.

\begin{lemma}[Lifting using the corrupted graph is ok]
    \label{lem:post-edit-stability}
    For every $\varepsilon>0$ there is $J_0(\varepsilon)$ such that for every $J\ge J_0$ there is $\eta>0$ such that for every sufficiently large fixed $s$, with high probability, every graph $F$ on $[n]$ with $|E(F)\triangle E(\bG)|\le\eta n$ satisfies
    \[
        \Bigl\|R_{F_J}\frac{A_{\bG}^{(s)}\bsigma^\star}{(d\lambda)^s}-R_{\bG}\frac{A_{\bG}^{(s)}\bsigma^\star}{(d\lambda)^s}\Bigr\|\le\varepsilon\sqrt n\,.
    \]
    Moreover, every graph $Q$ on $[n]$ with maximum degree at most $J$ satisfies $\|R_Q\|\le O_{d,\lambda,J}(1)$.
\end{lemma}

Finally, we will need some basic bounds on norms of $W^{(s)}$ and $A^{(s)}_{\bG} \bsigma^*$.

\begin{fact}[Norms of the walk scores]
    \label{fact:walk-score-norms}
    Let $(\bG,\bsigma^\star)\sim\SBM(n,d,\lambda)$ with $d\lambda^2>1$.
    There is $C=C(d,\lambda)$, not depending on $s$, such that for every fixed $s\ge1$, with high probability,
    \[
        \frac{\sqrt n}C\le\|W^{(s)}\|\le C\sqrt n
        \quad\text{and}\quad
        \Bigl\|\frac{A_{\bG}^{(s)}\bsigma^\star}{(d\lambda)^s}\Bigr\|\le C\sqrt n\,.
    \]
\end{fact}
These follow from \Cref{fact:local-law} and \cite[Proposition 6]{SM22}; see also \cite[Proposition 10]{SM22}.

\begin{proof}[Proof of \Cref{lem:robust-postprocessing}]

We elide the $\min_{a \in \{ \pm 1\}}$ through this proof for simplicity.
Fix $\eps > 0$.
It will be more convenient to work with a rescaling of $y$ given by $z = \| A_{\bG}^{(s)} \bsigma^* / (d \lambda)^s \| \cdot y$.
With $\delta$ to be chosen later, the algorithm gets as input $z$ such that
\[
\| z - A_{\bG}^{(s)} \bsigma^* / (d \lambda)^s \| \leq \delta \| A_{\bG}^{(s)} \bsigma^* / (d \lambda)^s \| \leq C\delta\sqrt n
\]
with high probability, where $C=C(d,\lambda)$ is the constant of \Cref{fact:walk-score-norms}.

We use this to bound the difference between the lifted vector $R_{F_J} z$ and the second eigenvector, with some convenient scalings chosen to create cancelations.

We have
\begin{align*}
\Bigl\|\frac{(d\lambda)^2}{(d\lambda)^2-1}R_{F_J}z-\|W^{(s)}\|\bxi\Bigr\|
& \le \frac{(d\lambda)^2}{(d\lambda)^2-1}\Bigl\|R_{F_J}\Bigl(z-\frac{A_{\bG}^{(s)}\bsigma^\star}{(d\lambda)^s}\Bigr)\Bigr\|+\frac{(d\lambda)^2}{(d\lambda)^2-1}\Bigl\|R_{F_J}\frac{A_{\bG}^{(s)}\bsigma^\star}{(d\lambda)^s}-R_{\bG}\frac{A_{\bG}^{(s)}\bsigma^\star}{(d\lambda)^s}\Bigr\|\\
& \quad+\Bigl\|\frac{(d\lambda)^2}{(d\lambda)^2-1}R_{\bG}\frac{A_{\bG}^{(s)}\bsigma^\star}{(d\lambda)^s}-W^{(s)}\Bigr\|+\bigl\|W^{(s)}-\|W^{(s)}\|\bxi\bigr\|\,.
\end{align*}

For the first term, $\|R_{F_J}\|\le O_{d,\lambda,J}(1)$ by \Cref{lem:post-edit-stability} and $\bigl\|z-A_{\bG}^{(s)}\bsigma^\star/(d\lambda)^s\bigr\|\le C\delta\sqrt n$, so it is at most $O_{d,\lambda,J}(\delta)\sqrt n$.

For the second term, \Cref{lem:post-edit-stability} bounds it by $\tfrac{(d\lambda)^2}{(d\lambda)^2-1}\,\varepsilon\sqrt n$, choosing $J$ large enough and then $\eta$ small enough.

For the third term, \Cref{lem:post-lift-identity} bounds it by $\varepsilon\sqrt n$ once $s$ is large.

For the fourth term, $\bigl\|W^{(s)}-\|W^{(s)}\|\bxi\bigr\|=\|W^{(s)}\|\cdot\bigl\|W^{(s)}/\|W^{(s)}\|-\bxi\bigr\|\le C\sqrt n\cdot o_{s\rightarrow\infty}(1)$ by \Cref{fact:walk-score-norms,lem:eigenvector-walk-score}.

All together, the left-hand side is at most $\bigl(O_{d,\lambda,J}(\delta)+O_{d,\lambda}(\varepsilon)+o_{s\rightarrow\infty}(1)\bigr)\sqrt n$.
Since $\|W^{(s)}\|\bxi$ has norm $\|W^{(s)}\|\ge\sqrt n/C$, the inequality $\bigl\|x/\|x\|-x'/\|x'\|\bigr\|\le2\|x-x'\|/\|x'\|$ gives $\|\widehat\xi(F)-\bxi\|\le2C\bigl(O_{d,\lambda,J}(\delta)+O_{d,\lambda}(\varepsilon)+o_{s\rightarrow\infty}(1)\bigr)$, which is small once $\varepsilon$ and then $\delta$ are small and $s$ is large.
\end{proof}

It remains to prove \Cref{lem:post-lift-identity,lem:post-edit-stability}.

\begin{proof}[Proof of \Cref{lem:post-lift-identity}]
We will show that $R_{\bG} A_{\bG}^{(s)} \bsigma^*$ is $W^{(s)}_{u \rightarrow v}$ up to rescaling and some explicit low-order corrections.

Fix a directed edge $u\to v$ of $\bG$.
$(A_{\bG}^{(s)} \bsigma^*)_v$ is a sum over nonbacktracking walks beginning at $v$.
We can split that sum according to whether the first step from $v$ goes to $u$ or not, to get
\[
    (A_{\bG}^{(s)}\bsigma^\star)_v=(d\lambda)^sW^{(s)}_{u\to v}+(d\lambda)^{s-1}W^{(s-1)}_{v\to u}\, \ .
\]
Exchanging the roles of $u$ and $v$,
\[
    (A_{\bG}^{(s)}\bsigma^\star)_u=(d\lambda)^sW^{(s)}_{v\to u}+(d\lambda)^{s-1}W^{(s-1)}_{u\to v}\,.
\]
Substituting both into $(R_{\bG}x)_{u\to v}=x_v-x_u/(d\lambda)$,
\begin{align*}
    \Bigl(R_{\bG}\frac{A_{\bG}^{(s)}\bsigma^\star}{(d\lambda)^s}\Bigr)_{u\to v}
    &=W^{(s)}_{u\to v}-\frac{W^{(s-1)}_{u\to v}}{(d\lambda)^2}-\frac{W^{(s)}_{v\to u}-W^{(s-1)}_{v\to u}}{d\lambda}\\
    &=\Bigl(1-\frac1{(d\lambda)^2}\Bigr)W^{(s)}_{u\to v}+\frac{W^{(s)}_{u\to v}-W^{(s-1)}_{u\to v}}{(d\lambda)^2}-\frac{W^{(s)}_{v\to u}-W^{(s-1)}_{v\to u}}{d\lambda}\,.
\end{align*}
By triangle inequality,
\[
    \Bigl\|\frac{(d\lambda)^2}{(d\lambda)^2-1}\,R_{\bG}\frac{A_{\bG}^{(s)}\bsigma^\star}{(d\lambda)^s}-W^{(s)}\Bigr\|
    \le\frac{1+|d\lambda|}{(d\lambda)^2-1}\,\bigl\|W^{(s)}-W^{(s-1)}\bigr\|\,.
\]
By \Cref{lem:bound-consecutive-scores}, with high probability $\|W^{(s)}-W^{(s-1)}\|\le\bigl(O_{d,\lambda}((d\lambda^2)^{-s/2})+o(1)\bigr)\sqrt n$, which is at most $\varepsilon\sqrt n\cdot((d\lambda)^2-1)/(1+|d\lambda|)$ once $s$ is large.
\end{proof}

\begin{proof}[Proof of \Cref{lem:post-edit-stability}]
We first establish the spectral norm bound in the ``moreover''.
Let $Q$ be any graph on $[n]$ with maximum degree $J$; let $x \in \R^n$.
Then
\[
\sum_{u \sim v} (R_Q x)_{u \to v}^2 \leq O_{d,\lambda}(1) \cdot \sum_{u \sim v} (x_u^2 + x_v^2) \leq O_{d,\lambda}(J) \|x\|^2 \, .
\]

Now we turn to the main statement of the lemma.
Henceforth in this proof let $x = A_{\bG}^{(s)} \bsigma^* / (d \lambda)^s$, so our goal is to bound $\|R_{F_J} x - R_{\bG} x\|$.
For any edge $u,v$ in both $\bG$ and $F_J$ the vector $(R_{F_J} x - R_{\bG} x)_{u \to v}$ is zero.
So
\[
    \|R_{F_J}x-R_{\bG}x\|^2\le O_{d,\lambda}(1)\sum_{\{u,v\}\in E(F_J)\triangle E(\bG)}(x_u^2+x_v^2)\,.
\]
Let $m(v)$ be the number of edges of $E(F_J)\triangle E(\bG)$ at $v$.
By Cauchy--Schwarz,
\[
    \sum_{\{u,v\}\in E(F_J)\triangle E(\bG)}(x_u^2+x_v^2)=\sum_{v\in[n]}m(v)\,x_v^2\le\Bigl(\sum_{v\in[n]}m(v)^2\Bigr)^{1/2}\|x\|_4^2\,.
\]
The vector $x$ is the local statistic $\bsigma^{(s)}$ of \Cref{sec:bethe-hessian-negative-spectrum}, so by \Cref{lem:bh-transfer} and \Cref{lem:bh-tree}(iii), with high probability $\|x\|_4^4\le Cn$ for a constant $C=C(d,\lambda)$ not depending on $s$.
So we just need to bound $\sum_{v \in [n]} m(v)^2$ with high probability.

Since $E(\bG_J)\subseteq E(\bG)$, every edge of $E(F_J)\triangle E(\bG)$ lies in $E(\bG)\setminus E(\bG_J)$ or in $E(F_J)\triangle E(\bG_J)$.
So $m(v)\le g_J(v)+m'(v)$, where $g_J(v)$ is the number of edges of $E(\bG)\setminus E(\bG_J)$ at $v$ and $m'(v)$ is the number of edges of $E(F_J)\triangle E(\bG_J)$ at $v$, and
\[
    \sum_{v\in[n]}m(v)^2\le2\sum_{v\in[n]}g_J(v)^2+2\sum_{v\in[n]}m'(v)^2\,.
\]
The first sum is the error from the degree cutoff, and the second is the error from the corruptions.

For the first sum, $g_J(v)$ is the number of edges at $v$ touching a vertex of degree more than $J$, so $g_J^2$ is a local statistic and by \Cref{lem:bh-transfer}, $\frac1n\sum_{v\in[n]}g_J(v)^2\to\|g_J\|_{L^2}^2$.
On the broadcast tree, $g_J(o)\le\mathrm{deg}(o)$ and $g_J(o)\to0$ as $J\to\infty$, so $\|g_J\|_{L^2}\to0$.

For the second sum, both $F_J$ and $\bG_J$ have maximum degree at most $J$, so $m'(v)\le2J$ and
\[
    \sum_{v\in[n]}m'(v)^2\le2J\sum_{v\in[n]}m'(v)=4J\,|E(F_J)\triangle E(\bG_J)|\,.
\]
We claim that $|E(F_J) \triangle E(\bG_J)| \leq O(J \eta n)$.
An edge of $E(\bG_J)\setminus E(F_J)$ was either removed by the adversary, or it is still in $F$ but has an endpoint whose degree the adversary pushed above $J$.
There are at most $2\eta n$ endpoints of corrupted edges, and each has at most $J$ edges in $\bG_J$, so $|E(\bG_J)\setminus E(F_J)|\le(2J+1)\eta n$.
Symmetrically, $|E(F_J)\setminus E(\bG_J)|\le(2J+1)\eta n$.

Putting these together, with high probability, simultaneously for every $F$ with $|E(F)\triangle E(\bG)|\le\eta n$,
\[
    \frac1n\sum_{v\in[n]}m(v)^2\le4\|g_J\|_{L^2}^2+O(J^2\eta)\,.
\]
Choosing $J$ large enough and then $\eta$ small enough makes this as small as desired, which concludes the proof.
\end{proof}


\section{AI methodology}

In May 2026, the authors proved, without any AI assistance, that initializing belief propagation at the second eigenvector succeeds at optimal recovery for any set of parameters above the KS threshold, along with a robust algorithm based on the Bethe Hessian matrix to estimate the second eigenvector in the presence of adversarial corruptions when the degree $d\to\infty$ sufficiently above the KS threshold.

In July 2026, the authors extensively interacted with GPT-5.6 Sol to upgrade the robust algorithm to estimate the second eigenvector for all $d > 1$ and all SNR above the KS threshold based on a degree-truncated nonbacktracking matrix power.
The analysis was highly technical, and required proving spectral norm estimates on asymmetric random matrices with correlated entries.
To abet verification, the authors used GPT-5.6 Sol to produce a Lean certificate.

However, this paper ultimately did not use that proof in favor of a shorter and more human-verifiable proof---in September 2026, the authors fed their earlier research notes to GPT-6 Astra with instructions to upgrade the Bethe Hessian-based algorithm to all parameter regimes, and in a single shot it produced a robust algorithm for second eigenvector estimation based on a suitably ``corrected'' Bethe Hessian matrix presented here.

The authors interacted extensively with frontier models (GPT-6 Astra, GPT-6.1 Sol, Opus 5.5, and Fable 5.1) to streamline proofs by finding simplifications, and to identify relevant literature containing quantitative estimates used in our proofs.
These models were used as writing and editing aids in generating this manuscript.


\section*{Acknowledgements}
The authors thank Prasad Raghavendra for many conversations about robust community estimation.

\clearpage
\phantomsection
\addcontentsline{toc}{section}{References}
\begingroup
\small
\bibliographystyle{alpha}
\bibliography{refs}
\endgroup

\clearpage

\crefalias{section}{appendix}

\appendix
\begingroup
\renewcommand*{\theHtheorem}{approxeigen.\arabic{theorem}}
\renewcommand*{\theHlemma}{approxeigen.\arabic{theorem}}
\renewcommand*{\theHfact}{approxeigen.\arabic{theorem}}
\renewcommand*{\theHcorollary}{approxeigen.\arabic{theorem}}

\section{Proof of \texorpdfstring{\Cref{lem:bh-approx-eigen}}{Lemma 6.9}}
\label{app:bh-approx-eigen}

In this section, we establish \Cref{lem:bh-approx-eigen}.

\lemsigmasapproxeigenvecofCs*

Denote $\rho = d\lambda^2 > 1$. We start by proving (a) and (b).

\begin{proof}[Proof of (a,b)]
    We begin with $s'=s$. For succinctness, denote $\mu=d\lambda$.
    For a vertex $v$, denote by $Z_t(v)$ the number of descendants of $v$ that are $t$ levels lower, and by $S_t(v)$ the sum of spins of descendants of $v$ that are $t$ levels lower.
    Let $\mathcal F_j$ reveal the tree and its spins through depth $j$.

    For simplicity, denote
    \[
        Q_s=((C^{(s)})^2\bx)(o)
        \quad\text{and}\quad
        R_s=(C^{(s)}(D-I)C^{(s)}\bx)(o) \mcom
    \]
    Let us denote
    \[
        M_t = (C^{(t)}\bx)(o) = \mu^{-t}S_t(o),
        \qquad
        F_s = \sum_{j=0}^{s-1} \rho^{-j} M_{s-j}.
    \]
    We shall show that
    \begin{align}
        \E(Q_s-F_s)^2
            &\lesssim\rho^{-s},
            \label{eq:qs-fs-bound}\\
        \E(R_s-dF_s-M_s)^2
            &\lesssim\rho^{-s},
            \label{eq:rs-fs-ys-bound}\\
        \E(F_s-cM_s)^2
            &\lesssim\rho^{-s}.
            \label{eq:fs-ys-bound}
    \end{align}
    Given this, (a) and (b) for $s'=s$ follow immediately.
    
    The proof of the above bounds is relatively simple---the expression for $((C^{(s)})^2\bx)(o)$ involves a sum over vertices reachable from $o$ by the concatenation of two nonbacktracking paths, each of length $s$.
    The depth at which a given concatenation ends depends solely on the size of the intersection of these two paths, that is, the part that is ``retraced.''

    Let us focus on \eqref{eq:qs-fs-bound}.
    Summing over the length of the retraced part of the path,
    \begin{equation}
        \mu^{2s}Q_s
            =S_{2s}(o)
             +\sum_{j=1}^{s-1}\sum_{|z|=s-j}
                    L_{j-1,s-j-1}(z)
             +Z_s(o)\bx_o,
        \label{eq:qs-exact}
    \end{equation}
    where $|z|=\dist(o,z)$.
    
    The martingale argument from the proof of \Cref{lem:bh-shells}, gives, for $b \ge a$,
    \begin{equation}
        \E(M_b-M_a)^2
            =\sum_{k=a+1}^{b}\rho^{-k}
            \le\frac{\rho^{-a}}{\rho-1},
        \qquad
        \sup_t\E M_t^2\le c.
        \label{eq:bh-martingale-tail}
    \end{equation}
    This allows us to replace $S_{2s}(o)$ with $\mu^{s} S_{s}(o)$.

    Simplifying the other terms requires calculations that have essentially already been performed in \cite{BLM15}---let us describe the upshot thereof.
    Denote
    \[
        L_{p,q}(z)
            =\sum_{\substack{a,b\text{ children of }z\\a\ne b}}
                  Z_p(a)S_q(b).
    \]

    The proof of \cite[Theorem~25]{BLM15} gives
    \begin{align*}
        \E[L_{p,q}(z)\mid\bx_z]
            &=d^{p+1}\mu^{q+1}\bx_z,\\
        \E[L_{p,q}(z)^2\mid\bx_z]
            &\lesssim d^{2p}\mu^{2q}.
    \end{align*}
    It follows that, for $1\le j<s$,
    \[
        \E\left[
            \mu^{-2s}\sum_{|z|=s-j}L_{j-1,s-j-1}(z)
            \,\middle|\,\mathcal F_{s-j}\right]
            =\mu^{-2s}d^j\mu^{s-j}S_{s-j}(o)
            =\rho^{-j}M_{s-j}.
    \]
    On the other hand, using independence of the subtrees of
    different vertices at the same depth, we may replace the
    conditional variance of the sum with the sum of the conditional
    variances. That is,
    \begin{align*}
        &\E\left(
            \mu^{-2s}\sum_{|z|=s-j}L_{j-1,s-j-1}(z)
            -\rho^{-j}M_{s-j}\right)^2\\
        &\qquad
            =\mu^{-4s}\E\left[
                \operatorname{Var}\left(
                    \sum_{|z|=s-j}L_{j-1,s-j-1}(z)
                    \,\middle|\,\mathcal F_{s-j}\right)\right]\\
        &\qquad
            \lesssim\mu^{-4s}d^{s-j}d^{2j}\mu^{2(s-j)}
            =\rho^{-(s+j)}.
    \end{align*}
    We also have $\E(\mu^{-2s}Z_s(o))^2\lesssim\rho^{-2s}$.
    Taking square roots of the variance bounds and summing over
    $j$ gives a total error
    \[
        O\left(\sum_{j=1}^{s-1}\rho^{-(s+j)/2}\right)
            =O(\rho^{-s/2}).
    \]
    Substituting these bounds in \eqref{eq:qs-exact} allows us
    to conclude that
    \[
        \E\left(
            Q_s-\underbrace{\sum_{j=0}^{s-1}
                               \rho^{-j}M_{s-j}}_{F_s}
            \right)^2
            \lesssim\rho^{-s},
    \]
    as desired.

    The second bound \eqref{eq:rs-fs-ys-bound} is very similar.
    Indeed, observing that $Z_1(u)S_s(u)=S_s(u)+L_{0,s-1}(u)$,
    \begin{align*}
        \mu^{2s}R_s
            &=\sum_{|u|=s}Z_1(u)S_s(u)
              +\sum_{j=1}^{s-1}\sum_{|z|=s-j}
                    L_{j,s-j-1}(z)
              +Z_{s+1}(o)\bx_o\\
            &=S_{2s}(o)
              +\sum_{j=0}^{s-1}\sum_{|z|=s-j}
                    L_{j,s-j-1}(z)
              +Z_{s+1}(o)\bx_o.
    \end{align*}
    We may therefore perform essentially the same manipulations to prove \eqref{eq:rs-fs-ys-bound}. We omit the details.

    Finally, let us prove \eqref{eq:fs-ys-bound}.
    By \eqref{eq:bh-martingale-tail} and the triangle inequality,
    \begin{align*}
        \E(F_s-cM_s)^2
            &=\E\left(
                \sum_{j=0}^{s-1}\rho^{-j}(M_{s-j}-M_s)
                -\frac{\rho^{-s}}{1-\rho^{-1}}M_s
                \right)^2\\
            &\lesssim
                \left(
                    \sum_{j=0}^{s-1}\rho^{-j}
                        \sqrt{\frac{\rho^{-(s-j)}}{\rho-1}}
                    +\rho^{-s}
                \right)^2
             \lesssim\rho^{-s}.
    \end{align*}
    This proves (a) and (b) for $s'=s$.

    When $s'\in\{s-1,s+1\}$, by \Cref{lem:bh-shells}, $\|\bsigma^{(s')}-\bsigma^{(s)}\|\to0$.
    Using the Cauchy--Schwarz inequality allows us to transfer this estimate over to conclude that $\E \left\| C^{(s)} \left( \bsigma^{(s')} - \bsigma^{(s)} \right) \right\|^2 \rightarrow 0$ and $\E \left\| C^{(s)} (D-I) \left( \bsigma^{(s')} - \bsigma^{(s)} \right) \right\|^2 \rightarrow 0$. Combining this with the case $s'=s$ proves (a) and (b).
\end{proof}

It remains to prove (c). This will require us to prove the following lemma, which yields a loose $O(1)$ bound when looking at $P\bx$ for $P$ being not just $C^{(s)}$, but the products of many such matrices.

\begin{lemma}
    \label{lem:poisson-sphere-moments}
    Let $P$ be a product of $p$  matrices, each of which is either one of the $\{C^{(j)} : j \ge 0\}$ or one of $\{D , D - I\}$.
    Then, there is a constant $K = K_{p,d,\lambda} < \infty$ such that $\|P\bx\| \le K$.
\end{lemma}

\begin{proof}
    We are interested in $\E\left[ (P\bx)(o)^2 \right]$.
    For starters, we may assume that all the matrices are either $D - \Id$ or one of the $\{C^{(j)}\}$, by replacing any occurrences of $D$ with $(D-\Id) + \Id$.
    Let us understand what $(P\bx)(o)$ looks like.
    For example, consider $(C^{(2)} \cdot (D - \Id) \cdot C^{(3)} \bx)(o)$. We can then express this as a summation over concatenations of two non-backtracking paths, the first of length $2$ and the second of length $3$. For each such concatenation, the summand is the product of the spin at the terminal vertex with the number of children of the intermediate vertex.
    
    More generally, $(P\bx)(o)$ is a summation over concatenations of non-backtracking paths of some prescribed length, with each summand being the products of the terminal spin, with the number of children of some subcollection of intermediate vertex.
    
    When looking at $(P\bx)(o)^2$, we multiply two such terms. Since each degree term appears a constant number of times (with the constant depending on $p$), and the marginal degree distribution of each vertex is $\Poi(d)$, we can essentially replace the contributions of each of these terms with an $O_{p}(d)$, resulting in at most a $d^{O(p)}$ term that we may absorb into $K$---in other words, we may in fact assume that all the matrices in $P$ are one of the $\{C^{(j)}\}$. On the other hand, the expectation of the product of two spins at distinct vertices $v,w$ is given by $\lambda^{\dist(v,w)}$.

    Let us carry out this calculation for $P = C^{(s)}$. The two walks involved in the expectation are then length $s$ walks down the tree having some intersection size $j$ (for some parameter $j$ that we will eventually sum over)---in expectation, there are $d^{s} \cdot d^{s-j}$ such pairs. The normalization of $C^{(s)}$ adds an additional factor of $(d\lambda)^{-2s}$. Finally, the two endpoints being distance $2(s-j)$ apart results in a $\lambda^{2s-2j}$ multiplicative term. That is, $\E (P\bx)(o)^2$ is, up to a constant, $\sum_{j \ge 0} d^{2s-j} \cdot \lambda^{2s-2j} \cdot (d\lambda)^{-2s} = \sum_{j \ge 0} \rho^{-j} < \infty$. Such calculations are very similar to those executed in \cite{BLM15}.

    The general computation when $P = \prod_{i=1}^{p} C^{(s_i)}$ is very similar. In this case, we can sum over $(r_j)_{1 \le j \le 2p}$, where $r_i$ is the number of edges in the $i$th walk that are contained in the part of the tree that have already been ``exposed'' so far by the previous walks. The number of such $p$-tuples of vertices is then $d^{2 \sum_{i=1}^{p} s_i - \sum_{j=1}^{2p} r_j}$. The normalization of the $C^{(s_i)}$ yields an additional $(\lambda d)^{- 2 \sum_{i=1}^{p} s_i}$ term. Finally, the spin correlation can be bounded from above by $\lambda^{2 \sum_{i=1}^{p} s_i - 2 \sum_{j=1}^{2p} r_j}$. Summing over $(r_j)$ yields
    \begin{align*}
        \E (P\bx)(o)^2 &\le \sum_{(r_j)} d^{2 \sum_{i=1}^{p} s_i - \sum_{j=1}^{2p} r_j} \cdot (\lambda d)^{- 2 \sum_{i=1}^{p} s_i} \cdot \lambda^{2 \sum_{i=1}^{p} s_i - 2 \sum_{j=1}^{2p} r_j} \\
            &= \sum_{(r_j)} \rho^{-\sum_{j=1}^{2p} r_j} = \left(\sum_{r \ge 0} \rho^{-r}\right)^{2p} < \infty \mper
    \end{align*}
    We omit the details of the proof when $P$ involves $D-I$ terms---as mentioned earlier, this only results in an additional $d^{O(p)}$ multiplicative term.
\end{proof}

Given the above, proving (c) is relatively easy.

\begin{proof}[Proof of \Cref{lem:bh-approx-eigen}(c)]
    Set
    \[
        e_s = C^{(s)}\bsigma^{(s)}-c\bsigma^{(s)} = \bigl((C^{(s)})^2-cC^{(s)}\bigr)\bx.
    \]
    By (a), $\|e_s\|\to0$.
    By \Cref{lem:bh-transfer}, the operators $C^{(j)}$, $D$, and $H$ are self-adjoint. As a result,
    \[
        \|C^{(s)}He_s\|^2
            =\iprod{e_s,H(C^{(s)})^2He_s}
            \le \|e_s\| \cdot \|H(C^{(s)})^2He_s\|.
    \]
    Note that the second vector is some bounded linear combinations of vectors $Px$, where $P$ is a product of finitely many terms of the form $C^{(j)}$, $D$, or $D-I$, exactly as in \Cref{lem:poisson-sphere-moments}.
    Therefore, we may bound the second term by $O(1)$, uniformly in $s$, and conclude that
    \[ \|C^{(s)}He_s\|^2 \lesssim \|e_s\|\to 0 \mper \qedhere \]
\end{proof}

\endgroup

\section{Deferred proofs for fast implementation}
\label{app:sdp-fast-deferred}


In this appendix, we prove multiple lemmas used in the SDP implementation: \Cref{thm:sdp-seven-entropy,lem:sdp-seven-exponential-action,lem:sdp-seven-exponential-density-sketch,lem:sdp-seven-oracle}.


\thmsdpsevenentropy*

The following proof is adapted from the matrix multiplicative-weights potential argument of Arora, Hazan, and Kale~\cite[Section~5]{AHK12}.

\begin{proof}
Put $R=\max_t\|H_t\|_{\mathrm{op}}$, $a_t=\|H_t\|_{\mathrm{op}}$, and
$K_t=H_t+a_tI$, so $0\preceq K_t\preceq2a_tI$.
The scalar inequality $e^{-u}\le1-u+u^2/2$ for $u\ge0$ gives
$e^{-\eta K_t}\preceq I-\eta K_t+\eta^2K_t^2/2$.
Golden--Thompson and $\log x\le x-1$ give
\[
 \begin{aligned}
 \log\operatorname{Tr}e^{S_t-\eta H_t}-\log\operatorname{Tr}e^{S_t}
 &\le\eta a_t+\log\operatorname{Tr}(X_te^{-\eta K_t})\\
 &\le\eta a_t-\eta\operatorname{Tr}(X_tK_t)+\tfrac{\eta^2}{2}\operatorname{Tr}(X_tK_t^2)
 \le-\eta\langle H_t,X_t\rangle+2\eta^2a_t^2.
 \end{aligned}
\]
For every $U\in\mathcal S$ and symmetric $B$,
$\log\operatorname{Tr}e^B\ge\langle B,U\rangle$. Telescoping from $S_1=0$
therefore gives
\[
 \frac1T\sum_t\langle H_t,X_t-U\rangle
 \le\frac{\log n}{\eta T}+2\eta R^2.
\]
The support inequality at $\widetilde X_t$ is
\[
 F(\widetilde X_t)-F(U)\le\langle H_t,\widetilde X_t-U\rangle
 =\langle H_t,X_t-U\rangle+\langle H_t,\widetilde X_t-X_t\rangle.
\]
Sum, apply convexity at $\bar X$, and choose a minimizer $U$.
\end{proof}

\lemsdpsevenexponentialaction*

For a related Taylor-expansion tail estimate, see \cite[Proof of Lemma~4.2, p.~151]{SV13}.

\begin{proof}
\label{proof:sdp-fast-deferred-exponential}
Set $\delta=\alpha/3$ and $p(x)=\sum_{j=0}^{m-1}x^j/j!$.
If $L=0$, then $S=0$ and $Q=I$, so the sandwich is immediate.
For $L>0$ and $|x|\le L/2$, the scalar Taylor remainder gives
\[
 \frac{|e^x-p(x)|}{e^x}
 \le e^L\frac{(L/2)^m}{m!}
 \le e^L\left(\frac{eL}{2m}\right)^m
 \le e^{L-m}\le\delta,
\]
using $m!\ge(m/e)^m$, $m\ge8L$, $e/16<e^{-1}$, and
$m\ge L+\log(1/\delta)$.
By orthogonal diagonalization of $S$, the scalar bounds
$1-\delta\le p(x)/e^x\le1+\delta$ imply
\[
 (1-\delta)^2e^S\preceq Q^2\preceq(1+\delta)^2e^S.
\]
Since $0<\alpha<1$, $(1-\delta)^2\ge1-\alpha$ and
$(1+\delta)^2\le1+\alpha$, proving the claimed L\"owner-order sandwich.
Finally, compute $w_0=u$, $w_j=Sw_{j-1}/(2j)$ for $1\le j<m$,
and sum $Qu=\sum_{j=0}^{m-1}w_j$.
This costs $O(m(T_S+n))$ exact-real operations, as claimed.
\end{proof}

\lemsdpsevenexponentialdensitysketch*

\begin{proof}
Set $\delta=\alpha/4$ and $k=\lceil32/\alpha^2\rceil$.
Take $Q$ from \Cref{lem:sdp-seven-exponential-action} at accuracy
$\delta$, draw independent uniform signs $G\in\{-1,1\}^{n\times k}$,
and return $\mathbf Z=QG/\|QG\|_F$.
Write $B=Q^2$, $a=\operatorname{Tr}B$, $W=QGG^TQ/k$,
$b=\operatorname{Tr}W$, $C=B/a$, and
$Y=\mathbf Z\mathbf Z^T=W/b=\widetilde X$.
The L\"owner-order sandwich makes $Q$ invertible, so $\mathbf Z\ne0$ and $b>0$.
For fixed $A,P$, put $h=\|A\|_{\mathrm{op}}+P$ and
\[
 N(D)=|\langle A,D\rangle|+P\|\operatorname{diag}D\|_1
 =\sup_{\|d\|_\infty\le P}|\langle A+\operatorname{Diag}(d),D\rangle|.
\]
The equality holds pointwise, allowing $d$ to depend on $G$.
With $r=\sqrt{2/k}$, \Cref{lem:sdp-seven-diagonal}, applied with
$V=Q$ and summed over the coordinate basis, gives
$\mathbb E\|\operatorname{diag}(W-B)\|_1\le ra$ and
$\mathbb E|b-a|\le ra$.
Summing the same estimate over a fixed orthonormal eigenbasis of $A$
with weights equal to its absolute eigenvalues gives
$\mathbb E|\langle A,W-B\rangle|\le ra\|A\|_{\mathrm{op}}$.
Since $Y$ is a density matrix, $N(Y)\le h$; hence the identity
\[
 Y-C=(W-B)/a+(a-b)Y/a
\]
yields $\mathbb E N(Y-C)\le2rh$.
Put $E=e^S$ and $t=\operatorname{Tr}E$.
The sandwich and simultaneous diagonalization of $B,E$ give
$\|B-E\|_*\le\delta t$ and $|a-t|\le\delta t$.
Thus $C-X=(B-E)/t+(t-a)C/t$ implies
$\|C-X\|_*\le2\delta$, and consequently $N(C-X)\le2\delta h$.
Therefore $\mathbb E N(Y-X)\le2(r+\delta)h\le\alpha h$.
Finally, the degree bound in \Cref{lem:sdp-seven-exponential-action}
at accuracy $\delta$ gives the claimed product and additional-operation
counts for the $k$ columns, with $O(nk)$ further operations to compute
$\|QG\|_F$ and scale $QG$; neither $a$ nor $t$ is computed.
\end{proof}

\lemsdpsevenoracle*

\begin{proof}
    If $p \in \mathcal D_K$, return $y=0$.
    Otherwise find $t>0$ satisfying that, if we truncate coordinates of $p$ to magnitude at most $t$, the resulting vector $p^*$ has squared $\ell_2$ norm exactly $K/n$, i.e.,
    \[
        p_i^*=\operatorname{sgn}(p_i)\min\{|p_i|,t\}, \qquad
        \sum_i (p_i^*)^2=K/n.
    \]
    Let $y = p^*/t$, we will show that $y\in\partial f(p)$.

    Consider arbitrary $z\in\mathbb R^n$ and its closest point $z^*\in \mathcal D_K$.
    By the definition of $y$, we have $y^{\top} p^* \geq y^{\top} z^*$ and $\|y\|_{\infty} \leq 1$.
    Therefore, it follows that
    \[
        y^{\top} (z-p^*)
        \leq y^{\top} (z-z^*)
        \leq \|z-z^*\|_1
        = f(z).
    \]
    Since $y_i = \operatorname{sgn}(p_i)$ for $|p_i|>t$ and $p-p^*$ is supported on those coordinates,
    we have $y^{\top} (p-p^*) = \|p-p^*\|_1$.
    Hence,
    \[
        f(z)
        \geq y^{\top} (z-p^*)
        = y^{\top} (z-p) + y^{\top} (p-p^*)
        \geq y^{\top} (z-p) + \|p-p^*\|_1
        \geq y^{\top} (z-p) + f(p).
    \]
    To find $t$, we sort the coordinate magnitudes of $p$ and scan the sorted list to locate the clipping threshold.
    Sorting takes $O(n\log n)$ time, and the scan takes $O(n)$ time.
\end{proof}


\end{document}